\documentclass[12pt,preprint]{aastex}

\usepackage{epsfig}
\usepackage{subfig}
\usepackage{hyperref}
\hypersetup{
 colorlinks=true,
 pdfborder={0 0 0},
 linkcolor=cyan,
 citecolor=blue,
}

\def\eg{{\em e.g.\ }}

\newcommand{\zz}[1]{\texttt{BzZ{#1}}}
\newcommand{\zzw}[1]{\texttt{BzZ{#1}W}}
\newcommand{\zn}[1]{\texttt{BzN{#1}}}
\newcommand{\pn}[1]{\texttt{BpN{#1}}}
\newcommand{\zzr}[2]{\texttt{BzZ{#1}{#2}}}

\newcommand{\indeg}[1]{{#1}^{\circ}}
\newcommand{\ds}{\displaystyle}
\newcommand{\myfrac}[2]{\frac{\ds {#1}}{\ds {#2}}}

\def\div{{\nabla \cdot}}

\def\vv{{\mathbf{v}}}

\def\vb{{\mathbf{B}}}
\def\vFc{{\mathbf{F_{c}}}}
\def\vI{{\mathbf{I}}}
\def\dt{{\partial _{t}}}
\def\dphi{{\partial _{\phi}}}
\newcommand{\pd}[2]{\frac{\partial {#1}}{\partial {#2}}}

\newcommand{\citeinp}[1]{\citeauthor{#1} \citeyear{#1}}



\begin{document}

\title{Global Simulations of Accretion Disks I: Convergence and Comparisons with Local Models}

\author{Kareem A. Sorathia\altaffilmark{1,3}, Christopher~S.~Reynolds\altaffilmark{2,3}, James~M.~Stone\altaffilmark{4}, Kris~Beckwith\altaffilmark{5}}

\altaffiltext{1}{Department of Mathematics, University of Maryland, College Park, MD 20742-2421} 
\altaffiltext{2}{Department of Astronomy and the Maryland Astronomy Center for Theory and Computation, University of Maryland, College Park, MD 20742-2421} 
\altaffiltext{3}{Joint Space Science Institute (JSI), University of Maryland, College Park, MD 20742-2421}
\altaffiltext{4}{Department of Astrophysical Sciences, Princeton University, Princeton, NJ 08544}
\altaffiltext{5}{JILA, University of Colorado at Boulder, CO 80309-0440}

\begin{abstract}
Grid-based magnetohydrodynamic (MHD) simulations have proven invaluable for the study of astrophysical accretion disks.  However, the fact that angular momentum transport in disks is mediated by MHD turbulence (with structure down to very small scales) raises the concern that the properties of the modeled accretion disks are affected by the finite numerical resolution of the simulation.  By implementing an orbital advection algorithm into the Athena code in cylindrical geometry, we have performed a set of global (but unstratified) Newtonian disk simulations extending up to resolutions previously unattained.    We study the convergence of these models as a function of spatial resolution and initial magnetic field geometry.  The usual viscosity parameter ($\alpha$) or the ratio of thermal-to-magnetic pressure ($\beta$) are found to be poor diagnostics of convergence, whereas the average tilt angle of the magnetic field in the $(r,\phi)$-plane is a very good diagnostic of convergence.  We suggest that this is related to the saturation of the MHD turbulence via parasitic modes of the magnetorotational instability.  Even in the case of zero-net magnetic flux, we conclude that our highest resolution simulations (with 32-zones and 64-zones per vertical scale height) have achieved convergence.

Our global simulations reach resolutions comparable to those used in local, shearing box models of MHD disk turbulence.  We find that the saturation predictors derived from local simulations correspond well to the instantaneous correlations between local flux and stress found in our global simulations.  However, the conservation of magnetic flux implicit in local models is not realized in our global disks.  Thus, the magnetic connectivity of an accretion disk represents physics that is truly global and cannot be captured in any ab-initio local model.
\end{abstract}


\section{Introduction}

Current understanding of angular momentum transport in accretion disks has been revolutionized by the discovery of the magnetorotational instability (MRI; \citeinp{mri1}, \citeinp{bh98}).  \citet{mri1} demonstrated that while differentially-rotating Keplerian systems are hydrodynamically stable, the inclusion of a weak magnetic field introduces a rapidly-growing instability that leads to turbulence.  The resulting turbulence is highly anisotropic and this anisotropy transports angular momentum radially outwards which allows material to fall inwards.  While analytic treatments are able to probe the behavior of the linear instability and some aspects of the early non-linear behavior (\citeinp{gx94}), fully turbulent plasmas, however, are too far into the non-linear regime for analytic methods to be feasible.  As such, computational work is at the forefront of efforts to understand MRI-driven magnetohydrodynamic (MHD) turbulence in accretion disks.

Broadly speaking, accretion disk simulations can be divided into two types: local and global.  In local, or shearing-box (\citeinp{hgb}, \citeinp{brandenburg95}), models, a small patch of an accretion flow is simulated by solving the MHD equations with linearized shear terms in a small Cartesian system in the rotating frame.  Global models, by contrast, involve simulating a much larger region of the disk and include curvature terms associated with cylindrical or spherical coordinates.  The reduced cost of local, or shearing-box simulations, makes them an ideal model to explore the large parameter space of turbulence in accretion disks over very long timeframes.  While current supercomputers are capable of fully global accretion disk simulations, local models have the benefit of exploring resolutions far in excess of any comparable global model.  Consequently, much of our knowledge of accretion disk turbulence comes from local models.  

Understanding to what extent knowledge of accretion turbulence gained from local models can be carried over to global disks is of vital importance and will be a focus of this work.  In \citet{sra10} we began an exploration of these issues, in particular focusing on the idea of using volume decompositions of global models to make direct comparisons to local ones.  That local models with a net magnetic flux exhibit stress that scales with increasing flux is well known (\citeinp{hgb}, \citeinp{sits04}, \citeinp{pcp07}), however it was demonstrated in \citet{sra10} that saturation predictors for local models carry over to an instantaneous correlation between flux and stress in small patches of global models.        
 
Numerical simulations are necessary to study accretion disk turbulence, but they are no panacea.  Choices ones makes regarding the solution of the relevant equations and initial conditions can leave unintended and unphysical artifacts in the results.  Understanding how to disentangle numerical artifacts from genuine physical phenomenon is vital to constructing simulations that can make reliable predictions.  In particular, the discretization of the physical simulation domain, or resolution, will be of interest to us here.  Concern was raised when it was discovered by \citet{fp07} that local unstratified simulations without explicit dissipation, and without a net magnetic flux threading the domain, result in angular momentum transport that vanishes with increasing resolution.  This lack of convergence is not robust as the inclusion of stratification, explicit dissipation, or a net magnetic field lead to a converged value of momentum transport (\citeinp{dsp10}, \citeinp{fromang10}).  Further, \citet{sra10} observe that, to the extent that local models are predicated on being representative of a small patch of a global disk, the constraint of enforcing a zero magnetic flux over a small physical domain is likely unphysical, albeit an important theoretical pathology.  These results highlight the importance of understanding the sometimes delicate nature of numerical convergence when dealing with simulations of turbulence.

The primary goal of this paper is to study accretion disk turbulence induced by a variety of different initial field topologies in the context of global simulations of higher resolution than have previously been considered.  To reduce the significant expense associated with global, three-dimensional MHD simulations we use an implementation of orbital advection (\citeinp{fargo}, \citeinp{jgg08}), which results in an \emph{order of magnitude speed-up} for the simulations considered here.  These simulations are meant to be comparable with the surveys performed for local models by \citet{hgb} and for global by \citet{hawley01}.  As we will see, global simulations analogous to the local models considered by \citet{fp07} do converge to non-zero angular momentum transport.  Additionally, we consider the resolution requirements to attain convergence as well as explore the question of what convergence means in relation to non-dissipative simulations of turbulent systems.  We continue the work started in \citet{sra10} and \citet{bas11} and explore connections between local and global accretion disk models in a more controlled setting utilizing simulations tailored to the task.    

The plan of the paper is as follows.  Section \S\ref{sec:md} describes the simulations and diagnostics used in the rest of the paper as well as a brief description of orbital advection.  Section \S\ref{sec:evol} compares the disk evolution under the influence of varying initial magnetic field topologies and discusses this evolution in the context of reduced anomalous viscosity models.  Section \S\ref{sec:conv} explores the use of a selection of convergence metrics: physical metrics (Section \S\ref{sec:convphys}); numerical metrics (Section \S\ref{sec:convnum}); and spectral metrics (Section \S\ref{sec:convspec}).  Section \S\ref{sec:convta} presents results suggesting that the magnetic tilt angle is a robust indicator of convergence that is invariant of initial magnetic topology.  A discussion of the convergence study with an emphasis on a comparison with the recent work of \citet{hgk11} is presented in \S\ref{sec:convdis}.  Finally, Section \S\ref{sec:lvg} presents the results of our comparisons between local and global models including the evolution, flux distribution, and instantaneous flux-stress relation in the local ensemble.  Our conclusions and a general discussion is presented in Section \S\ref{sec:con}, and an Appendix follows detailing orbital advection and its implementation.

\section{Methodology and Diagnostics}
\label{sec:md}

The work presented here is based on a series of simulations exploring the behavior of global accretion disks under a variety of resolutions and initial seed magnetic fields.  Our simulations model an unstratified, isothermal, relatively cold Keplerian disk in a Newtonian potential.  These simulations are run in the context of 3-d ideal isothermal magnetohydrodynamics; the equations are integrated using the cylindrical coordinate extension \citep{athcyl} to the Athena code package \citep{athena}.  The equations of isothermal  MHD are, for reference:

\begin{mathletters}
\label{eqn:mhd}
\begin{eqnarray}
\dt \rho + \div (\rho \vv) &= 0, \\
\dt (\rho \vv) + \div (\rho \vv \vv - \vb \vb + P^{*}\vI) &= -\rho \nabla \Phi , \\
P = c_{s}^{2}\rho ,\\
\dt \vb  - \nabla \times (\vv \times \vb) &= 0 .
\end{eqnarray}
\end{mathletters}

Equations \ref{eqn:mhd}a-\ref{eqn:mhd}d are written in conservative form, where $\rho$ is the gas density, $P$ is the gas pressure and is defined through an isothermal equation of state, $\vv$ is the velocity vector, and $\vb$ is the magnetic field vector.  The total pressure, $P^{*} = P+|\vb|^{2}/2$, is the sum of the gas and magnetic pressure.

The equations are solved in cylindrical coordinates, denoted by $(R,\phi,z)$, with $r=\sqrt{R^{2}+z^{2}}$ referring to the standard spherical radius.  The accretion disks we evolve are Newtonian and unstratified, as in \citet{armitage98} and \citet{hawley01}, meaning that the gravitational potential is independent of the $z$ coordinate and given by $\Phi = -1/R$.  Neglecting the vertical dependence of the gravitational potential physically means that our simulations are meant to model the midplane of a realistic disk.  This eliminates the effects of magnetic buoyancy and allows a more pure probe of MRI-driven phenomena.  Numerically, it allows us to remove the vertical variation of zones per scale-height without resorting to the use of a more complex graded mesh with poorly understood grid-scale dissipation.

As one of the goals of this study is to explore issues of convergence, a word about resolution is in order.  We classify simulations based on the number of zones per vertical scale height, specifically $H_{0} / \Delta z$.  Given a choice of zones per scale height, the resolution is constrained by the condition that the aspect ratio is $1:1:1$ at the fiducial point, $R_{0} = 2$.  This results in the condition $\Delta z = \Delta R \approx R_{0} \Delta \phi$.  The ability to explore simulations with an aspect ratio of unity is a particular benefit of orbital advection, summarized below and detailed in the Appendix, and stands in contrast to the vast majority of the current literature on simulations of global accretion disks.    

The expense of running these simulations using the standard cylindrical integrator would be extreme, for instance the highest resolution simulation considered would take well over 2M CPU-hours.  To reduce the expense, the first step of this work was the implementation of an orbital advection algorithm (\citeinp{fargo}, \citeinp{jgg08}).  The particular method is based on the algorithm and code for the implementation of orbital advection in the shearing-box framework (\citeinp{athfargo}) of Athena.  The essence of orbital advection is the observation that in accretion disk simulations the dominant speed is the Keplerian velocity.  Consequently, $\Delta t \sim \min \{\Delta \phi / \Omega_{K}\}$.  One can instead solve the equations of MHD in the local rotating frame, and in this frame (assuming a subthermal magnetic field) $\Delta t \sim \Delta \phi / c_{s}$.  One can approximate the speed-up of orbital advection by considering the maximum Keplerian Mach number of the disk, $M_{k} = R \Omega_{K}/c_{s}$.  For the disks considered here the Keplerian Mach number peaks at the inner edge, and attains a value of $M_{k}(R=1) = 20$.  This predicted speed-up is very nearly attained, with the deviation attributed to a supersonic Alfven speed during the saturation of the MRI and the small expense of a separate advection routine that evolves the disk in accordance with the background Keplerian flow.  Details of the implementation and testing of the orbital advection algorithm used here are presented in Appendix~\ref{sec:oa}.

\begin{figure}
  \centering
  \subfloat[Volume rendering from simulation \zzw{64}.]{\label{fig:globstilla} \includegraphics[width=0.6\textwidth]{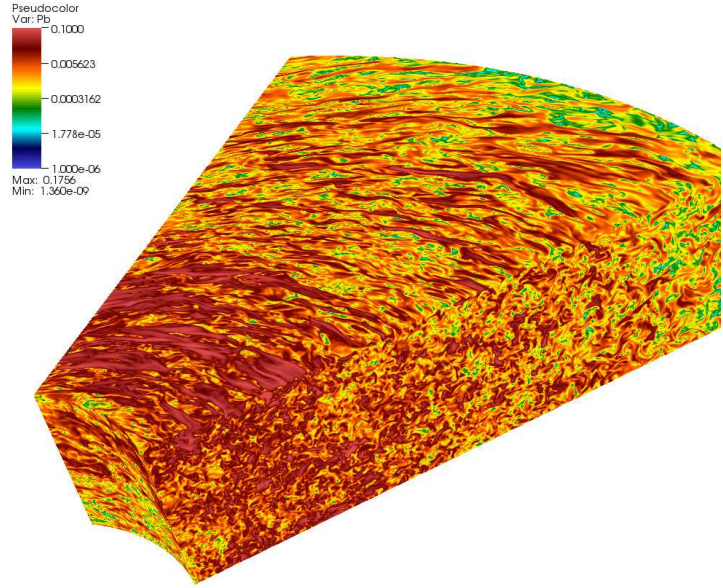}}\\                
  \subfloat[Plane $z=0$, from simulation \zz{32}.]{\label{fig:globstillb} \includegraphics[width=0.6\textwidth]{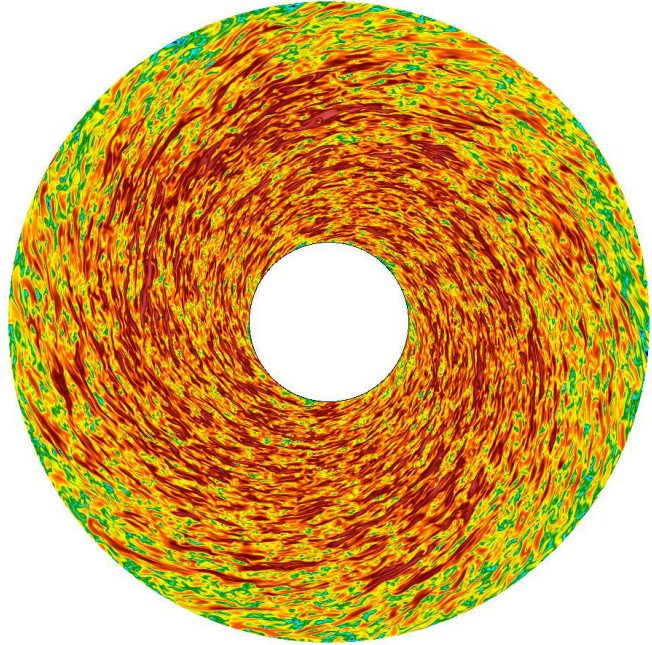}} 
  \caption{Still images of logarithmic magnetic pressure at orbit 75.  Both images use the same color table.}
  \label{fig:globstill}
\end{figure}

To give a sense of the simulations and their resolution, Figure~\ref{fig:globstill} shows two still images of the MRI-driven turbulence in the saturated state.  The first, Figure~\ref{fig:globstilla}, is a volume rendering of the logarithmic magnetic pressure in a simulation with a vertical resolution of $64$ cells per vertical scale height, denoted simulation \zzw{64} in the terminology introduced below.  The second, Figure~\ref{fig:globstillb}, shows the same variable in the $z=0$ plane at the same timestep from a simulation with a vertical resolution of $32$ cells per vertical scale height, denoted simulation \zz{32}.  

In cylindrical coordinates the physical domain spans $(R,\phi,z) \in [1,4] \times [0,2\pi] \times [-2H_{0},2H_{0}]$, where $H_{0} = 0.2$, with the exception of two reduced runs that use a smaller azimuthal domain.  The sound speed is set so that, $H(R_{0}) = H_{0}$, where $R_{0} = 2$ and
\begin{equation}
H(R) = \sqrt{2} c_{s} \Omega_{K}^{-1},
\end{equation}          
with $\Omega_{K} = R^{-1.5}$.  

The hydrodynamic variables are initialized with a constant density, $\rho = 100$, and with velocities initialized as
{\begin{mathletters}
\label{hydroinit}
\begin{eqnarray}
v_{R} (R,\phi,z)= 0, \\
v_{\phi} (R,\phi,z) = R\Omega_{K}(1+\delta), \\
v_{z} (R,\phi,z) = \delta,
\end{eqnarray}
\end{mathletters}}
where $\delta$ is a uniformly-distributed random perturbation such that $\delta \in [-10^{-2},10^{-2}]$.  


To explore the effects of varying initial seed topologies, we consider here three distinct initial field configurations; a zero net flux vertical field (BzZ), a net-flux vertical field (BzN), and a net-flux azimuthal field (BpN).  For simplicity we will often distinguish between the zero net flux (ZF) simulations and the net-flux simulations (NF).  For the net-flux vertical field runs, the initial magnetic field is set such that the fastest-growing unstable mode of the MRI is equal to $H(R_0)$, i.e., $\lambda_{MRI} \equiv 2\pi \sqrt{16/15}  \:v_{A,z}/\Omega=H(R_0)$ where $v_{A,z} = |B_{z}|/\sqrt{\rho}$.  In our lowest-resolution runs, this gives a fastest growing mode that is marginally resolved, $\lambda_{MRI}/\Delta z = 8$.   For the zero-net flux runs, this vertical field configuration is modulated by a sinusoidal function.   The toroidal field runs are constructed to ensure that the critical azimuthal wavenumber of the toroidal MRI, $m_{c} = R\Omega/v_{A}$ (where $v_{A} = |\vb|/\sqrt{\rho}$), is constant in radius with a value of 20.  Finally, the field strengths are tapered to zero close to the radial boundaries of the calculation.   These choices of seed fields are not meant to be representative of what may be present in an astrophysical accretion disk, but are chosen to allow controlled experiments to study MRI-driven disk turbulence.  

\begin{figure}
\centering
\includegraphics[width=0.8\textwidth]{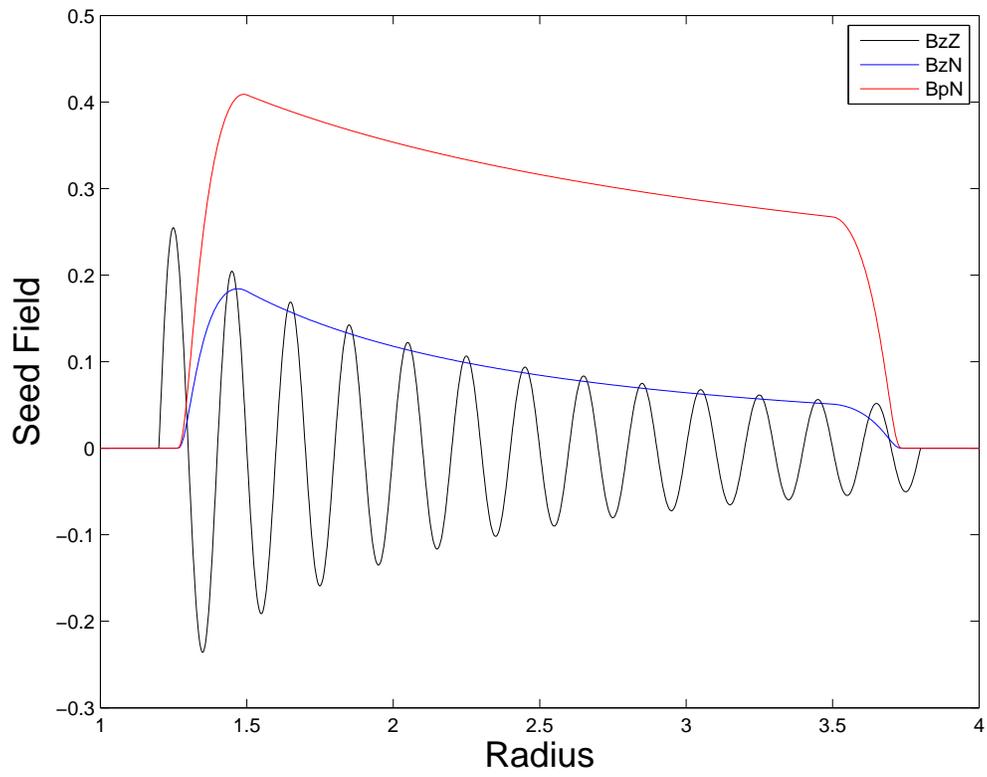}                
\caption{Radial profiles of initial magnetic field topologies.}
\label{fig:init}
\end{figure}

Formally, the fields for the three cases are given by
{\begin{mathletters}
\label{mhdinit}
\begin{eqnarray}
B_{z} (R)= A_{0}\myfrac{I(R)R_{*}\Omega_{K}(R_{*})}{R} \sin \left[ \myfrac{2\pi (R-R_{0})}{H_{0}} \right], B_r=B\phi=0\ {\rm (case\ BzZ)},\label{eqn:mhdinita}\\
B_{z} (R)= A_{N}S(R)\Omega_{K}(R), B_r=B\phi=0\ {\rm (case\ BzN)},\label{eqn:mhdinitb}\\
B_{\phi}(R) = \myfrac{S(R)\sqrt{\rho/R}}{M_{c}}, B_r=B_z=0\ {\rm (case\ BpN)}. \label{eqn:mhdinitc}
\end{eqnarray}
\end{mathletters}}
\noindent In the above definition, $A_{0}$ and $A_{N}$ represent scaling terms, $R_{*}$ is a piecewise constant function of radius that assumes the central radial value of each sinusoid, and $M_{c} = 20$.  The function $I(R)$ is an indicator function on the domain $R \in [1.2,3.8]$.  The function $S(R)$ is a mollifier function with value unity on the domain $R \in [1.5,3.5]$, zero on the complement  of the domain $R \in [1.5-H_{0},3.5+H_{0}]$, and smoothly interpolates between the two domains.   Figure~\ref{fig:init} shows the resulting radial profiles of the initial magnetic field strengths.

As the simulations described here are unstratified, periodic boundary conditions are appropriate in the azimuthal and vertical direction.  Radial boundary conditions are, as always, more difficult to implement appropriately.  The boundary conditions used in the simulations presented here are particularly simple: the values of physical quantities in the ghost zones are simply copied over from the last physical zone.  There are two exceptions to this: the radial velocity and the azimuthal velocity.  The radial velocity uses what we refer to as an enforced diode condition to ensure that material does not enter the simulation.  Formally, this means that if the radial velocity in the last physical zone is directed outwards then it is copied into the ghost values, otherwise the radial velocities in the ghost zone are set to zero.  For the azimuthal velocity a Dirichlet boundary condition is used in which the ghost zone values are set to the appropriate Keplerian value.  While some of the simulations described here are initialized to have a zero-net magnetic field, the boundary conditions are not constructed to enforce this flux constraint during the natural evolution of the disk.  To ensure that these choices of radial boundary condition do not affect the physical behavior of the simulations several potential boundary conditions were considered.  These include two choices for hydrodynamic variables and two choices for the magnetic field variables.  In addition to the simple copy with diode condition for the hydrodynamic variables used, a variant in which the perturbed azimuthal velocity (i.e., the azimuthal velocity in the local Keplerian frame) was held constant in the ghost zones with an analogous diode condition.  Instead of a simple copy of the magnetic field variables, an alternative in which the magnetic field in the ghost zone was constrained to only have a divergence-free radial component was tested.  These choices result in four possible radial boundary conditions, and all were tested to ensure that none resulted in an appreciable change in the evolution and behavior of the disk.  It is worth noting that replicating a magnetic field configuration from one cell to another in a curvilinear coordinate system will not, in general, guarantee that it is divergence free.  Indeed, the magnetic field in the ghost zones will not necessarily respect the solenoidal constraint.  However, the ghost zones are necessary merely for the reconstruction of the magnetic field inside the physical domain and the use of constrained transport will guarantee that the magnetic field in the simulation domain will satisfy the solenoidal constraint.

The details of the suite of simulations are given below in Table~\ref{tab:param}.  The simulations are classified according to their initial field topology and resolution (defined as $H_{0}/\Delta z$).  These simulations are run for a period of time measured by the orbital period at the inner radial edge of the simulation.  While the majority of the simulations presented use a full azimuthal domain and an aspect ratio of unity, extra control runs were performed to study the importance of these choices.  Wedge runs, utilizing a reduced azimuthal domain of $\pi/4$, were run to assess the importance of low-m modes and to access resolutions higher than were possible with the restrictive constraints of the other simulations.  These simulations are denoted \zz{32W} and \zz{64W}.  In additional to using a truncated azimuthal domain, it is common in the literature to use a reduced azimuthal resolution.  To assess the importance of azimuthal resolution, two runs (\zz{32R} and \zz{32RR}) were performed which halved and quartered, respectively, the number of azimuthal grid cells used in run \zz{32} while maintaining the vertical and radial resolution.  Also included in Table~\ref{tab:param} are a series of scalar metrics whose definitions are given in \S\ref{sec:diag}.   

\begin{table}[htdp]

\begin{center}
\begin{tabular}{|l|c|c|c|c|c|c|}
\hline
Simulation & Azimuthal & Resolution & Orbits & $<\alpha>_{QSS}$ & $<\beta>_{QSS}$ &  $T_{1/2}$ \\
 & Range & $(N_{R},N_{\phi},N_{z})$ &  & & & \\
\hline
BzZ8 &  $2\pi$ & (120,480,32) & 200 & 0.013 & 26.55 & 184.4 \\
BzZ16 &  $2\pi$ & (240,960,64) & 200 & 0.02 & 23.01 & 127.2 \\
BzZ32 &  $2\pi$ & (480,1920,128) & 200 &  0.018 & 27.67 & 116.1 \\
\hline
BzZ32W & $\pi/4$ & (480,240,128) & 100  & 0.023 & 23.35 & N/A \\ 
BzZ64W & $\pi/4$ & (960,480,256) & 100 & 0.024& 20.40 & 90.0 \\
\hline 
BzZ32R &  $2\pi$ & (480,960,128) & 200 & 0.021 & 24.03 & 109.6 \\
BzZ32RR &  $2\pi$ & (480,480,128) & 200 & 0.015 & 30.01 & 127.5 \\
\hline
\hline
BzN8 &  $2\pi$ & (120,480,32) & 200 & 0.058 & 8.18 & 20.0 \\
BzN16 &  $2\pi$ & (240,960,64) & 200 &  0.076 & 8.26 & 19.4 \\
\hline
\hline
BpN8 &  $2\pi$ & (120,480,32) & 200 & 0.055 & 6.37 & 56.2 \\
BpN16 &  $2\pi$ & (240,960,64) & 200 & 0.064 & 7.37 & 40.2 \\
BpN32 &  $2\pi$ & (480,1920,128) & 200 & 0.067 & 7.42 & 35.5 \\
\hline
\end{tabular}
\end{center}
\caption{Simulation Parameters}
\label{tab:param}
\end{table}

\subsection{Diagnostics}
\label{sec:diag}

As turbulence involves fluctuation quantities in both space and time, diagnostic quantities will invariably involve some combination of spatial and temporal averaging.  For simplicity, we define here the quantities we will use below.  The most common quantity is simple volume-integration and for a variable $X(R,\phi,z,t)$ and is defined as $<X>(t) = \int_{V} X dV$, where $V$ represents the full simulation domain.  Related to this is volume-averaging, defined as $\bar{X}(t) = <X>/|V|$.  

The fiducial timescale of these simulations is the orbital period at the inner edge of the disk, $P_{o} = 2\pi$, and for brevity is simply designated as an orbit without further qualification.  Often we would like to compute scalar values of quantities that are representative of a simulation as a whole.  In contrast to local simulations in which conserved quantities cannot change due to the closed nature of the simulation, global simulations involve significant evolution.  The initial growth of certain physical quantities, particularly the stress and magnetic energy, is exponential and driven by the linear phase of the MRI.  These quantities reach a peak value and quickly decrease as the linear phase of the MRI transitions into a fully turbulent state in which there is significantly reduced secular variation.  We will refer to this stage as the quasi-steady state (QSS), which we define for simulations initialized with a vertical field as between 50 orbits and the end of the simulation.  Due to the slower growth rate of the toroidal MRI we quantify the QSS as between 100 orbits and the end of the simulation.  Scalar representative values, like those in Table~\ref{tab:param}, are denoted $<X>_{QSS}$ (or $\bar{X}_{QSS}$) and are defined as the temporal average over the timeframe defined as the QSS of the volume-integral (or volume-average) of the quantity.  An alternative diagnostic to quantify disk evolution is the use of the measurement $T_{1/2}$, which refers to the number of orbits it takes for half of the total mass in the disk to leave the simulation domain.  The values of this quantity for all simulations considered, save for simulation \zzw{32} which did not accrete half its mass in the timeframe considered, are given in Table~\ref{tab:param}.  
 
The dynamics of accretion disk turbulence are of particular astrophysical interest, due to the anisotropic structure of the resultant turbulent magnetic field and its ability to drive angular momentum radially outward.  To study the efficacy of angular momentum transport, many of our diagnostics will focus on the stress that allows this transport to happen.  This stress is of two types: the Reynolds stress, $T_{R\phi} = \rho v_{R}\delta v_{\phi}$, and the Maxwell stress, $M_{R\phi} = -B_{R}B_{\phi}$, with the Maxwell stress generally dominating the Reynolds component.  While these two quantities are of the most direct physical relevance, it is common to scale the stress by the gas pressure resulting in the following diagnostics:

\begin{mathletters}
\begin{eqnarray}
\label{eqn:alpha}
\alpha = \myfrac{T_{R\phi} + M_{R\phi}}{P}, \\
\alpha_{M} = \myfrac{M_{R\phi}}{P}.
\end{eqnarray}
\end{mathletters}
Also of interest is the strength of the magnetic field in relation to the thermal energy of the gas, given by $\beta = P/P_{b}$, the ratio of gas and magnetic pressure.  Although an abuse of notation, for brevity we will use the notation $<\alpha>$ and $<\beta>$ to refer to the volume-integral of the numerator divided by the volume-integral of the denominator.   

In an effort to study how well resolved the MRI is, we consider the following diagnostics
{\begin{mathletters}
\label{eqn:resolvfrac}
\begin{eqnarray}
F_{z} = |V|^{-1} \int_{V} (\lambda_{MRI} \ge 8\Delta z)dV, \\
F_{\phi} = |V|^{-1} \int_{V} (\lambda_{C} \ge 8 R \Delta \phi) dV,
\end{eqnarray}
\end{mathletters}}
where in the above equations logical statements refer to indicator functions that assume the value of unity or zero based on the truth or falsity of the statement.  The characteristic wavelength of the toroidal field, $\lambda_{C}$, is defined analogously to $\lambda_{MRI}$ save for the use of the toroidal Alfven speed in place of the vertical.  These scaled integrals represent the fraction of the disk where the fastest-growing modes of the vertical and critical toroidal modes of the MRI are resolvable, using the often employed 8-zone criterion.  These are related to the quality factors, first used by \citet{nkh10}, given as

{\begin{mathletters}
\label{eqn:quality}
\begin{eqnarray}
Q_{z} = \lambda_{MRI} / \Delta z, \\
Q_{\phi} = \lambda_{c} / (R \Delta \phi),
\end{eqnarray}
\end{mathletters}}
where these values are in general functions of space and time.  Use of the quality factor as a diagnostic represents an important step in appreciating the numerical resolvability of the MRI, but we feel that the resolvability fraction diagnostic we employ is an even more stringent resolvability requirement.  

It is well-known that the anisotropy of the magnetic field is key to angular momentum transport, and indeed the correlation between the radial and azimuthal component of the magnetic field is an important diagnostic of angular momentum transport in disks.  An alternative measure of the anisotropy, called the magnetic tilt angle, was first discussed by \citet{ggsj09}.  This measure is defined as,

\begin{equation}
\label{eqn:ta}
\theta_{B} = \arcsin(\alpha_{M}\beta)/2.
\end{equation}
Physically, this can be thought of as an approximation to the angle between the planar magnetic field and the azimuthal axis assuming a weak vertical magnetic field.  An estimate of the tilt-angle is derived as $\theta_{B} \approx \indeg{15}$, in \citet{ggsj09}, by noting that all the local models considered satisfy the relationship $\alpha\beta \approx 1/2$.  Strictly speaking, $\alpha_{M} < \alpha$, and thus we expect this estimate to be an upper bound.

Studying the spectral structure of turbulent flows is often more natural than alternative diagnostics in physical space.  Our primary interest here will be studying the azimuthal structure of power spectra of physical quantities, in particular the density and magnetic pressure.  To this end, we begin by defining the subdomain 
\begin{equation}
\mathcal{S} = [R_{0}-2H_{0},R_{0}+2H_{0}] \times [0,2\pi] \times [-2H_{0},2H_{0}] .
\end{equation}
For simulations that do not model the full $2\pi$ radian azimuthal domain the subdomain definition is modified to include the full azimuthal range modeled.  We then consider, for a physical quantity $X(R,\phi,z)$, the azimuthal Fourier decomposition $X(R,m,z)$.  Further, we define $X(m)$ as the volume-weighted radial and vertical mean over the subdomain $\mathcal{S}$.  This removes spatial transients, but to remove temporal transients we also average between orbits 50 and 100, for simulations seeded with a vertical field, and between orbits 100 and 150 for simulations seeded with a toroidal field.  We refer to this reduced temporal domain as RQSS.  To ensure that these results are not adversely affected by secular trends within the subdomain related to the evolution of the disk we scale, at each timestep, by $<X>^{\mathcal{S}}$, the volume integral of the quantity over the subdomain.  Formally,
\begin{equation}
\label{eqn:pow}
\hat{X}(m) = \left < \myfrac{X(m)}{<X>^{\mathcal{S}}} \right >_{RQSS} .
\end{equation} 
To more directly study important azimuthal scales we will employ the transformation of the wavenumber, $m$, to an effective azimuthal wavevector given by
\begin{equation}
k_{\phi} = \myfrac{m}{2\pi R_{0}}.
\end{equation}  

One final tool we employ when studying the spectral structure of disk turbulence is the use of a fiducial power spectra.  To study the relative importance of large-scale versus small-scale features, we consider a fiducial power spectra given by, $P_{m} = m^{-1}$.  This fiducial spectra has the property that the inner and outer scales are equally important to the total integrated power over the accessible wavenumbers.  We diagnose the dominant azimuthal scale by considering the location of the peak value of the spectra defined in Eqn~\ref{eqn:pow} scaled by this fiducial power spectrum, namely $m\hat{X}(m)$.  

The above formalism will be employed to study the structure of density and magnetic energy, and while it is common to study the structure of stress in a similar manner we choose a slight variant.  Astrophysical interest in stress as a diagnostic is based on its ability to drive radial angular momentum transport.  We note, however, that when projecting this quantity onto the Fourier basis all but the $m=0$ mode will correspond to an azimuthal average of zero.  Instead, we choose to analyze the structure of contributions to stress that result in net radial transport.  Formally, we define

\begin{equation}
\tilde{\alpha_{M}}(m) = \left< \myfrac{-B_{R}(m)B_{\phi}(m)}{<P>^{\mathcal{S}}} \right>_{RQSS}
\end{equation}    
  
One of the goals of this paper will be to explicitly compare local and global models, towards this end we proceed in a similar fashion to \citet{sra10} and decompose our global simulation into a set of subvolumes through which we can calculate ``local'' statistics.  This is accomplished by considering subvolume of the physical domain $(R,\phi,z) \in \mathcal{S}$ and decomposing it into wedges of size $[H_{0},2\pi H_{0},H_{0}]$.  This yields, at each timestep, a set of 160 subvolumes in which relevant physical quantities can be volume-averaged.  To calculate statistics, we simply take the mean, denoted $[X](t)$, and when relevant, the standard deviation of these quantities at each timestep.

We also consider the instantaneous correlation of flux and stress both as a convergence criterion and as a means of comparing local and global models.  The flux-stress connection was first explored by \citet{hgb} in the context of saturation predictors and further refined to take into account numerical effects by \citet{pcp07}.  Saturation predictors map the initial flux threading a local model to the predicted stress after the initialization and saturation of MRI-driven turbulence.  It is noted in \citet{hgb} that saturated stress increases with the strength of the vertical field threading a simulation domain, specifically $\alpha \propto \lambda_{MRI}$.  This observation is augmented in \citet{pcp07} to include the numerical constraint that a sufficiently weak net field will be unresolvable and thus behave in a manner identical to a zero-net field.  Formally, the saturation predictor presented in \citet{pcp07} is given by
\begin{equation}
\label{eqn:mriscalez}
\alpha_{M}\Big(\myfrac{H}{L}\Big)^{(5/3)} = 0.61 \times \left\{
\begin{array}{ll}
\Delta / L & : \lambda_{MRI} \leq \Delta ,\\
\lambda_{MRI}/L & : \Delta < \lambda_{MRI} \leq L, \\
0 & : L < \lambda_{MRI} ,
\end{array}
\right.
\end{equation} 
where $L$ and $\Delta$ is the box size and grid-scale respectively.  Their saturation predictor includes three regions: an unresolved region in which $\lambda_{MRI}$ is unresolvable and consequently $\alpha \propto \Delta$; the resolved region studied in \citet{hgb}; and a stable region in which $\lambda_{MRI}$ exceeds the vertical domain of the simulation and turbulence is absent.  

To study the global analogue to these local saturation predictors, we proceed in a manner similar to \citet{sra10}.  We calculate the instantaneous flux and stress in each subvolume  at each timestep after orbit 50, to remove possible noise in the data from the linear growth rate of the MRI.  The resulting flux-stress pairs are logarithmically binned according to flux in order to diagnose trends.  In contrast to the local model and \citet{sra10}, where flux-stress pairs are scaled by $L$ and $(H/L)^{5/3}$ respectively, we proceed in a manner more appropriate to global simulations in which the size of the subdomain has no physical meaning.  The flux-stress pairs calculated here are of the form ($\lambda_{MRI}/\Delta z$,$\alpha_{M}$), as in \citet{bas11}.  Scaling the local flux to the grid scale is more directly meaningful and facilitates a closer inspection of the transition point discussed in \citet{sra10}.

In addition to studying the relationship between vertical flux and stress, we also consider the analogous connection in the presence of toroidal flux.  Arguably, the presence of strong net toroidal field is more astrophysically relevant (\citeinp{vb89}).  The evolution of the toroidal MRI is more complex than its vertical counterpart, and there is no simple form for its most unstable mode.  We proceed, as in \citet{hgb}, by considering the toroidal flux defined by $\lambda_{c} = 2\pi \sqrt{16/15}  \:v_{A,\phi}/\Omega$.  The saturation predictor there is given as

\begin{equation}
\label{eqn:mriscalep}
\alpha_{M} \propto \myfrac{L_{y}}{H^{2}} \lambda_{c},
\end{equation}  
where $L_{y}$ is the effective azimuthal length of the shearing box.  To test this in the global context we consider flux-stress pairs of the form ($\lambda_{c}/R\Delta \phi$, $\alpha_{M}$) and perform the same logarithmic binning as described for the vertical flux-stress pairs.

\section{Evolution of Global Disks}
\label{sec:evol}

Prior to a full discussion of the resolution dependance of the simulations that will be presented in Section \S\ref{sec:conv}, we focus on the field topology dependance of several fiducial runs.  For this we choose the highest resolution simulation of each field topology that was run for the full 200 orbits, these are: \zz{32}, \zn{16}, and \pn{16}.  The most significant difference between global and local simulations is the secular evolution of global simulations.  Open boundary conditions allows mass to be accreted off the grid and the total magnetic flux to evolve in a dynamical manner.  The development of radial structure of the mass profile adds radial pressure gradients to the dynamics of the turbulence.     

\begin{figure}
  \centering
  \subfloat[Mass fraction]{\label{fig:globevola} \includegraphics[width=0.5\textwidth]{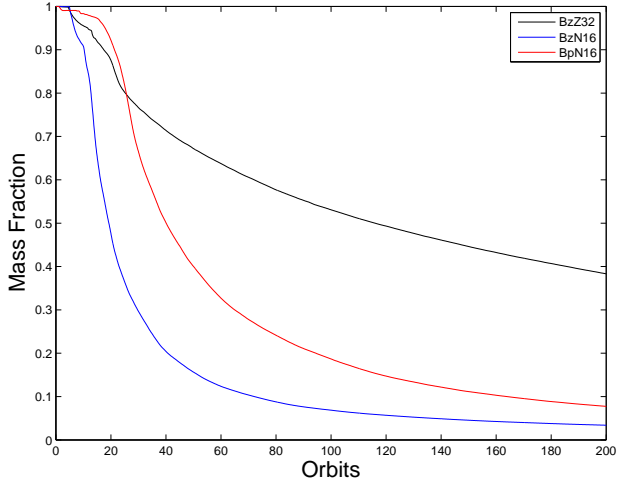}}                
  \subfloat[Accretion efficiency ($\alpha_{M}$)]{\label{fig:globevolb} \includegraphics[width=0.5\textwidth]{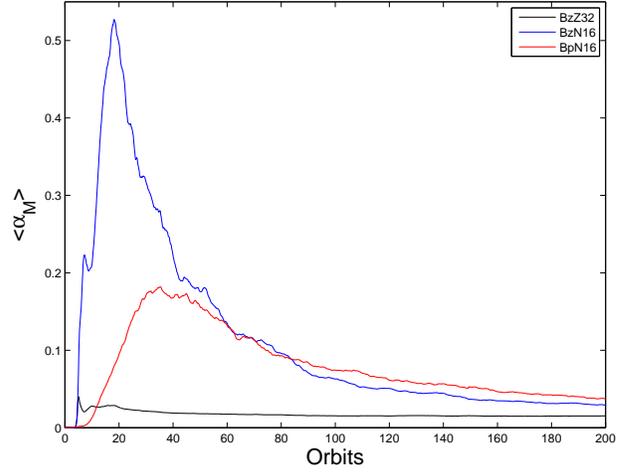}} \\
  \subfloat[Dimensionless magnetic energy ($\beta^{-1}$)]{\label{fig:globevolc} \includegraphics[width=0.5\textwidth]{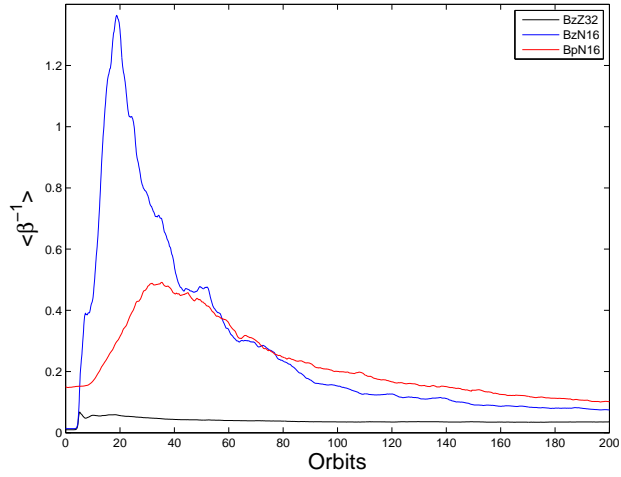}}   
  \caption{Comparison of the evolution of the fiducial simulations.}
  \label{fig:globevol}
\end{figure}

To provide a sense of the evolution, Figure~\ref{fig:globevol}, shows the temporal variation of several global quantities.  The evolution of the mass fraction, $<\rho>(t)/<\rho>(t=0)$, is shown in Figure~\ref{fig:globevola}, and Figures~\ref{fig:globevolb} and~\ref{fig:globevolc} show the evolution of $<\alpha_{M}>$ and $<\beta^{-1}>$ respectively.  Most striking in all of these figures is the significant accretion and field amplification caused by the presence of a net magnetic field in the initial growth phase.  Contrasting this, is the comparatively similar behavior of these quantities in the QSS.  In this initial phase the net field runs exhibit magnetic fields with energies comparable to and even in excess of the thermal energy of the disk and accretion efficiencies an order of magnitude above those normally associated with zero-net field disks.  Indeed, a majority of the mass is accreted during the initial growth phase. 

The significant accretion of mass is in contrast to what one would expect from a simple estimate of the viscous timescale, $T_{\nu} = R^{2}/\nu$, where the viscosity takes the common form $\nu = \alpha_{SS}c_{s}H$.  Calculation of the viscous timescale at the fiducial radius, $R_{0} = 2$, and converting to orbits yields $T_{\nu} \approx 63.6/\alpha_{SS}$.  Comparison with the mass evolution (Figure~\ref{fig:globevola}) and $T_{1/2}$ (Table~\ref{tab:param}) suggests a value of $\alpha_{SS}$ considerably in excess of the value of $\alpha$ measured in the QSS (Table~\ref{tab:param}) or even during the initial evolution (Figure~\ref{fig:globevolb}).  For instance, equating $T_{\nu}$ and $T_{1/2}$ suggests a value of $\alpha_{SS} \approx 3$ for run \zn{16} ($<\alpha>_{QSS} = 0.076$) and $\alpha_{SS} \approx 0.5$ for \zz{32} ($<\alpha>_{QSS} = 0.018$).         

The viscous timescale is, of course, a crude estimate.  Formally, it should be valid only in the case of a radially localized mass distribution with a viscosity exhibiting minimal radial dependance.  Attempting to apply this estimate to global disks of the type described here stretches the approximation far beyond its area of applicability.  Understanding the evolution of global disks in the context of an anomalous viscosity model requires a more sophisticated treatment including the radial structure.  Towards this end, we compare our simulated disks with a 1-dimensional reduced model for the time-evolution of the surface density $\Sigma = \int \rho dz$ based upon an anomalous viscosity 
\begin{equation}
\pd{\Sigma}{t} = \frac{3}{R}\pd{}{R} \left[ \sqrt{R} \pd{(\nu \Sigma \sqrt{R})}{R} \right] ,
\label{eqn:alphamod}
\end{equation}
(e.g., see \citeinp{pringle81}).  We solve Eqn~\ref{eqn:alphamod}, with $\nu = \alpha_{SS}c_{s}H$, using the values of $\alpha$ calculated from the full simulation.    We use the vertically and azimuthally averaged values of $\alpha$, temporally spaced every tenth of an orbit, in place of $\alpha_{SS}$ in the formulation of the turbulent viscosity.  We note that $\alpha_{SS}$ and $\alpha$ are related by a factor of $3/\sqrt{2}$.  The radial dependance of the viscosity is important, as the evolution of the surface density depends on its spatial derivatives.  The early evolution of the disk is characterized by small-scale structure in the stress that is likely to form an important contribution to the evolution of the surface density, which itself exhibits weak spatial variability during the initial phase of the evolution.  The evolution equation is solved utilizing a simple implicit, finite-differencing scheme.  For simplicity, the boundary conditions are set by the constraint that the mass profile agrees with the vertically and azimuthally averaged mass in the full simulation at the inner and outer boundary.

\begin{figure}
  \centering
  \subfloat[Evolution of mass fraction.]{\label{fig:alphamoda} \includegraphics[width=0.5\textwidth]{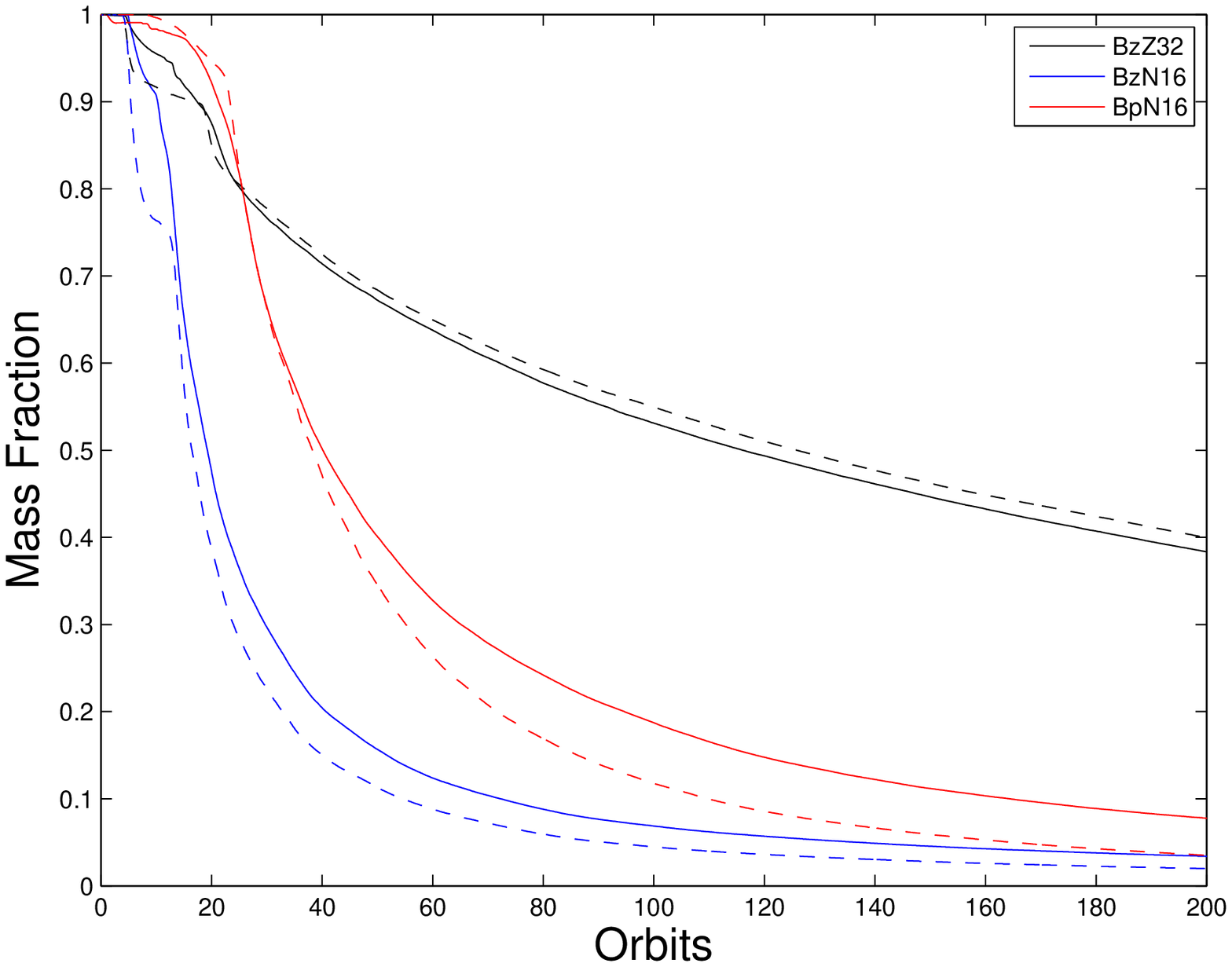}}                
  \subfloat[Radial mass profile at orbit 75.]{\label{fig:alphamodb} \includegraphics[width=0.5\textwidth]{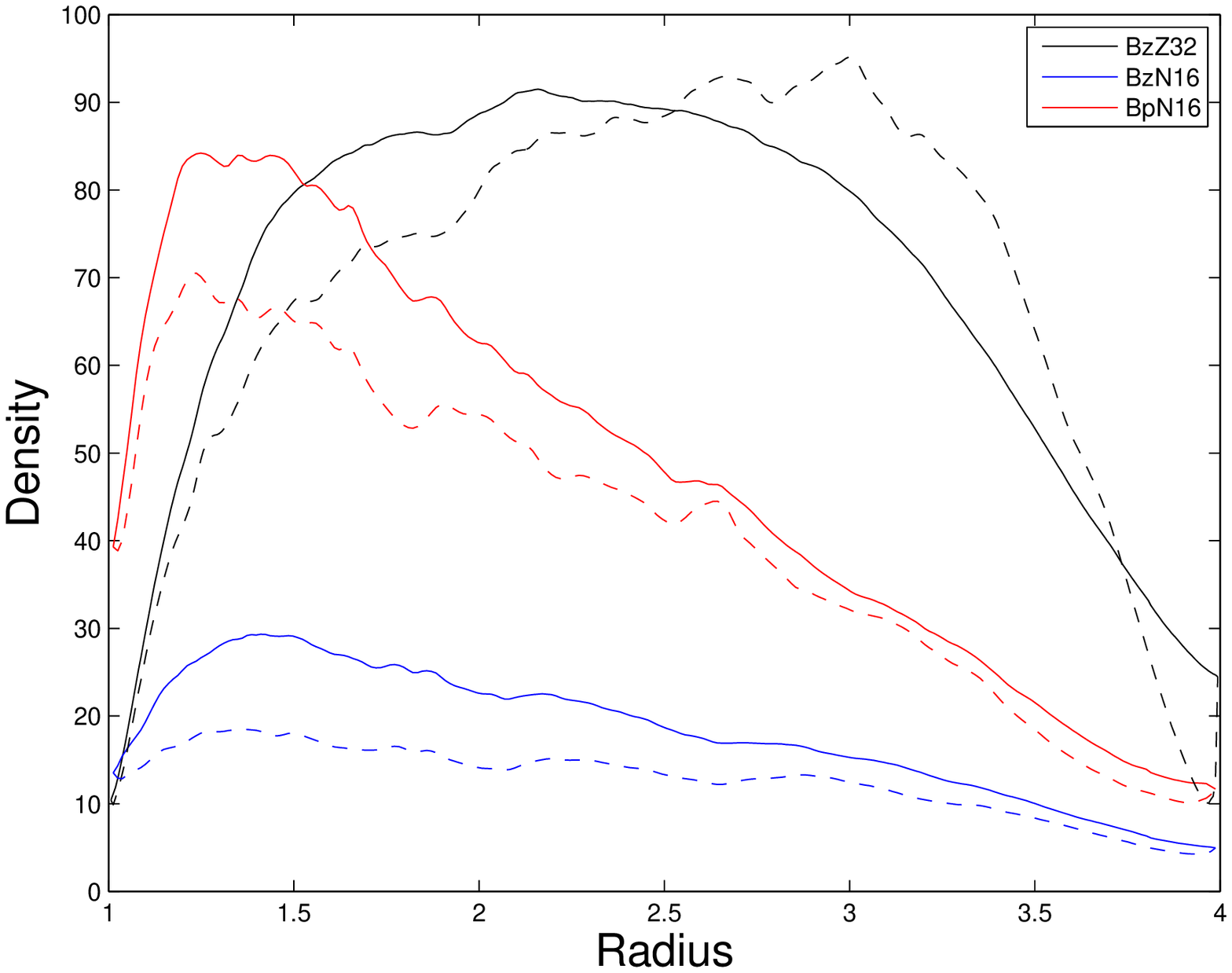}}   
  \caption{Comparison of simulations (solid) and reduced model (dashed).}
  \label{fig:alphamod}
\end{figure}

Figure~\ref{fig:alphamod} compares the evolution predicted by the reduced model to that obtained in the full MHD simulation.  A comparison of the broad features of the evolution, specifically the mass fraction, is given in Figure~\ref{fig:alphamoda}.  In all cases the reduced model accurately predicts the mass evolution of the disk.  More detail is given in Figure~\ref{fig:alphamodb}, in which the radial mass profile computed by the reduced model is compared to that from the full simulation at orbit 75.  The overall features of the profile are  captured, although we notice a larger discrepancy than observed in the mass fraction.  The observed discrepancy appears to be largely caused by the limitations of the temporal discretization used in which values of $\alpha$ from the simulation are calculated ten times every orbit.  The initial growth of the MRI induces fluctuations with significant variability in both space and time that are not adequately captured by the temporal discretization.  The model showed in Figure~\ref{fig:alphamod} must agree with the full simulation if the latter conserves mass and angular momentum in the limit as the temporal spacing of $\alpha$ from the simulation vanishes.  Indeed, taking as the initial condition the density profile after the saturation of the turbulence and evolving this profile using the one-dimensional model and data from the simulation removes much of the discrepancy.
      
While this section has focused on a formal comparison of our MHD simulations with reduced models, these results may have more direct astrophysical implications.   The most direct way of determining $\alpha$ in real accretion disks is the analysis of dwarf nova outbursts and X-ray transient outburst.   As pointed out by \citet{kpl07}, these estimates suggest $\alpha \sim 0.1-0.4$ whereas numerical simulations (including those presented here) typically obtain steady-state values of $\alpha$ that are an order of magnitude smaller.   In the light of our results, we note that an accretion disk which has just entered an outburst state may well go through a period of field-growth that resembles the early transients seen in our simulations.   During these transients, the effective value of $\alpha$ is substantially enhanced, and spatio-temporal gradients in $\alpha$ further enhance the angular momentum transport.  Thus, it is interesting to conjecture that the large values of $\alpha$ inferred from outburst systems correspond to these transient phenomena.  These issues will be explored in a future publication.  

\section{Convergence of Global Disk Simulations}
\label{sec:conv}
Standard tests of convergence rely on running simulations with increasing resolution while leaving the underlying physical problem unchanged.  Convergence in the case of turbulent non-explicitly-dissipative systems is inherently ill-defined for two reasons.  The first is that when increasing resolution there will invariably be minor differences in the seed perturbations that feed the instability, and as a result we cannot expect precise agreement between simulations.  The second and more fundamental reason is that in ideal MHD the dissipation scale is set by the grid scale, and thus when increasing resolution we are not leaving the underlying physical problem unchanged.  Further, changes in resolution alters the evolution of the disk; changes in mass accretion and mass distribution results in a fundamentally different disk.  In light of these complexities, we will hereafter take the notion of convergence to mean that an increase in resolution will leave relatively unchanged spatially- and temporally-averaged measurements.  

We consider three broad categories of convergence metrics: physical, numerical, and spectral.  The physical metrics, $\alpha_{M}$ and $\beta$, are perhaps the most natural and directly physically relevant and therefore have the longest history of use as a diagnostic of accretion disk turbulence.  However, as we will demonstrate these metrics are often ambiguous and display non-monotonic resolution dependance.  Numerical metrics, the resolvability fractions and quality factors, directly measure how well the linear MRI is resolved and were a focus of the convergence study described by \citet{hgk11}.  The spectral metrics we employ study the azimuthal structure of the turbulent flow, and specifically seek to identify the dominant azimuthal scale and its dependance on resolution.  We will demonstrate that all of these metrics are useful diagnostics towards studying the nature of the turbulent flow in accretion disks, however as a convergence criterion the magnetic tilt angle appears unambiguous and robust.

\subsection{Physical Metrics}
\label{sec:convphys}

\begin{figure}
\centering
\subfloat[Accretion efficiency ($\alpha_{M}$), ZF runs.]{\label{fig:physcona} \includegraphics[width=0.5\textwidth]{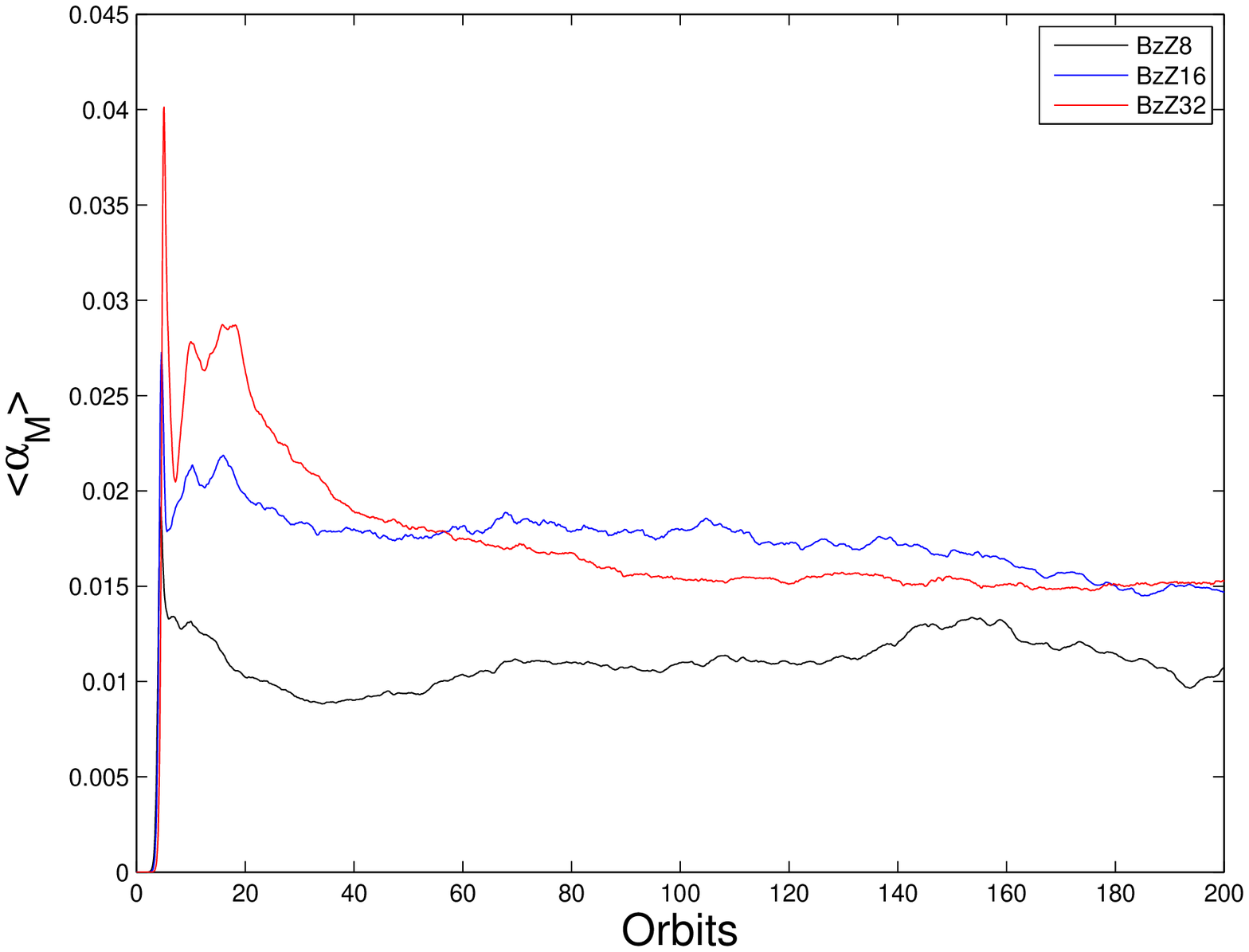}}
\subfloat[Magnetic energy ($\beta^{-1}$), ZF runs.]{\label{fig:physconb} \includegraphics[width=0.5\textwidth]{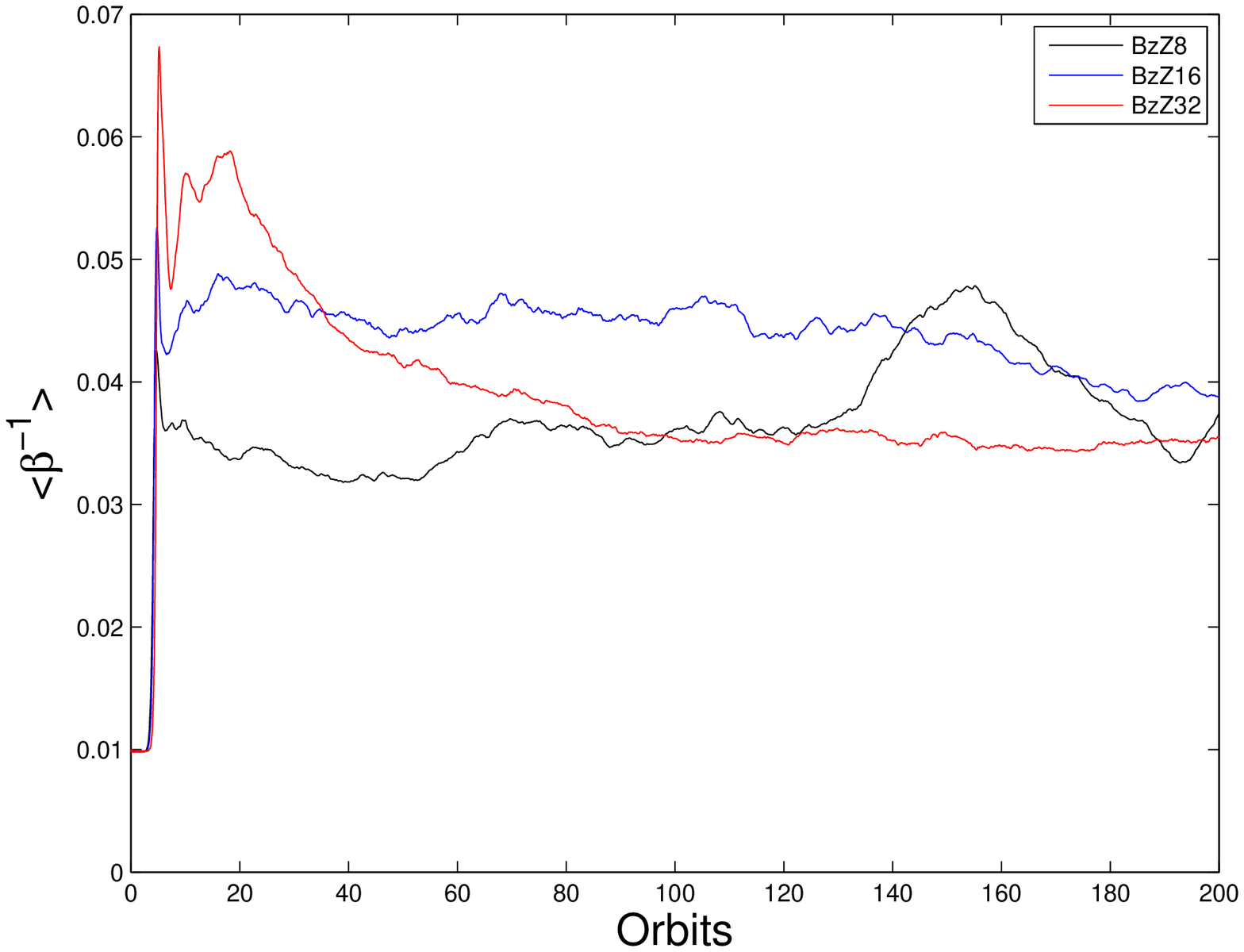}} \\
\subfloat[Accretion efficiency ($\alpha_{M}$), NF runs.]{\label{fig:physconc} \includegraphics[width=0.5\textwidth]{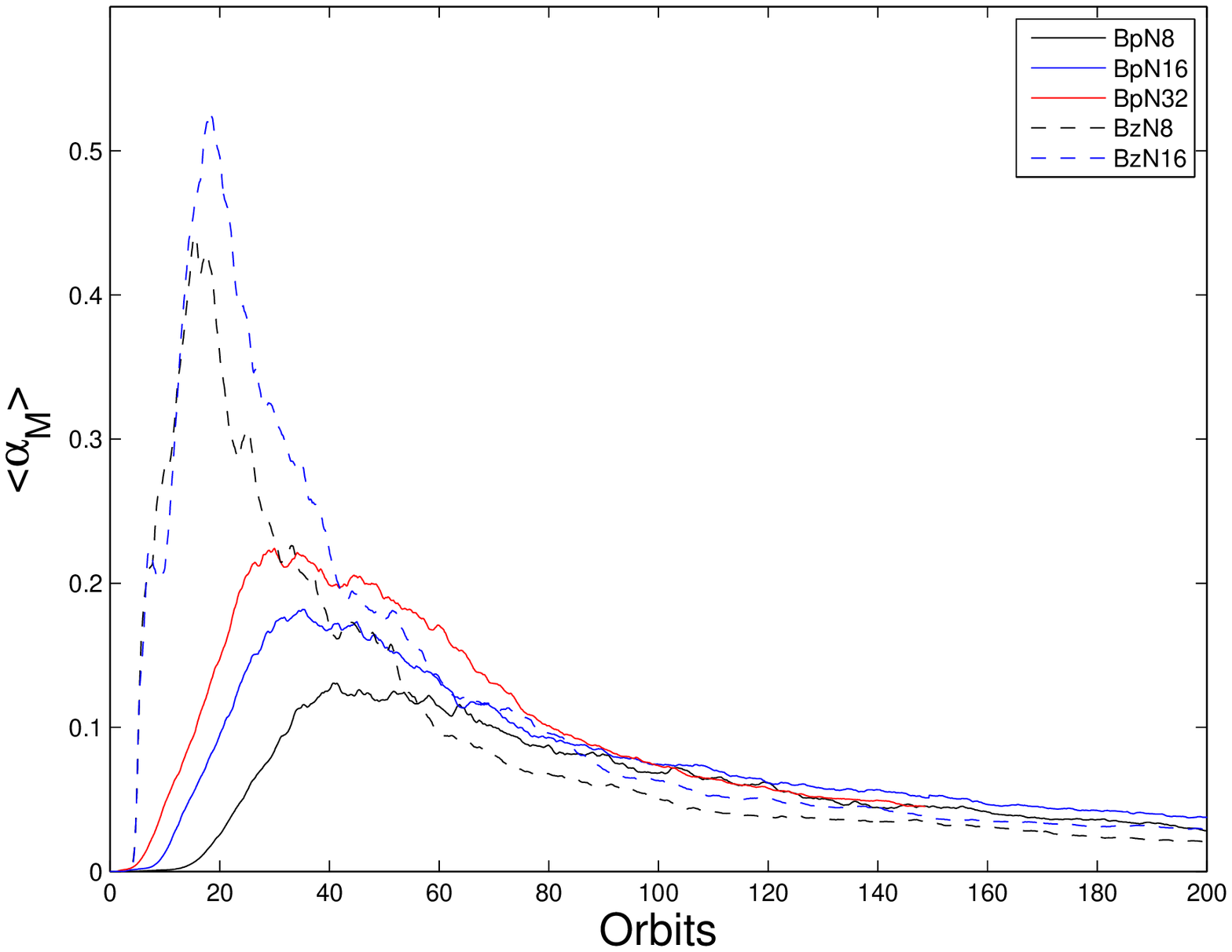}}
\subfloat[Magnetic energy ($\beta^{-1}$), NF runs.]{\label{fig:physcond} \includegraphics[width=0.5\textwidth]{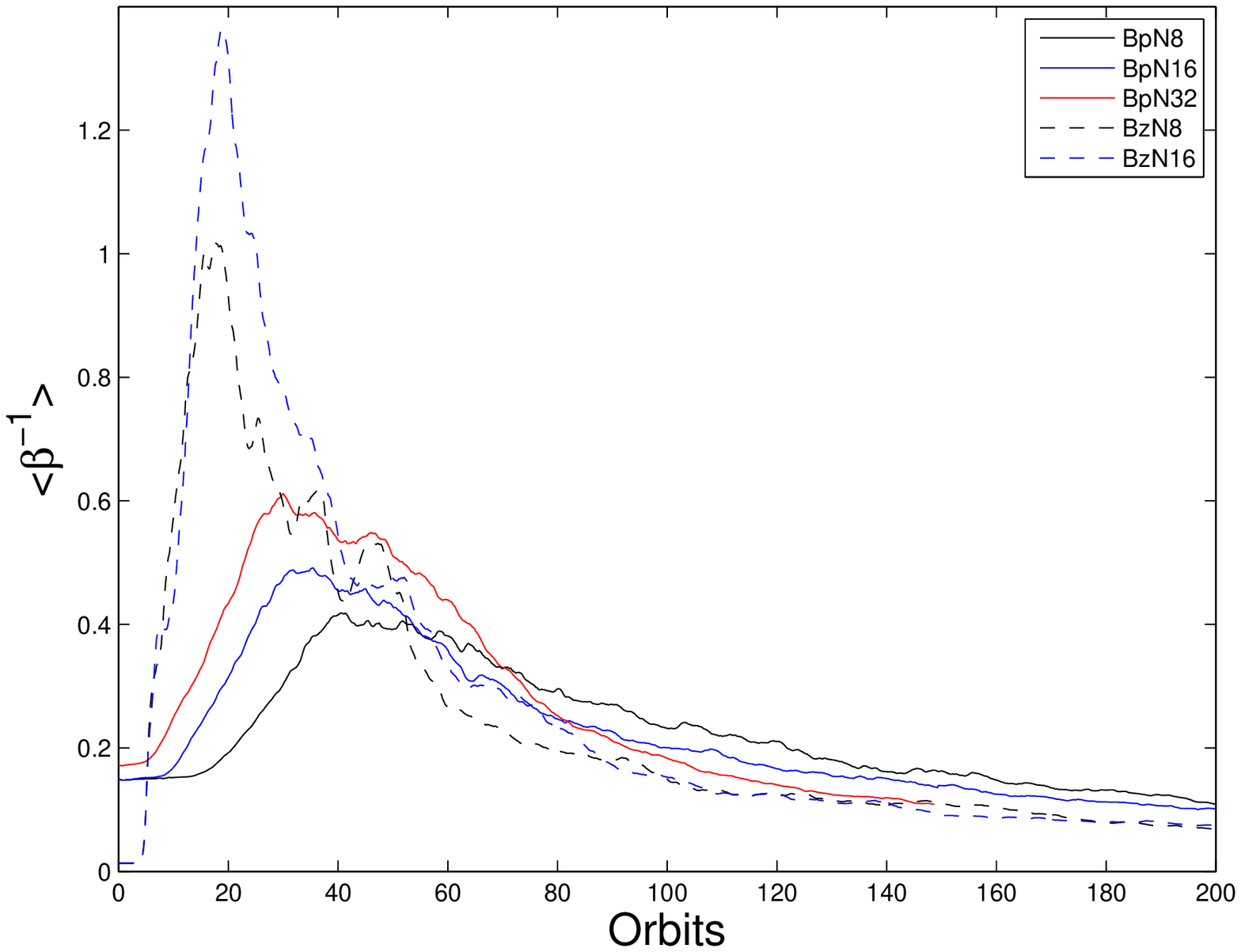}}  
\caption{Physical convergence metrics, ZF and NF models.}
\label{fig:physcon}
\end{figure}

The simplest and most astrophysically relevant convergence criterion is accretion efficacy as measured by the dimensionless stress $\alpha_{M}$.  Figure~\ref{fig:physcona} shows the evolution of the Maxwell stress over the course of the simulation for all three standard ZF models.  As is expected, initial peaks in the stress associated with the linear growth of the vertical MRI are resolution dependent.  While the initial magnetic field is constructed so that the most unstable mode, $\lambda_{MRI}$, is resolvable in each simulation there are other slower-growing unstable modes whose resolvability will vary depending on the resolution of the simulation.  The true test of convergence is the behavior of the stress  in the saturated quasi-steady state.  The lowest resolution simulation, \zz{8}, exhibits $<\alpha_{m}>_{QSS} \approx 0.01$, while both higher resolution simulations exhibit a comparable stress to each other that is roughly 50\% greater than \zz{8}.
  
While this fundamental criterion of convergence in stress is satisfied, other diagnostics paint a more subtle picture.  Figure~\ref{fig:physconb} illustrates the evolution of the scaled magnetic energy, $\beta^{-1}$.  Again, as expected, we see that the initial field amplification is monotonic with resolution and dominated by the growth of the toroidal magnetic field, however the saturation and transition into the fully non-linear state is more complex than expected.  While \zz{32} peaks at a higher value due to the larger number of resolvable, unstable MRI modes the lower resolution simulations, \zz{8} and \zz{16}, maintain a stronger magnetic field proportionally.  These lower resolution simulations also lack the steep drop-off in magnetic field energy often associated with the saturation of the MRI.  This may be a consequence of the inability to resolve the parasitic instabilities associated with saturation of the linear MRI.  Also of note, is the late-time field growth associated with \zz{8}.  The nature of this growth is unclear, but may be suggestive of a very low-frequency temporal behavior.  

Our analysis of the physical metrics of models initialized with a seed field possessing net flux proceeds in much the same way as our analysis of the ZF models.  As above, we begin by considering the time evolution of the global quantities, $\alpha_{M}$ and $\beta^{-1}$.  The results are shown in Figures~\ref{fig:physconc} and \ref{fig:physcond} respectively.  The evolution of the linear MRI displays a significant resolution dependence, as would be expected, but in all cases the values of $\alpha_{M}$ in the saturated state are comparable for the same initial field topologies.  Due to the significantly higher values of $\alpha_{M}$ in the net field simulations we see a corresponding significant increase in mass loss of the disk (Table~\ref{tab:param}) compared to the ZF runs.  Field amplification for the runs initialized with a vertical field is monotonic with resolution.  However, regarding field amplification the net toroidal runs behave more analogously to the net-zero field runs discussed above, in which lower-resolution simulations seem to exhibit greater field amplification in the QSS.  The presence of a net field, regardless of topology, results in order of magnitude increases in the peak values of $\alpha_{M}$ and $\beta^{-1}$ compared to the ZF simulations considered above.  Insofar as angular momentum transport is concerned, there is only a weak dependance on resolution in the saturated state even for the most poorly resolved simulations.  The evidence from these global metrics suggests that net-flux simulations converge more quickly, however we'll see through the consideration of the other metrics that the picture is more subtle.

Overall, we see that while the behavior of $\alpha_{M}$ paints a simple picture and is suggestive of convergence, the magnetic energy is far more volatile and does not follow a clear pattern in its dependance on resolution.  Further, as a convergence criterion $\alpha_{M}$ suffers from its dependance on initial field topology.  

\begin{figure}
\centering
\subfloat[Accretion efficiency ($\alpha_{M}$)]{\label{fig:physreda} \includegraphics[width=0.5\textwidth]{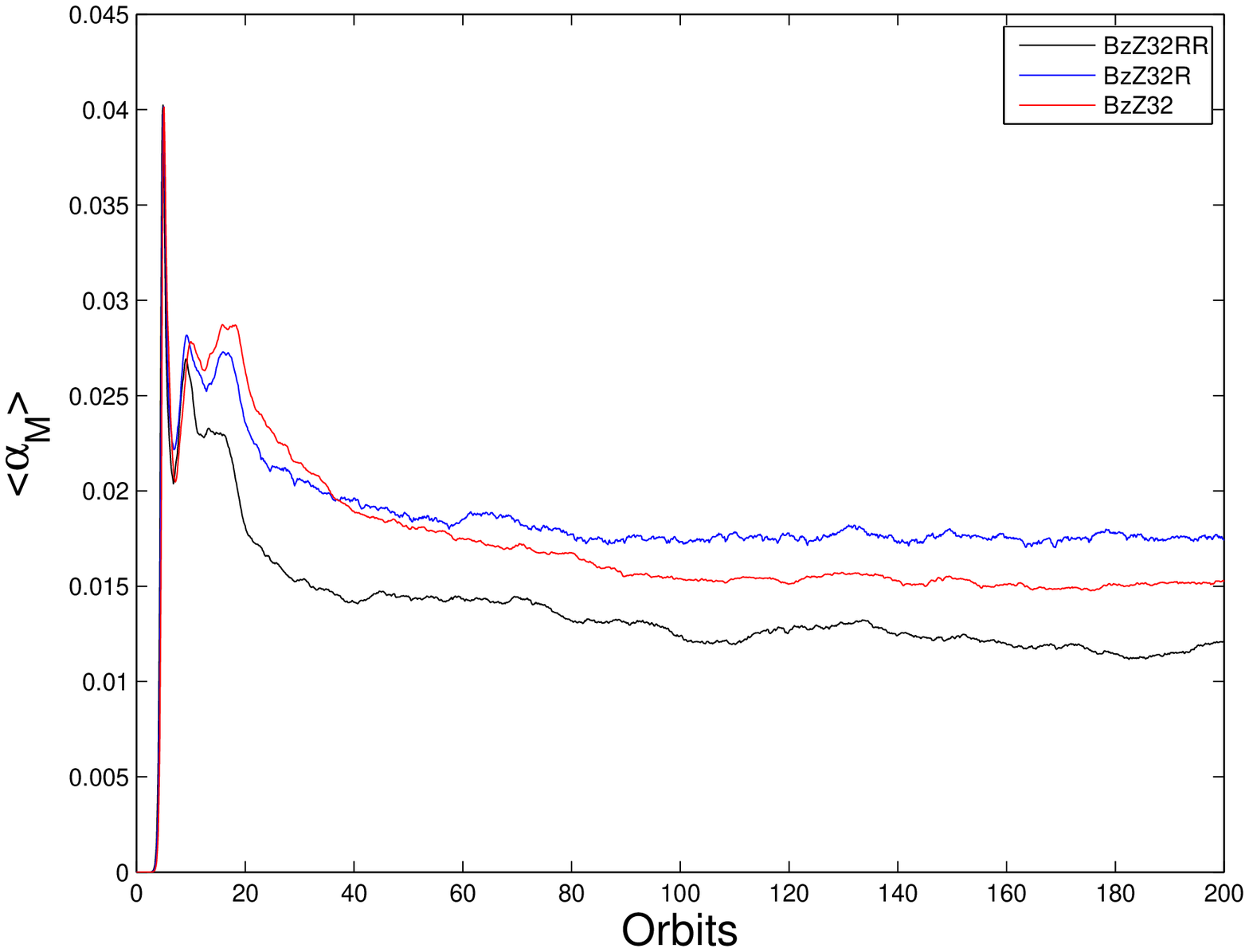}}                
\subfloat[Magnetic energy ($\beta^{-1}$)]{\label{fig:physredb} \includegraphics[width=0.5\textwidth]{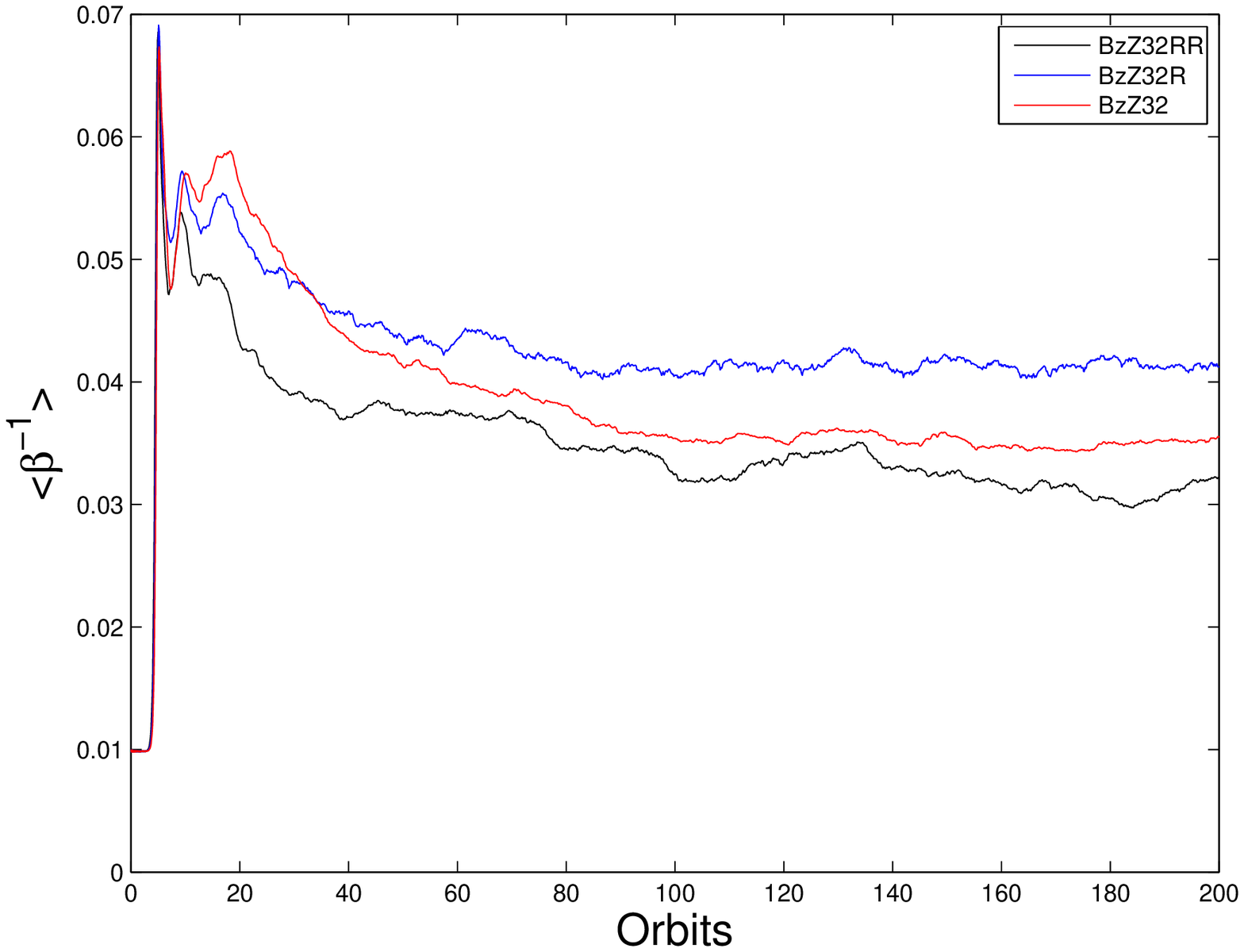}} \\
\caption{Physical convergence metrics for reduced azimuthal resolution runs.}
\label{fig:physred}
\end{figure}

The use of orbital advection allows us a unique opportunity of exploring isotropic resolutions in a cost-effective manner.  Without orbital advection, the timestep constraint is set by $v_{K}/\Delta_{\phi}$ and thus doubling azimuthal resolution results in a steep price, specifically a quadrupling of the necessary operations from the doubled number of cells to time-advance and the half-timestep being used.  The result is that in the majority of the literature, azimuthal resolution is compromised for computational expediency.   The goal of runs \zzr{32}{R} and \zzr{32}{RR} are to study the importance of azimuthal resolution in a controlled way in comparison to run \zz{32} and deduce a maximum aspect ratio from which converged turbulence is guaranteed.  Figures~\ref{fig:physreda} and~\ref{fig:physredb} shows the evolution of $\alpha_{M}$ and the scaled magnetic energy for these runs.  We note that in both cases we see non-monotonic behavior, specifically reducing the resolution by a factor of two results in increased accretion efficacy and toroidal field amplification.  Further reduction results in a significant drop-off in both of these quantities.  While increased field amplification with reduced resolution is expected from Figure~\ref{fig:physconb}, that a reduction in resolution could increase angular momentum transport is unexpected.  Assessing convergence from these global quantities is murky, at best.  For now, we merely note the importance of azimuthal resolution by pointing out that reducing azimuthal resolution can severely impact accretion efficiency.  The convergence of these reduced azimuthal runs is returned to in \S\ref{sec:convta} where it is demonstrated that run \zzr{32}{R} is converged, whereas \zzr{32}{RR} is not.      

\subsection{Numerical Metrics}
\label{sec:convnum}

\begin{figure}
\centering
\subfloat[Vertical resolvability, ZF runs.]{\label{fig:resolvfraca} \includegraphics[width=0.5\textwidth]{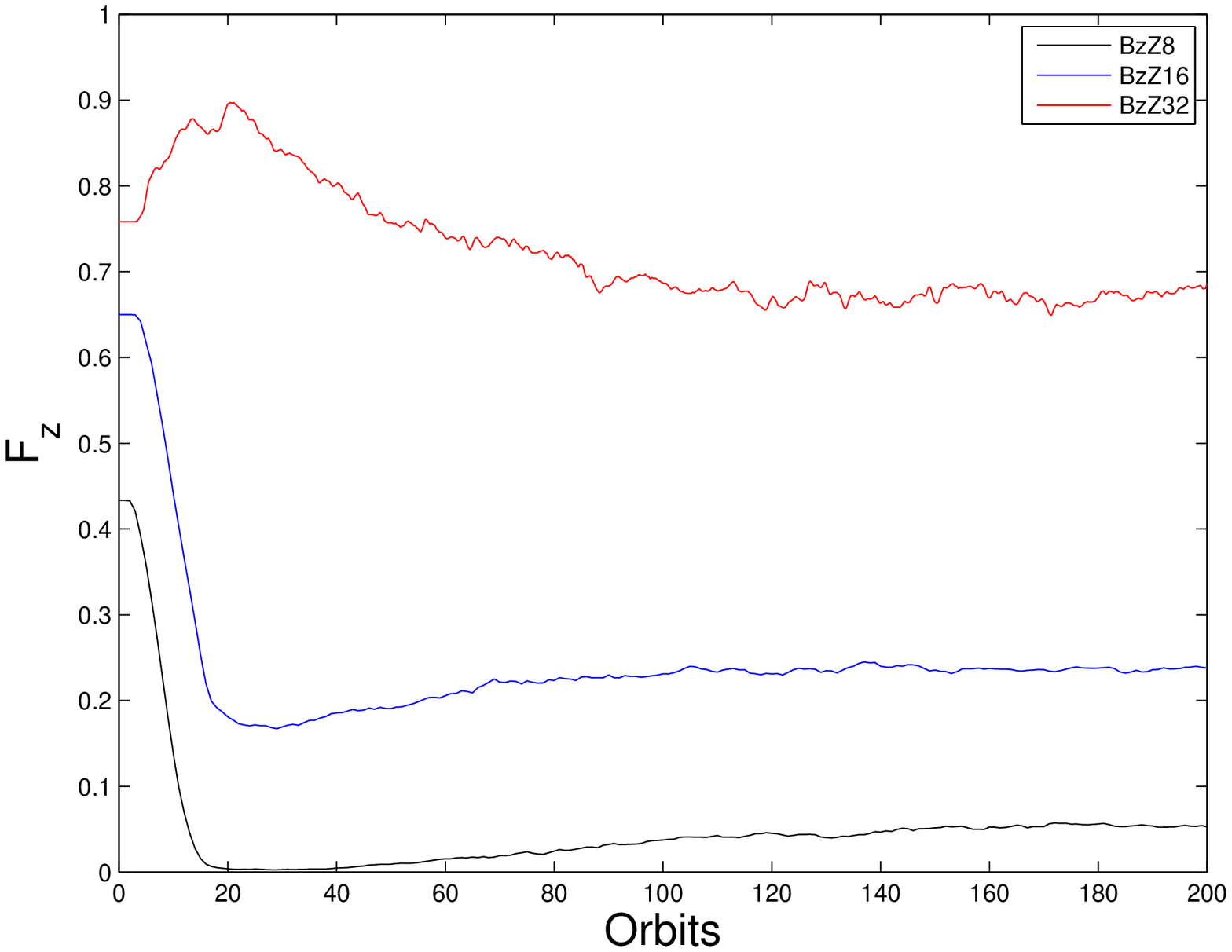}}
\subfloat[Toroidal resolvability, ZF runs.]{\label{fig:resolvfracb} \includegraphics[width=0.5\textwidth]{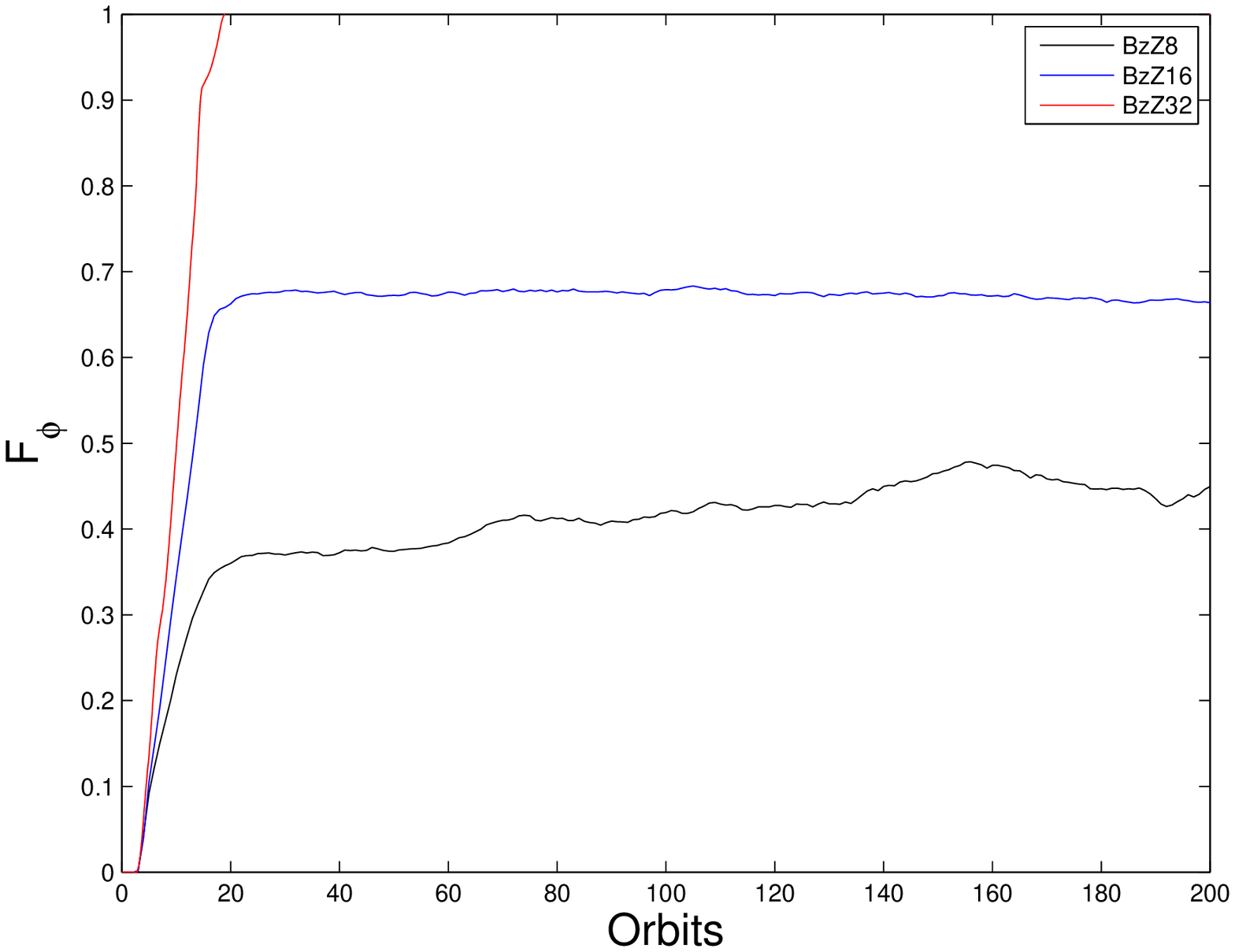}} \\
\subfloat[Vertical resolvability, NF runs.]{\label{fig:resolvfracc} \includegraphics[width=0.5\textwidth]{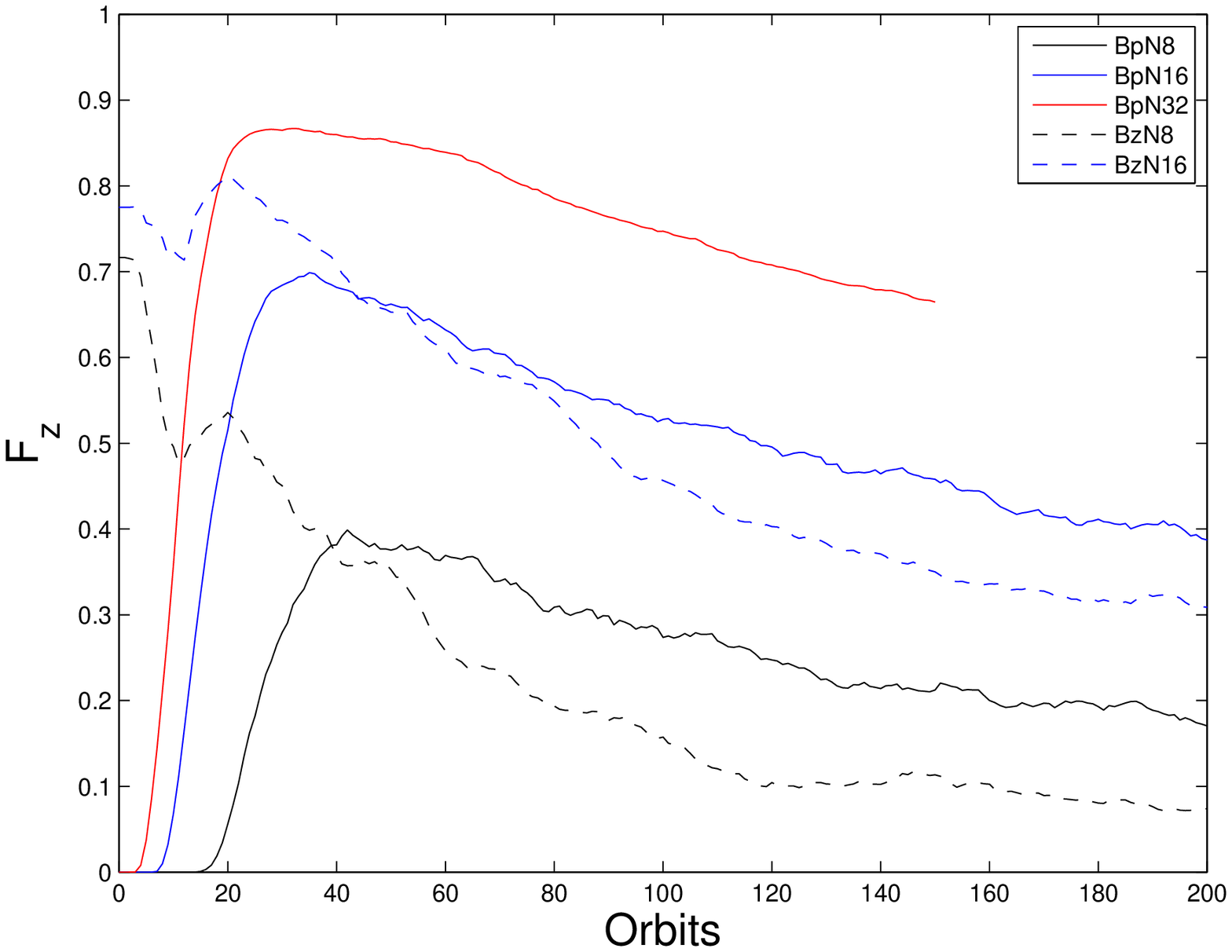}}
\subfloat[Toroidal resolvability, NF runs.]{\label{fig:resolvfracd} \includegraphics[width=0.5\textwidth]{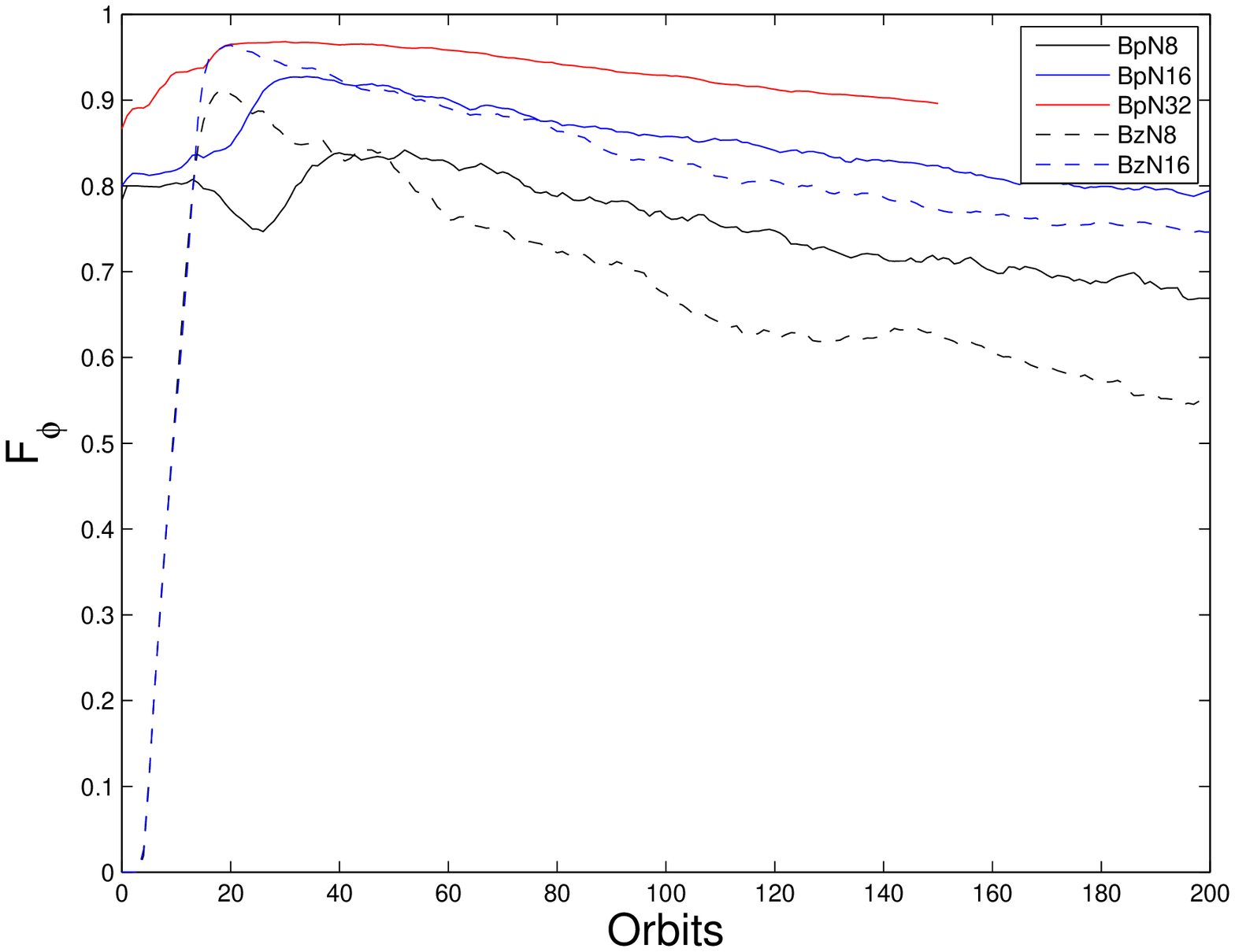}}  
\caption{Vertical and toroidal resolvability fractions for ZF and NF simulations.}
\label{fig:resolvfrac}
\end{figure}

Next, we consider a comparison of the resolvability fractions defined in Equation~\ref{eqn:resolvfrac} and shown in Figures~\ref{fig:resolvfraca} and~\ref{fig:resolvfracb}.  The initial radial profile of the vertical seed field is constructed so that $\lambda_{MRI}/\Delta z \geq 8$ for simulation \zz{8} in a small neighborhood around each maximal value of the sinusoid.  The variations in the initial values of $F_{z}$ are due to the size of the neighborhood about each maximal value of the initial sinusoid in which the resolvability criterion is satisfied.  Of all the diagnostics considered, the most startling behavior is seen when considering $F_{z}$.  The two lower resolution simulations, \zz{8} and \zz{16}, both show a significant initial drop in the resolvability of the vertical MRI correlated with the linear growth and saturation phase of the evolution.  Following the transition, both of these simulations show a slight increase in the resolvability fraction but clearly in most of the disk the vertical MRI is not adequately resolved.  In contrast to this, \zz{32} seems to exhibit substantially differing behavior and shows a value of $F_{z}$ that is roughly constant during the full evolution of the simulation.  The evolution of $F_{\phi}$, shown in Figure~\ref{fig:resolvfracb} is by comparison much simpler and monotonic in nature.  As the toroidal field grows due to shear amplification driven by the vertical MRI, the toroidal MRI becomes resolvable throughout a significant portion of the disk.  

The behavior of the resolvability fractions paints an interesting picture.  Runs \zz{16} and \zz{32} exhibit similar saturated values of stress, but their ability to resolve the vertical MRI seem to be significantly different.  This suggests the intriguing possibility that run \zz{16}, while initially seeded with a vertical field is actually reliant on the resolvability of the toroidal field to reach a comparable stress to run \zz{32}.  However, if these two simulations are actually taking differing routes to turbulence there then must be some additional mechanism that accounts for the similar values of stress achieved in the quasi-steady state.  The resolvability fractions suggest that run \zz{32} has reached a categorically different state in its resolvability of the vertical MRI, and the toroidal MRI is resolvable throughout almost all of the disk.  

Next we consider the resolvability fractions of the NF runs, given in Figures~\ref{fig:resolvfracc} and~\ref{fig:resolvfracd}.  The general resolvability of the vertical MRI is, at first glance, generally somewhat poor, as demonstrated in Figure~\ref{fig:resolvfracc}.  Of all the simulations considered, only \pn{32} resolves the vertical MRI in a majority of the disk over the full evolution.  As was the case with the ZF runs considered, all of the simulations resolve the toroidal MRI in the majority of the disk (Figure~\ref{fig:resolvfracd}).  From these resolvability fractions, only \pn{32} is clearly well-resolved.  Of interest is that the resolvability fraction in the saturated state appears to have stronger dependence on resolution than initial field topology.  

\begin{figure}
\centering
\subfloat[Vertical quality factors, ZF runs.]{\label{fig:quala} \includegraphics[width=0.5\textwidth]{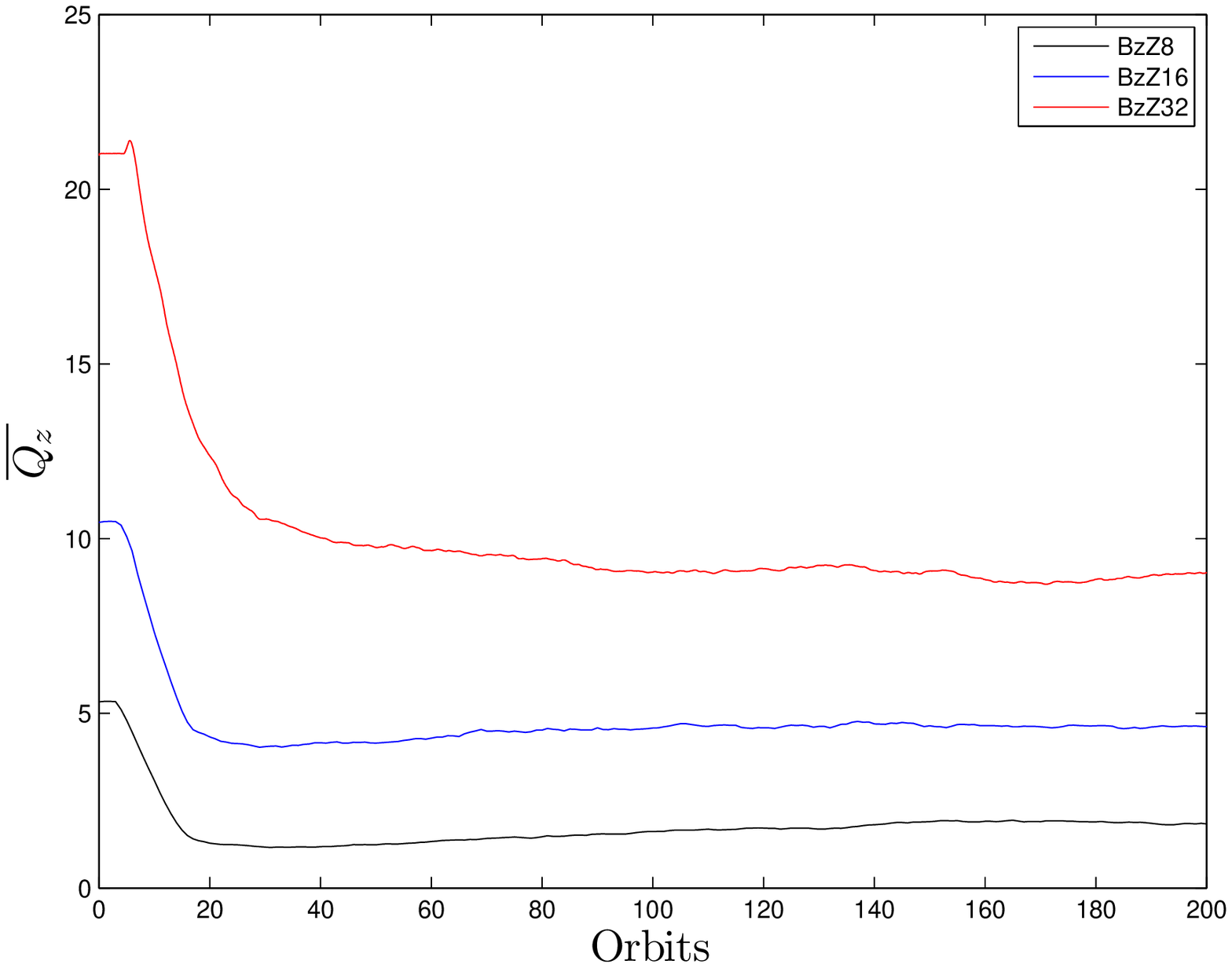}}
\subfloat[Toroidal quality factors, ZF runs.]{\label{fig:qualb} \includegraphics[width=0.5\textwidth]{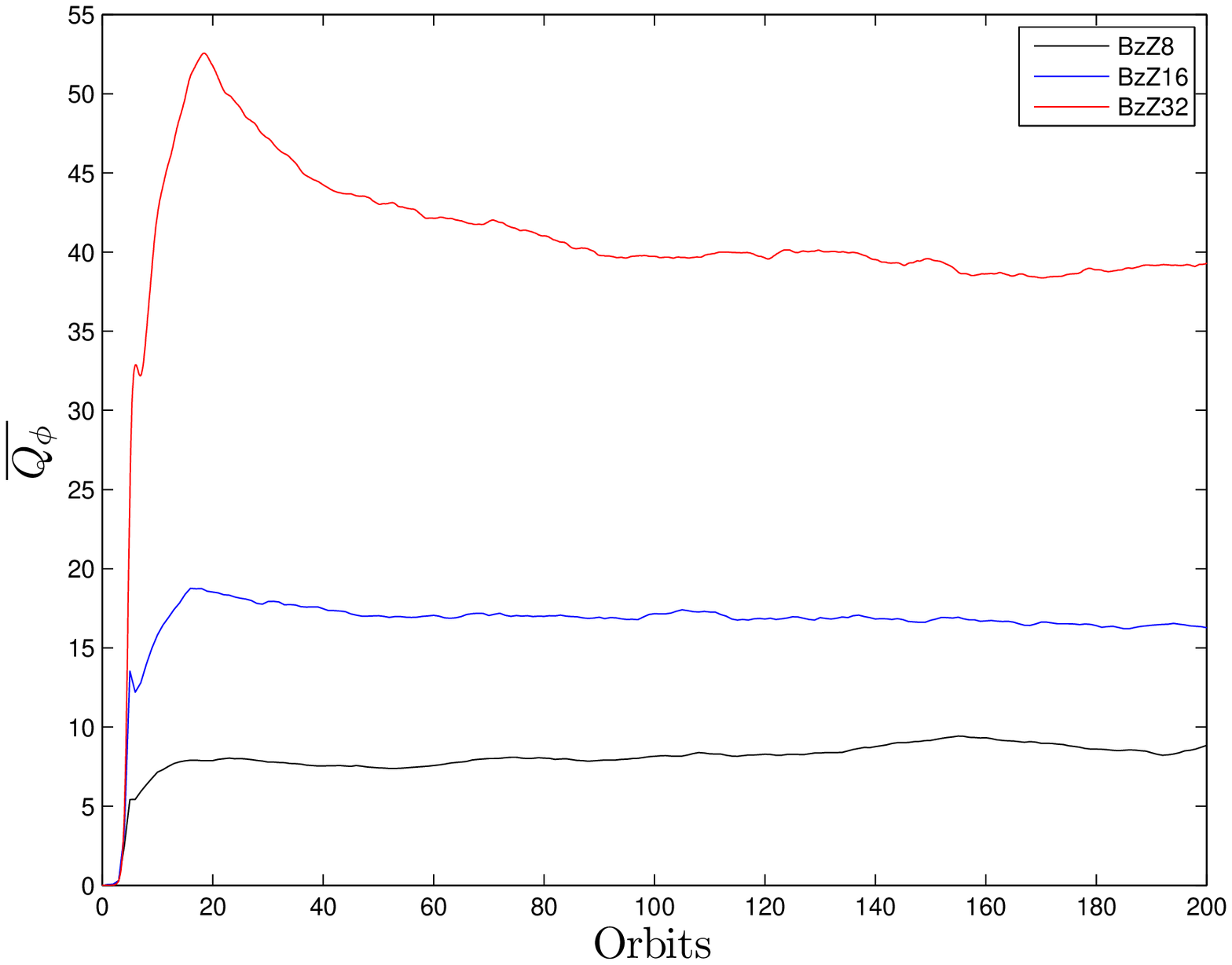}} \\
\subfloat[Vertical quality factors, NF runs.]{\label{fig:qualc} \includegraphics[width=0.5\textwidth]{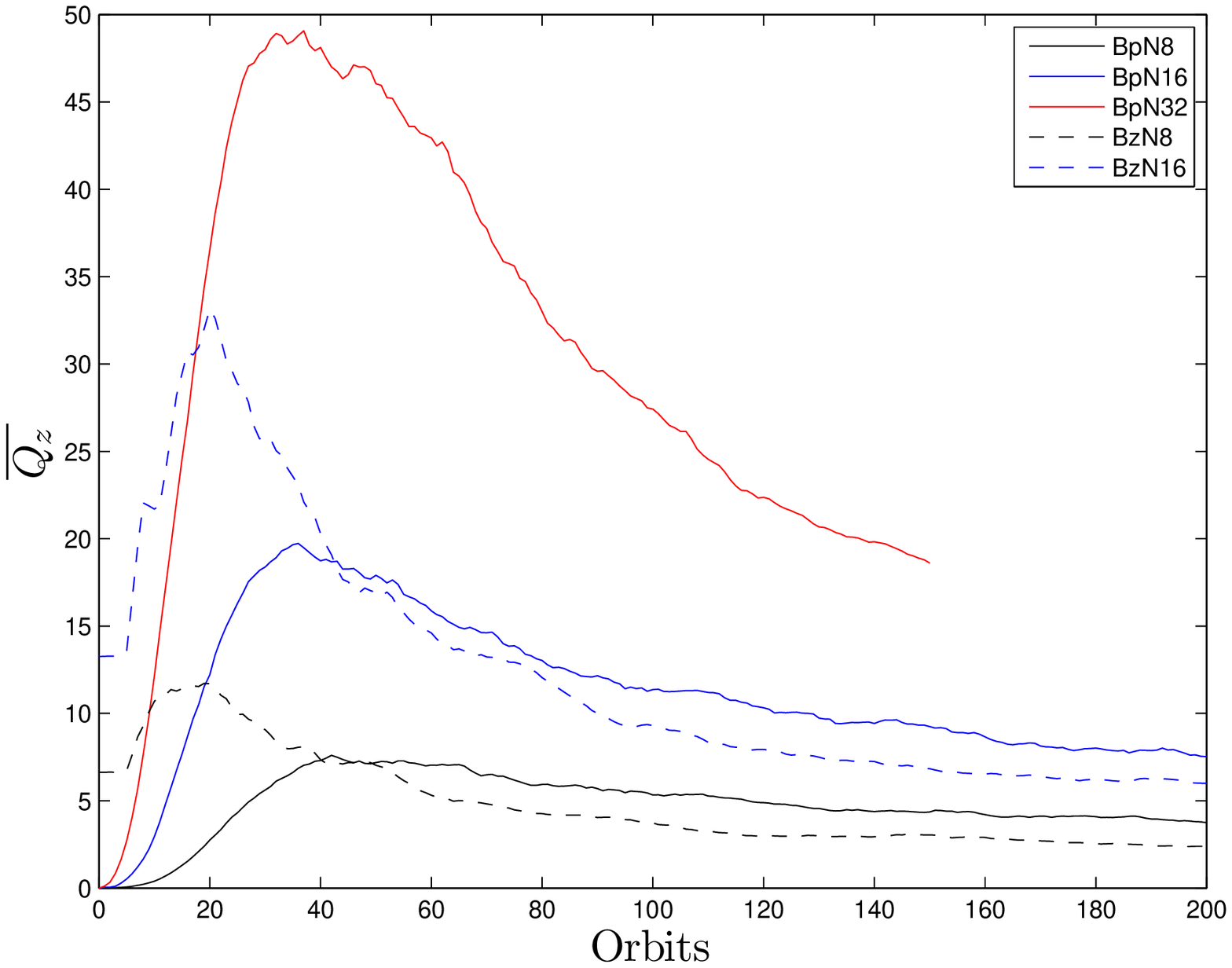}}
\subfloat[Toroidal quality factors, NF runs.]{\label{fig:quald} \includegraphics[width=0.5\textwidth]{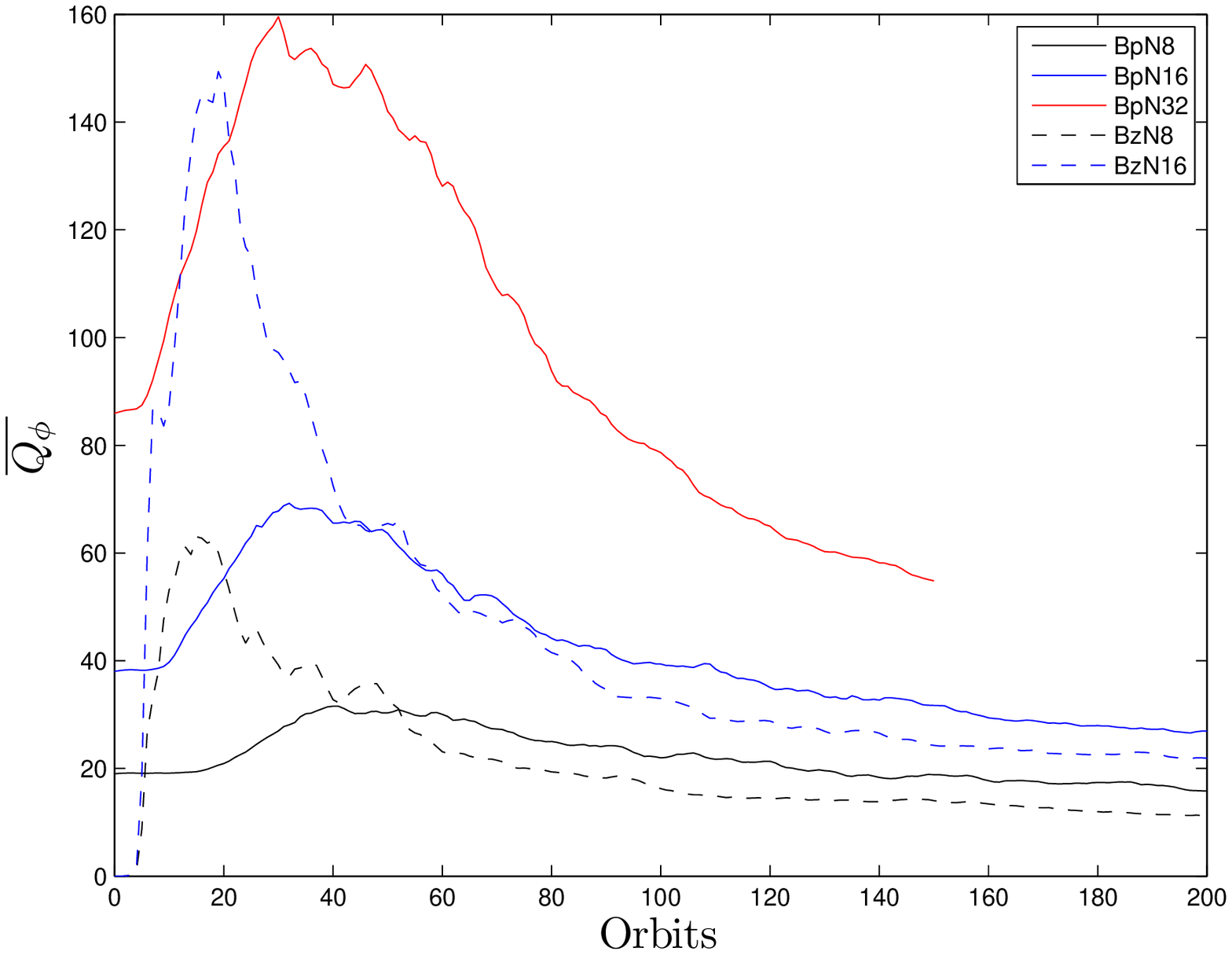}}  
\caption{Temporal evolution of vertical and toroidal quality factors for ZF and NF simulations.}
\label{fig:qual}
\end{figure}

The resolvability fractions employ the often used 8-zone resolvability criterion; however this is somewhat arbitrary.  They allow us to get a sense of how much of the disk is ``well''-resolved.  Now we employ the quality factors to understand how ``well''-resolved the disk is.  Figures~\ref{fig:quala} and~\ref{fig:qualb} illustrate the evolution of the vertical and toroidal quality factors, respectively, for the ZF runs.  The vertical quality factors roughly double with corresponding resolution doublings, starting at $Q_{z} \approx 2$ for the lowest resolution simulation.  This results in a situation in which run \zz{16} is crudely resolved with an average of four zones per $\lambda_{MRI}$ and \zz{32} is demonstrably resolved.  Consideration of the toroidal quality factor results in a reiteration of the point made regarding the toroidal resolvability fraction, $F_{\phi}$, specifically that the two highest resolution simulations clearly are resolving the toroidal MRI.  In particular, run \zz{32} has an average of 40 zones per critical wavelength.  

Consideration of the quality factors for the NF runs (Figures~\ref{fig:qualc} and~\ref{fig:quald}) reveal a similar resolution dependance as that seen in the resolvability fractions.  With the exception of \pn{32}, all of the net-flux runs evolve to a point with an average of a vertical quality factor less than eight, the fiducial criterion.  The toroidal quality factors, in contrast, are quite large and the simulations with all but the lowest resolution simulations exhibit toroidal quality factors above 20.  

Overall, consideration of the resolvability fractions and quality factors are taken as evidence that runs \zz{32} and \pn{32} are resolved whereas \zz{16} is barely resolved.

\subsection{Spectral Metrics}
\label{sec:convspec}

While integrated global quantities are undoubtedly important when diagnosing turbulence, it is often the case that the spectral structure of turbulence is more amenable to study than is the physical structure.  Convergence in this domain is again inherently ill-defined due to the non-dissipative nature of these simulations.  Increasing the resolution of a simulation increases the number of modes in which power can reside, and it would be unphysical to expect that these newly opened modes would remain free of power.  We adopt as a definition of convergence that the mode at which power peaks remains constant with increasing resolution.  To probe the spectral structure at the smallest scales, we include in consideration the wedge runs, \zzw{32} and \zzw{64}, where the function of the former is primarily as a control to ensure that the small-scale behavior is not altered by the reduction of azimuthal domain. 

\begin{figure}
\centering
\subfloat[Density]{\label{fig:zfpowa} \includegraphics[width=0.5\textwidth]{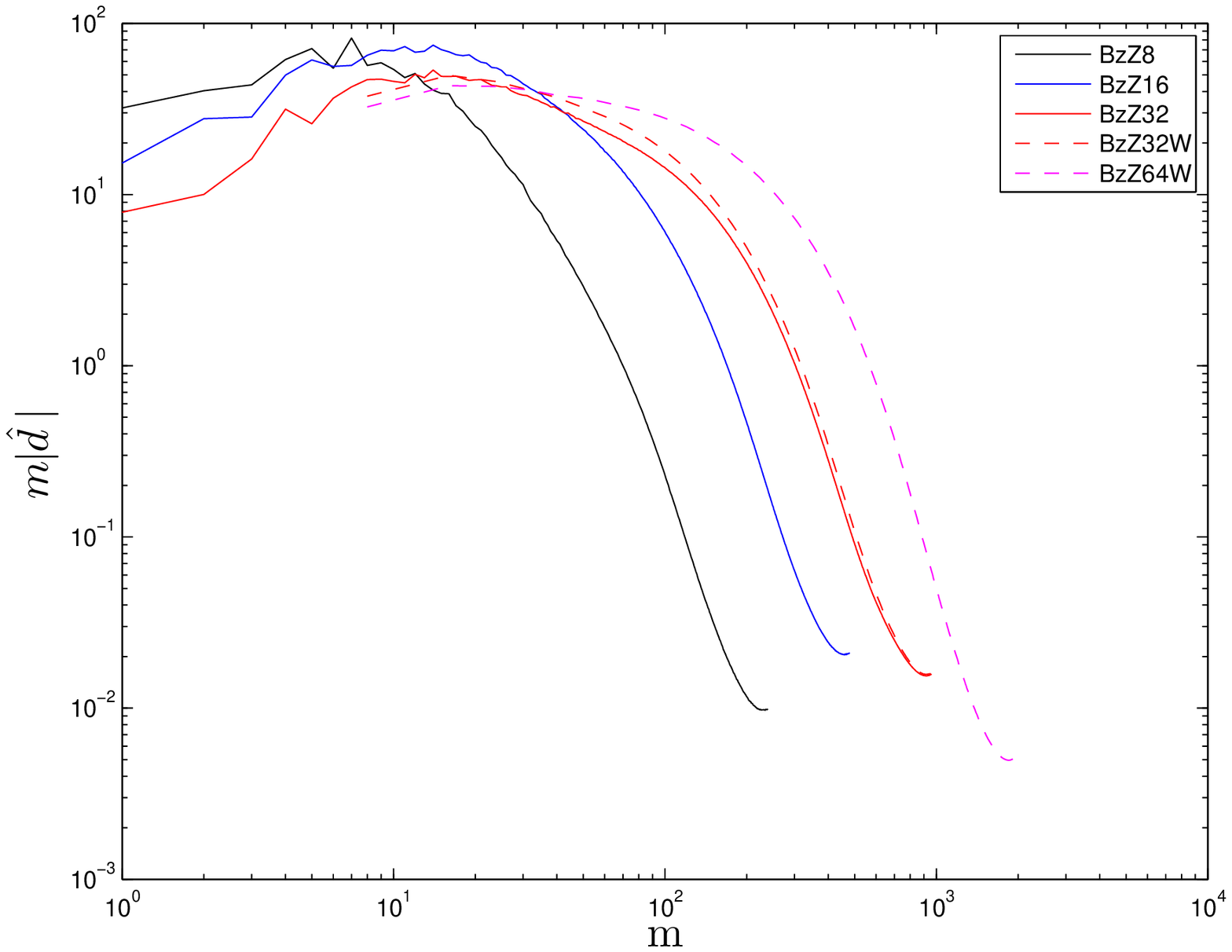}}                
\subfloat[Magnetic Pressure]{\label{fig:zfpowb} \includegraphics[width=0.5\textwidth]{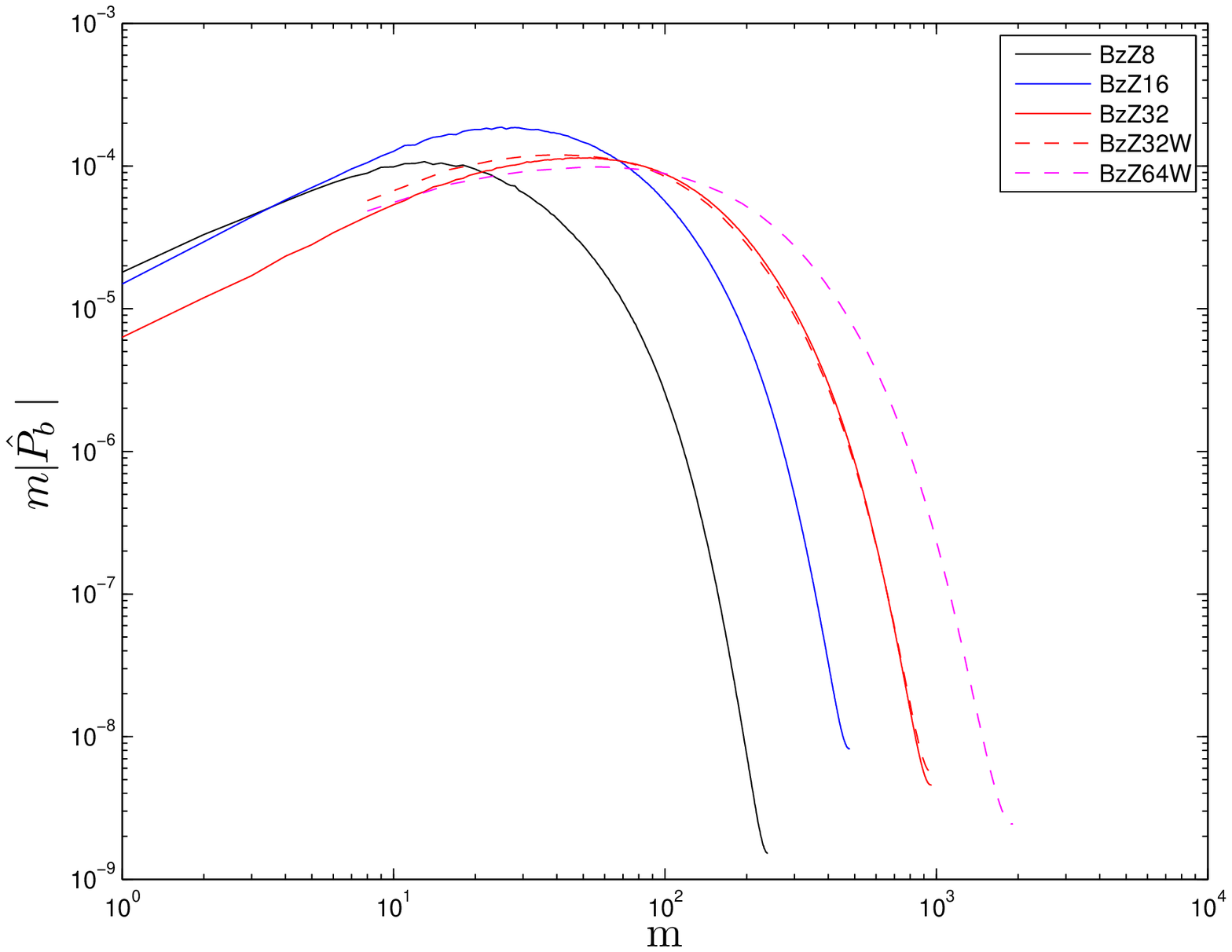}}\\
\subfloat[Stress ($M_{R\phi}$)]{\label{fig:zfpowc} \includegraphics[width=0.5\textwidth]{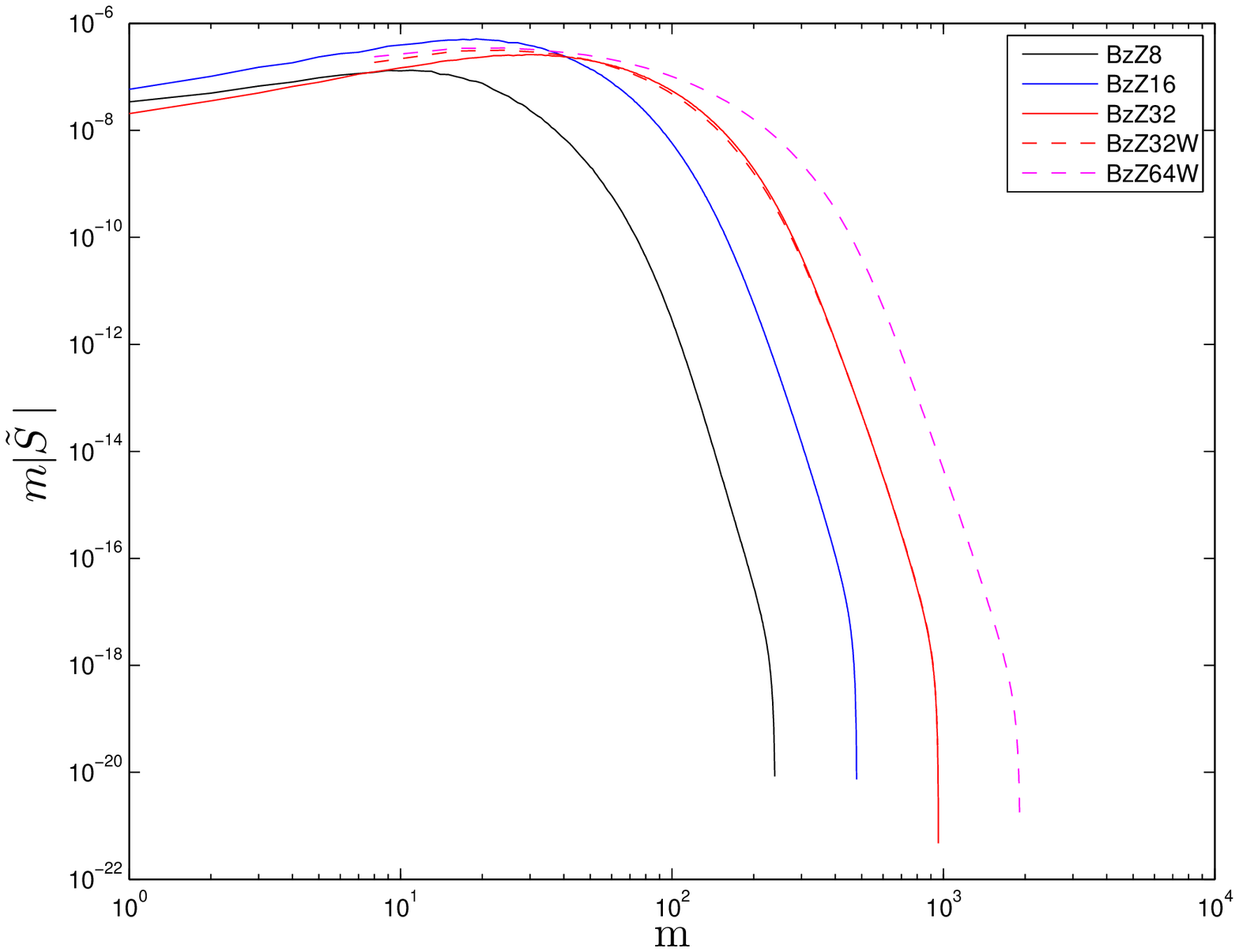}}
\caption{Azimuthal distribution of power, ZF simulations.}
\label{fig:zfpow}
\end{figure}

Figure~\ref{fig:zfpow} shows the time-averaged azimuthal power spectra of the density, magnetic pressure, and stress scaled to remove the secular evolution and temporally averaged for 50 orbits beginning at orbit 50.  It is clear from these figures that reducing the azimuthal domain of the simulation, as done in \zzw{32}, does not significantly change the small-scale distribution of power.  The importance of large-scale structure to the overall density profile of the disk is clear from Figure~\ref{fig:zfpowa}.  Comparing the structure of the magnetic pressure and stress, Figures~\ref{fig:zfpowb} and~\ref{fig:zfpowc} respectively, suggests that the magnetic pressure is clearly dominated by intermediate scales while the stress is more equally distributed at large and intermediate scales.  From a visual inspection, it is clear that all of the power spectra are self-similar and, for resolutions above 32 zones per scale height, all peak at approximately the same scale.  At small scales the power spectra look quite like those of the non-converged, zero net-flux models presented by \citet{fp07}.  The large-scale behavior, however, is quite different as the power at large scales is at most weakly dependent on resolution in stark contrast to the non-converged models of \citet{fp07}.

\begin{figure}
\centering
\subfloat[Density]{\label{fig:nfpowa} \includegraphics[width=0.5\textwidth]{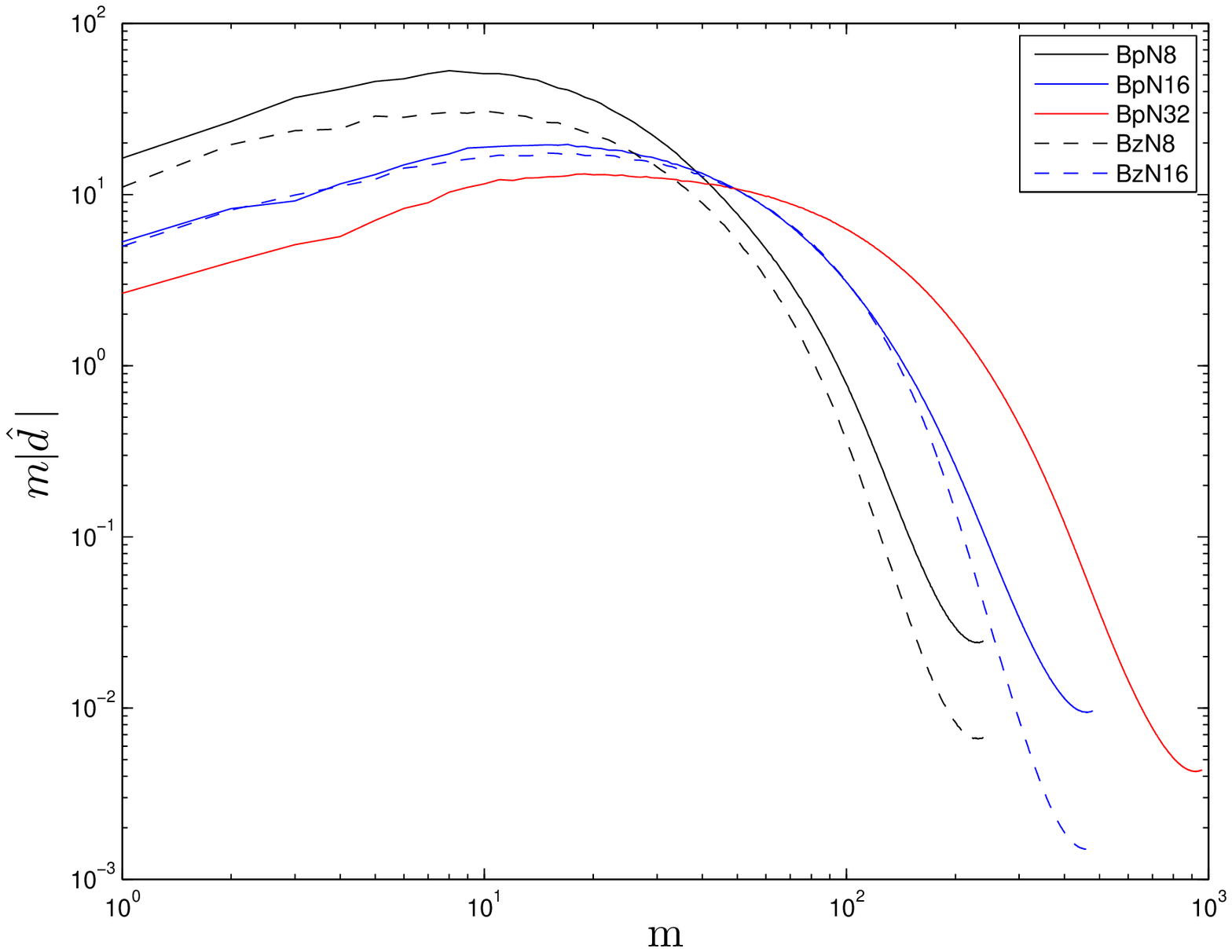}}                
\subfloat[Magnetic pressure]{\label{fig:nfpowb} \includegraphics[width=0.5\textwidth]{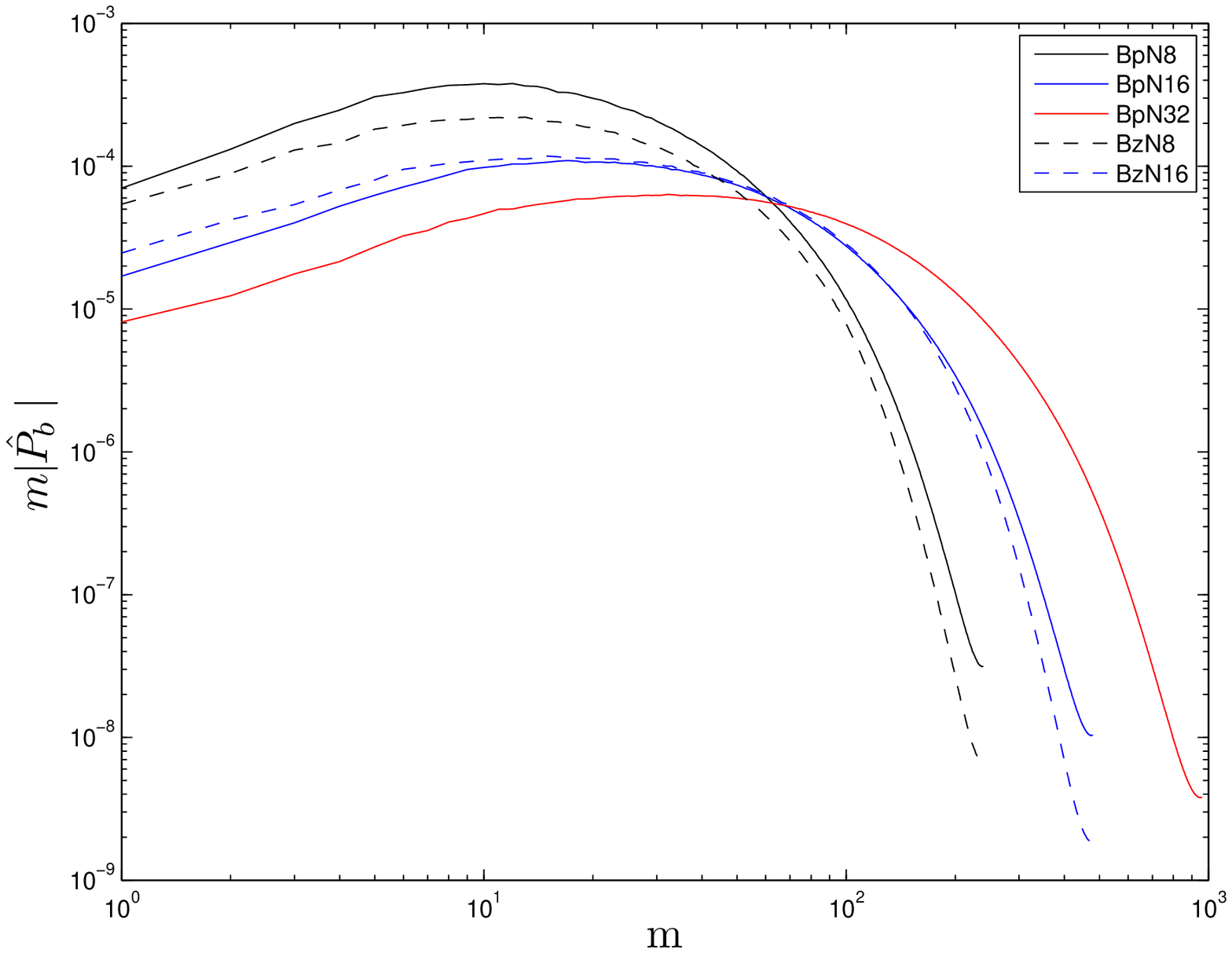}}\\
\subfloat[Stress ($M_{R\phi}$)]{\label{fig:nfpowc} \includegraphics[width=0.5\textwidth]{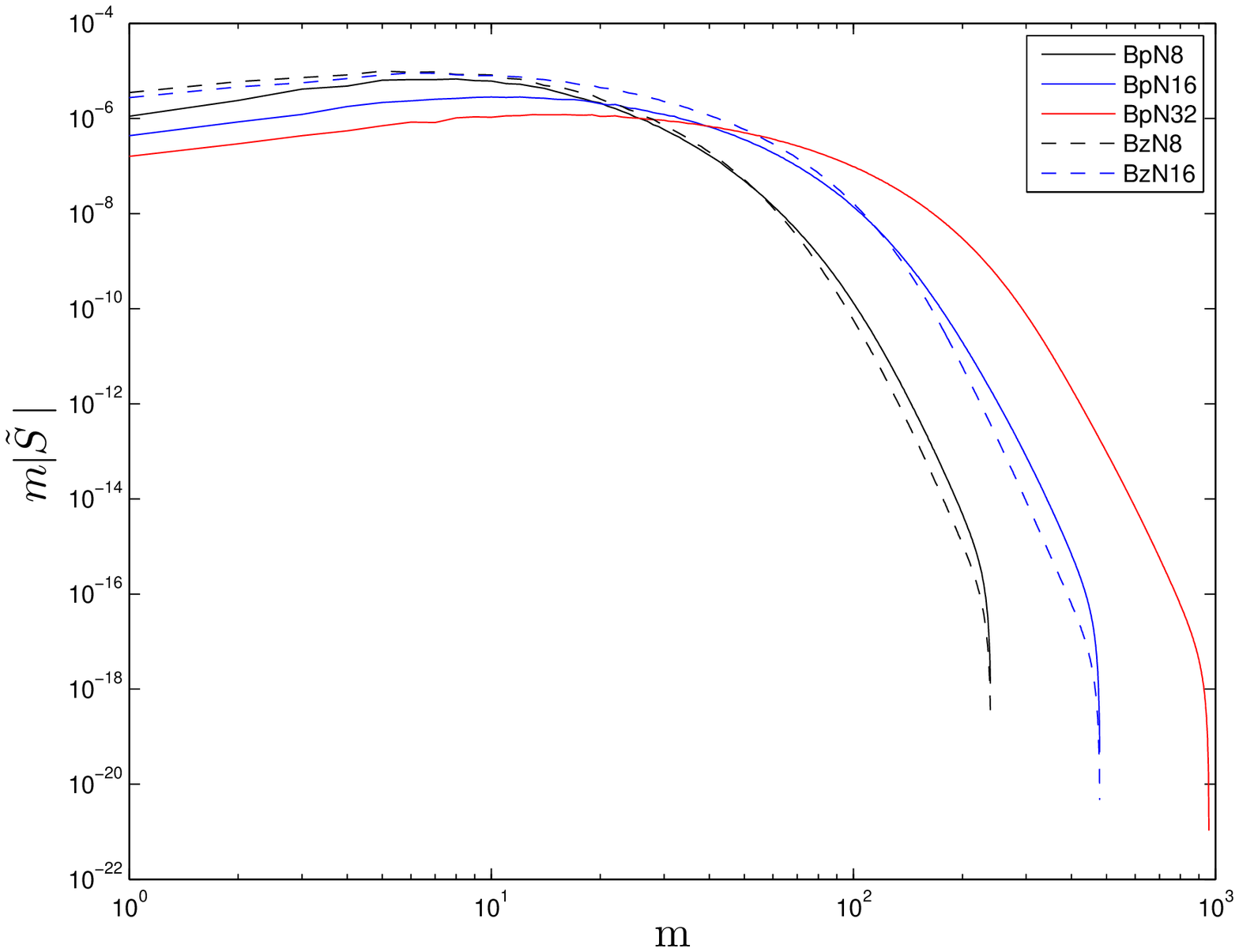}}
\caption{Azimuthal distribution of power, NF simulations.}
\label{fig:nfpow}
\end{figure}

Next, we consider the azimuthal spectral structure of the turbulence for the NF runs.  As for the ZF runs we consider the density (Figure~\ref{fig:nfpowa}), the magnetic pressure (Figure~\ref{fig:nfpowb}), and the stress (Figure~\ref{fig:nfpowc}).  Broadly, we note again that the structure is dependent largely on resolution and only minimally on field topology.  Simulations of lower resolution are more dependent on large-scale features, and as resolution increases we see a more even distribution of power between the intermediate and small scales.  While the density and pressure have a clear peak at an intermediate scale, the scale-distribution of stress is more evenly distributed at large to intermediate scales.  

To quantify the dominant scale, we consider in Table~\ref{tab:pow} the wavenumber at which the wavenumber-scaled power peaks for each simulation and for each of the variables considered.  For the eventual comparison of these results with local models, it is useful to consider this measurement both as wavenumber, appropriate for global simulations, and as an effective azimuthal wavevector, $k_{\phi}H_{0}$, which is more directly analogous to local models.  The dominant density scale is much larger than the corresponding dominant magnetic scales.  All of the simulations considered exhibit a dominant density scale of approximately $4H_{0}$, and in this regard resolution does not appear to play a significant role save for the absolute lowest resolution simulation.  Reducing the azimuthal domain results in a increase of the effective scale associated with the magnetic quantities.  While the magnetic quantities do not clearly converge with increasing resolution, the peak of the highest resolution simulation, \zzw{64}, is almost precisely displaced in wavenumber by eight.  The reduced azimuthal domain simulations use only an eighth of the full $2\pi$ domain, which means that the accessible wavenumbers are limited to modes that are integer multiples of eight.  In this regard, the peak of runs \zzw{32} and \zzw{64} are within one accessible wavenumber of the peak of \zz{32}.  With this in mind, we again take this as evidence of the convergence of \zz{32}.  

The dominant density scale is roughly halved when going from run \pn{8} to \pn{16}, however the transition from \pn{16} to \pn{32} results in only a minor $(\sim 10\%)$ decrease and suggests that this latter run is near convergence.  This is contrasted with the dominant scales of the magnetic energy and stress which decrease much more significantly in the transition from \pn{16} to \pn{32}.  While the dominant density scale of run \zz{32} is twice that of the dominant stress scale, the toroidal field runs display a much more comparable value of these two scales.  Overall, we take this evidence as suggesting that run \pn{32} is approaching convergence but not fully so.

The most striking feature of the net vertical field simulations is the significantly larger dominant stress scale $(\sim 10H_{0})$.  Also larger than expected is the dominant magnetic energy scale, which is again in excess of the other topologies considered.  These large scales may suggest the existence of a memory of the initial field configuration.    In particular, this large structure may be a remnant of the linear shearing phase in the initial growth of the MRI.  This linear shearing phase is particularly strong in the simulations initialized with a net vertical field, and exhibit magnetic pressures in excess of the gas pressure.  

\begin{table}[htdp]
\begin{center}
\begin{tabular}{|c|c|c|c|c|c|c|}
\hline 
Simulation & \multicolumn{2}{|c|}{Density Peak} & \multicolumn{2}{|c|}{$P_{b}$ Peak} &\multicolumn{2}{|c|}{Stress Peak} \\
\hline
 & m & $k_{\phi}H_{0}$ & m & $k_{\phi}H_{0}$ & m & $k_{\phi}H_{0}$\\
\hline
 \zz{8} & 7 & 0.111  & 13 & 0.207 & 11 & 0.176 \\
 \zz{16} & 14 & 0.223 & 25 & 0.398 & 19 & 0.302 \\
 \zz{32} & 14 & 0.223 & 48 & 0.764 & 33 & 0.525 \\
 \zzw{32} & 16 & 0.255 & 40 & 0.637 & 24 & 0.382 \\
 \zzw{64} & 16 & 0.255 & 56 & 0.891 & 24 & 0.382 \\
\hline
 \pn{8} & 8 & 0.127  & 12 & 0.191 & 8 & 0.127\\
 \pn{16} & 17 & 0.271 & 17 & 0.271 & 12 & 0.191\\
 \pn{32} & 19 & 0.302 & 32 & 0.509 & 16 & 0.255\\
\hline
 \zn{8} & 10 & 0.160 & 13 & 0.207 & 5 & 0.080\\
 \zn{16} & 15 & 0.234 & 15 & 0.234 & 6 & 0.095 \\
\hline

\end{tabular}
\end{center}
\caption{Comparison of dominant azimuthal mode in the power spectra of density, magnetic pressure, and stress.}
\label{tab:pow}
\end{table}%

\subsection{Tilt Angle}
\label{sec:convta}

\begin{figure}
\centering
\subfloat[ZF Runs]{\label{fig:tacompa} \includegraphics[width=0.5\textwidth]{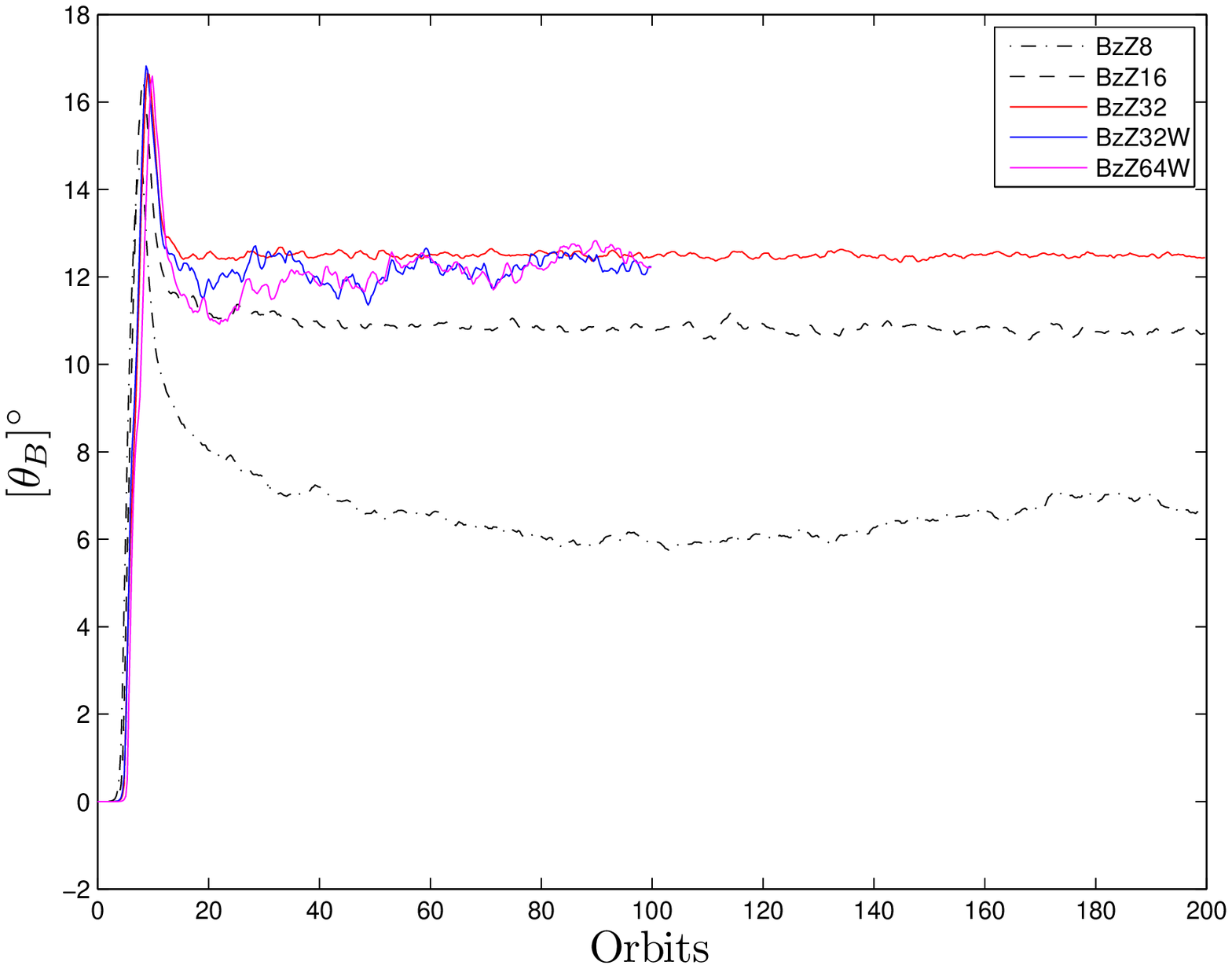}}                
\subfloat[NF Runs]{\label{fig:tacompb} \includegraphics[width=0.5\textwidth]{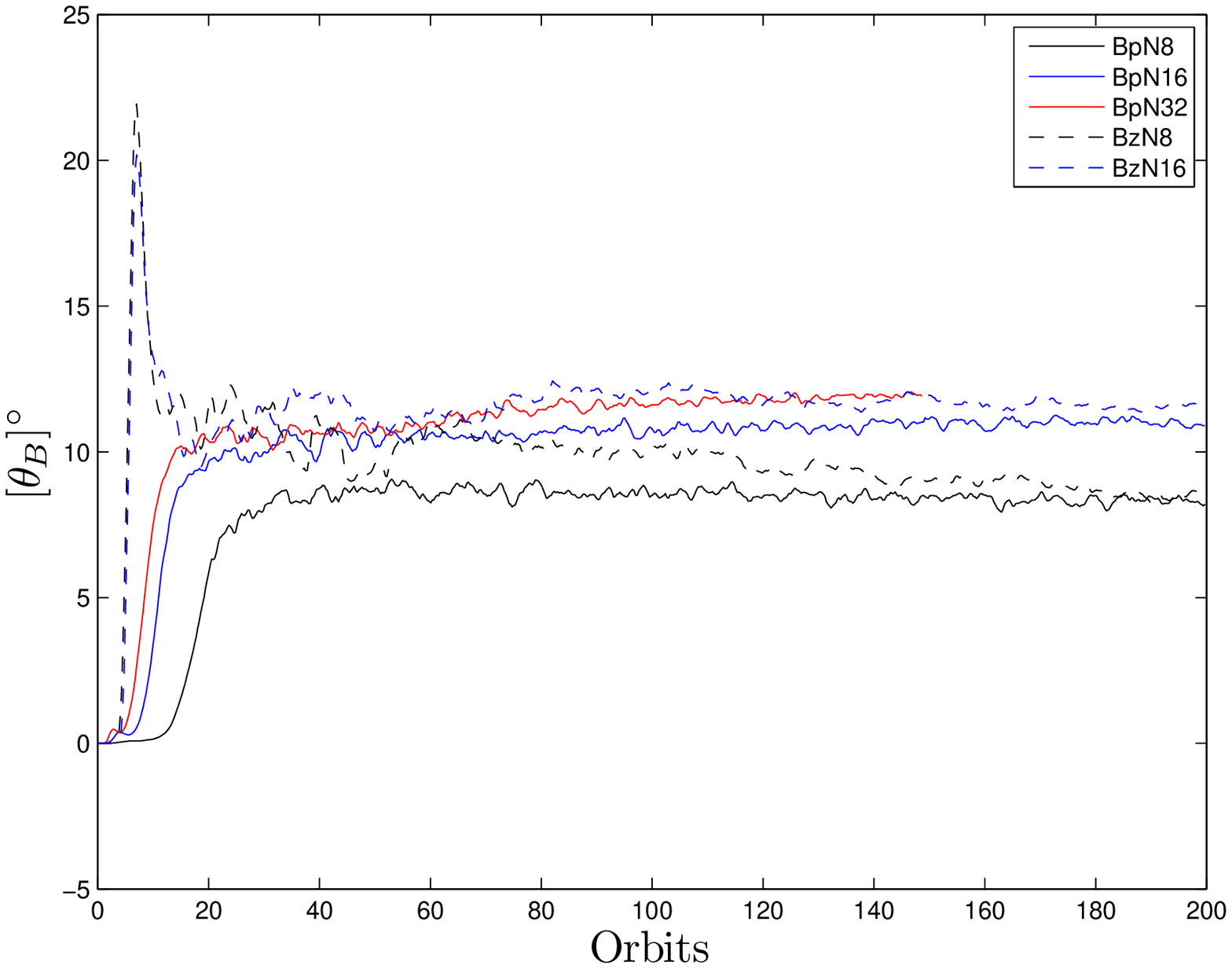}}\\
\subfloat[Reduced Azimuthal Runs]{\label{fig:tacompc} \includegraphics[width=0.5\textwidth]{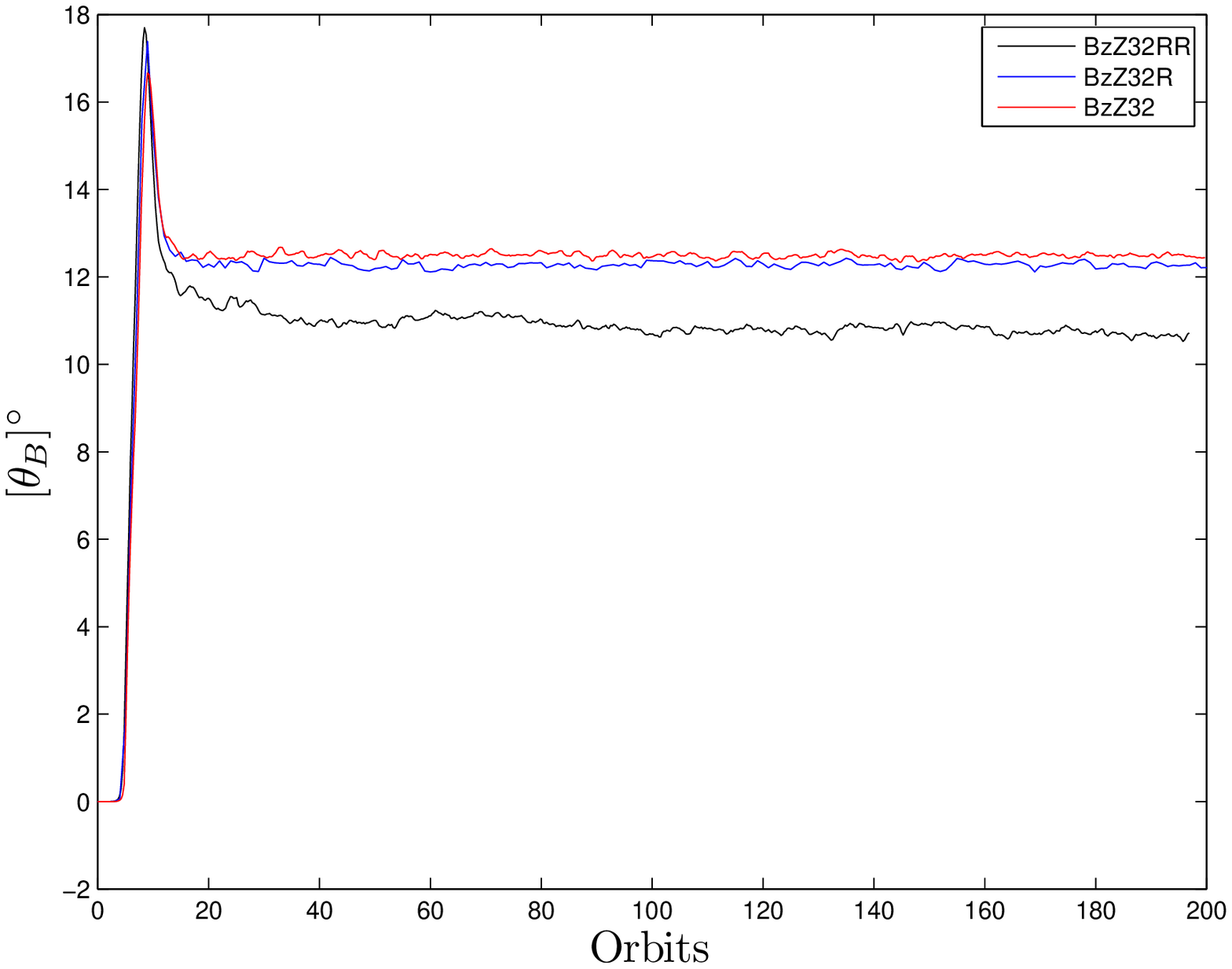}}                
\subfloat[Comparison of fiducial models]{\label{fig:tacompd} \includegraphics[width=0.5\textwidth]{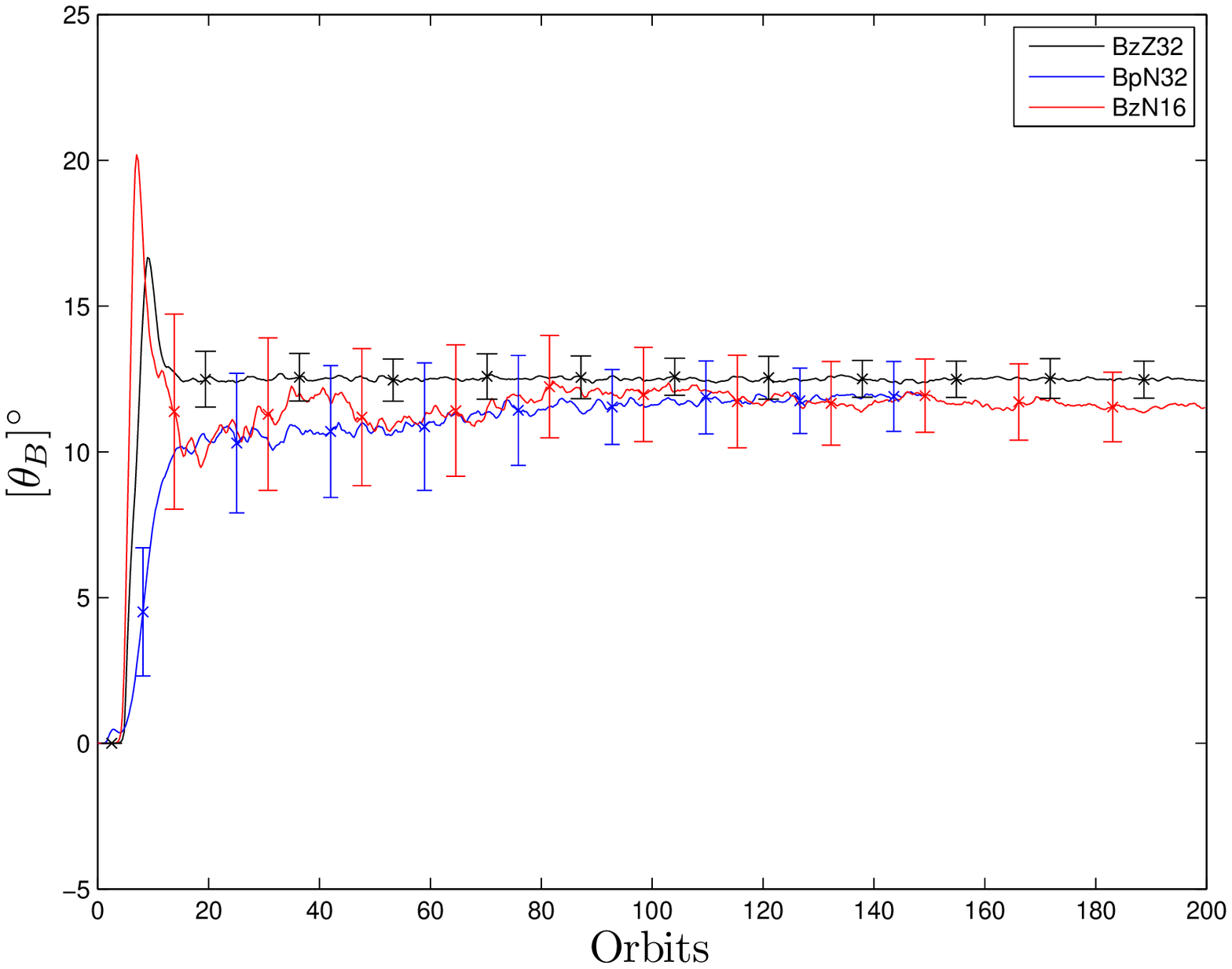}}\\  
\caption{Temporal evolution of the magnetic tilt angle.}
\label{fig:tacomp}
\end{figure}

The metrics discussed thus far are useful tools to measure convergence, but they suffer from the limitation that their behavior must be compared against other simulations using identical field topologies.  Meaningful convergence studies can be computationally quite expensive and therefore a more robust indicator of convergence that is independent of field topology would be useful.  The evidence we will present here suggests that the magnetic tilt-angle, defined in Eqn~\ref{eqn:ta}, may indeed be this diagnostic.  As a precursor to this, we consider the evolution of the tilt angle in all of the ZF simulations (Figure~\ref{fig:tacompa}).  The magnetic tilt angle reaches an approximately steady-state value that is almost identical for all of the higher resolution simulations (above 32 zones/$H_{0}$).  This value, $\theta_{B} \approx 13^{\circ}$, is remarkably close to the estimated value of $15^{\circ}$ (\citeinp{ggsj09}).  Also of note is that while the initial peaks in both stress and magnetic energy are strongly resolution dependent, the tilt angle, a function of the product of the two, has an initial peak that is seemingly independent of resolution.  This may suggest that the transition from the linear growth of the MRI into saturated turbulence depends on a precise relationship between the quantities $\alpha_{M}$ and $\beta$.                    

Next, we consider the behavior of the magnetic tilt angle in the NF runs, given in Figure~\ref{fig:tacompb}.  As resolution increases, we see a corresponding increase in the tilt angle.  Regarding the toroidal field simulations, we note the strong similarity between runs \pn{16} and \pn{32}.  This suggests that a further doubling in resolution may indeed prove convergence.  The vertical field simulations display a similar behavior and ambiguity.  Unfortunately, with the simulations available we are not able to conclusively demonstrate convergence for these net field runs, however we do argue that \pn{32} is likely converged.  

At this point, we recall the difficulty of defining convergence using the physical metrics in the context of the reduced azimuthal resolution runs, \zzr{32}{R} and \zzr{32}{RR}.  The physical metrics are ambiguous but, when we consider the tilt angle (Figure~\ref{fig:tacompc}), we see a clear monotonicity with resolution.  A reduction in the azimuthal resolution by a factor of two results in a very minor change in the tilt angle, however further reduction of the azimuthal resolution results in a much more significant alteration.  We take this as evidence that run \zzr{32}{R} is converged, whereas \zzr{32}{RR} is not.  This demonstrates the importance of azimuthal resolution, and indeed suggests that treating azimuthal resolution on nearly equal footing with vertical resolution is of vital importance towards ensuring simulations that are numerically converged. 

The results we have presented indicate that the magnetic tilt angle is a powerful diagnostic tool towards demonstrating convergence of MRI-driven accretion disk simulations.  The fact that it is monotonic with resolution, exhibits minimal variation in the saturated state, and appears to converge suggest that fiducial values of converged tilt angle can be computed and compared against simulations.  Additionally, as demonstrated in Figure~\ref{fig:tacompd}, the value in the saturated state appears to be almost independent of initial field topology and that there may exist a single fiducial value of tilt angle.  We conjecture that higher resolution simulations initialized with a net field will indeed definitively demonstrate the existence of a single scalar value of the magnetic tilt angle that characterizes accretion disk turbulence.  Assuming this can be done, this would prove a powerful tool towards verifying convergence without resorting to the significant expense of running higher resolution control simulations.  Further, the existence of a single fiducial tilt angle would imply the ability to unify the often disparate phenomenology associated with varying initial field topologies.

The natural question raised by the existence of a single scalar that appears to characterize the saturated, fully non-linear state of MRI-driven MHD turbulence is whether this can be derived from theory.  The relationship, $\alpha \beta \approx 1/2$, was first observed by \citet{hgb} and discussed at length by \citet{Blackman:2008fe}.  The latter work includes a heuristic argument that $\alpha$ and $\beta$ should be related by a constant, however predicting that constant is beyond the means of the dimensional analysis used.  Recent work by \citet{pg09} provides the potential for a more quantitative understanding of this relationship.  Through an analysis of the growth and subsequent saturation of a single MRI mode by parasitic instabilities \citet{pg09} relate the values of $\alpha$ and $\beta$ in the saturated state to the magnetic tilt angle $\theta_{B}$.  Additionally, this analysis suggests the dependence of the magnetic tilt angle on dissipative coefficients.  While \citet{pg09} provides a theoretical foundation as to the importance of the tilt angle, the manner in which this analysis can be extended to a fully nonlinear simulation in which mode-mode interaction is likely important is unclear and beyond the scope of this paper. 

\subsection{Discussion}
\label{sec:convdis}

\citet{hgk11} survey a series of local simulations to test various proposed convergence metrics, and then compare these metrics with a series of pseudo-Newtonian simulations of thick accretion disks.  Our comparison to their work begins with their choice of metrics: $\alpha_{Mag} = M_{R\phi}/P_{B}$; the quality factors, $Q_{z}$ and $Q_{\phi}$ (though the definitions given there differ from ours by a numerically negligible factor of $\sqrt{16/15}$); and the correlations, $<B_{R}^{2}/B_{\phi}^{2}>$ and $<B_{z}^{2}/B_{R}^{2}>$.  We note that the measure $\alpha_{Mag}$ used in \citet{hgk11} is not the same as our metric $\alpha_{M}$; while both involve a pressure scaling of the magnetic stress the former uses the magnetic pressure whereas the latter uses the gas pressure.  We also note that two of the metrics proposed are in an information content sense equivalent to the magnetic tilt angle, $\theta_{B}$.  Due to the nature of the magnetic tilt angle being a measure of anisotropy in the planar field, we can rewrite $\alpha_{Mag} = \alpha_{M}\beta$ which implies $\sin(2\theta_{B}) = \alpha_{Mag}$.  Similarly, the correlation $B_{R}^{2}/B_{\phi}^{2} = \tan^{2}(\theta_{B})$.  Regarding the latter correlation, $B_{z}^{2}/B_{R}^{2}$, the authors note that this term does not appear to exhibit a strong trend with increasing resolution and as such we will not consider it further.  To aid in our comparison we include a summary of convergence metrics in Table~\ref{tab:conv}.

For the local models reviewed in \citet{hgk11}, simulations of 64 zones per scale-height result in quality factors of $Q_{z} \approx 10$ and $Q_{\phi} \approx 40$.  These are fairly comparable to each other, and run \zz{32} (Table~\ref{tab:conv}), despite the disparate initial field topologies.  A zero-net vertical field is used in \citet{dsp10} whereas a net toroidal field is used in \citet{shb11}.  Of note, is that the net toroidal field model considered here, \pn{32}, shows significantly higher quality factors at lower resolutions.  While \citet{dsp10} also consider a model utilizing 128 zones per scale-height which results in an increase in the quality factors by approximately 150\%, this does not result in a change in $\alpha_{Mag} = 0.36$ from the 64 zone per scale-height run.  This value of $\alpha_{Mag}$ corresponds to $\theta_{B} \approx \indeg{10.5}$.  The simulations discussed in \citet{shb11} result in a largest value of tilt angle of $\theta_{B} \approx \indeg{11.8}$.  These are both in contrast to the values seen in our global runs with a slightly larger value of $\theta_{B} \approx \indeg{12} - \indeg{13}$.    

In their discussion of stratified local models, \citet{hgk11} also note the importance of azimuthal resolution which is bolstered by the work presented here.  In \S\ref{sec:convta} we present evidence that when the azimuthal resolution is reduced by a factor of four from an aspect ratio of unity, the resultant simulations exhibit significantly lower accretion and magnetic tilt angles.  Also discussed, is the ability for large toroidal quality factors to compensate for a poorly resolved vertical MRI.  Indeed, this likely explains the results for simulation \zz{16} in which a comparable value of $\alpha$ in the steady-state is achieved despite the significant discrepancy between the resolvability fractions of simulations \zz{16} and \zz{32}.            

The global simulations presented by \citet{hgk11} are diagnosed to have lower quality factors and smaller tilt angles than the ones presented here and to the local models they consider.  However, we observe that the stratified local models also seem to involve a smaller tilt angle than what may be expected from our unstratified global models.  \citet{bas11} measure a tilt angle of approximately $\indeg{9}$ in a stratified global simulation.  This suggests that stratification itself may suppress tilt angle somewhat and a future resolution study of stratified global disks using orbital advection would be useful.  Overall, the work suggests that $Q_{z} \gtrsim 10-15$ for poorly resolved toroidal quality factors ($Q_{\phi} \approx 10$), and that larger values of toroidal quality factor ($Q_{\phi} \gtrsim 25$) can alleviate the constraint on the vertical quality factor.  Based on these constraints, we find that all of the simulations utilizing a resolution exceeding or equal to 32 zones per scale-height are converged as well as the simulations seeded with a net field and above 16 zones per scale-height.  The criterion presented here of a converged tilt angle would not consider\ the runs \zn{16} and \pn{16} converged, but would consider run \zzr{32}{R} converged in contrast to the quality factor criterion. 

\begin{table}[htdp]
\begin{center}
\begin{tabular}{|c|c|c|c|}
\hline
Simulation & $ <\theta_{B}>_{QSS} $ & $ \bar{Q_{z}}_{,QSS} $ & $\bar{Q_{\phi}}_{,QSS}$ \\
\hline
\zz{8} & $\indeg{8.61}$ & 1.68 & 8.38 \\
\zz{16} & $\indeg{11.44}$ & 4.57 & 16.82 \\
\zz{32} & $\indeg{12.86}$ & 9.91 & 41.22 \\
\zzw{32} & $\indeg{12.75}$ & 9.98 & 33.95 \\
\zzw{64} & $\indeg{12.76}$ & 24.51 & 73.77 \\
\zzr{32}{RR} & $\indeg{11.12}$ & 6.64 & 7.38 \\
\zzr{32}{R} & $\indeg{12.59}$ & 9.08 & 16.13 \\
\hline
\pn{8} & $\indeg{8.08}$ & 4.46 & 18.86 \\
\pn{16} & $\indeg{10.98}$ & 9.23 & 31.92 \\
\pn{32} & $\indeg{12.09}$ & 22.2 & 64.22 \\
\zn{8} & $\indeg{9.37}$ & 3.54 &16.24 \\
\zn{16} & $\indeg{11.83}$ & 8.97 & 32.41 \\
\hline
\end{tabular}
\end{center}
\caption{Convergence Metrics}
\label{tab:conv}
\end{table}%

\section{Comparing Local and Global Models}
\label{sec:lvg}

The restricted geometric domain of local models makes them ideal to explore accretion disk turbulence at resolutions far in excess than what is feasible in global models.  However ensuring that these local models accurately represent the small-scale dynamics of accretion disk turbulence is a vital validation of the wealth of results learned from local models.  The simulations presented here provide a unique opportunity to accomplish this as the resolutions of these global models are not only comparable to, but greater, than the vast majority of local simulations in the literature.

\begin{figure}
\centering
\subfloat[Azimuthal velocity ($v_{y}/c_{s}$)]{\label{fig:sbstilla} \includegraphics[width=0.6\textwidth]{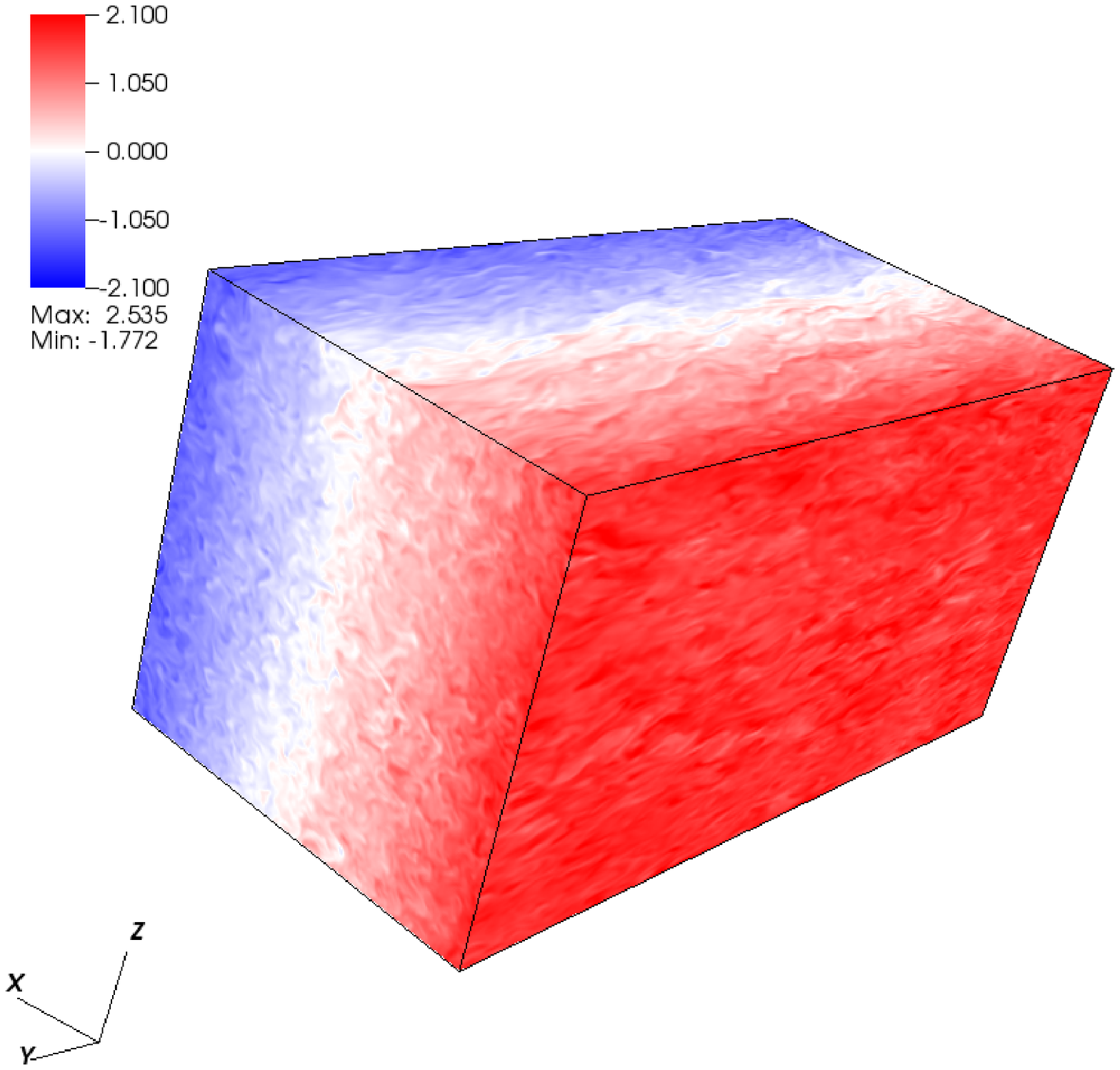}}\\                
\subfloat[Azimuthal magnetic field]{\label{fig:sbstillb} \includegraphics[width=0.6\textwidth]{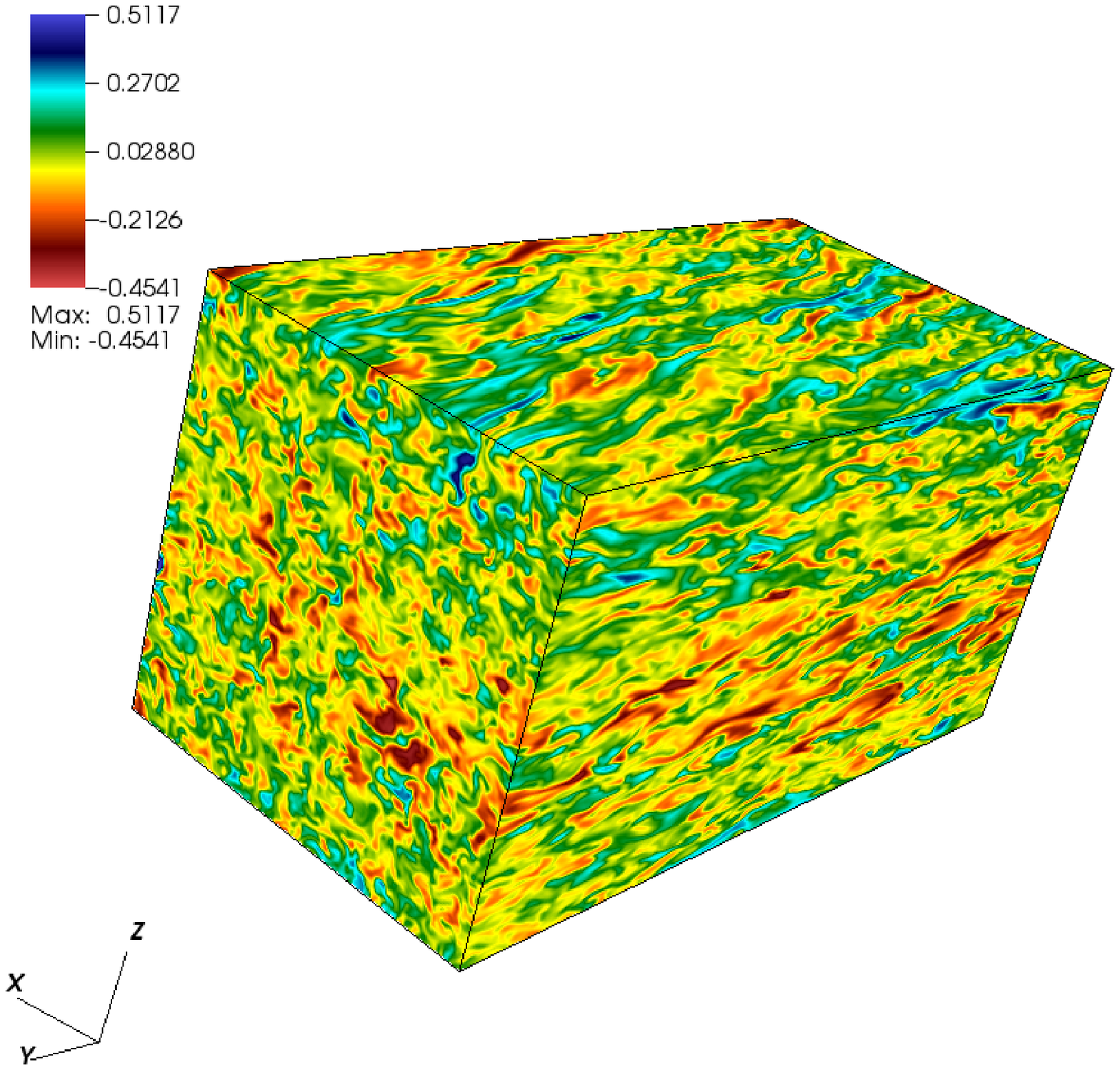}} 
\caption{Still images of subvolume of simulation \zzw{64} at orbit 100 mapped to the local Cartesian frame.}
\label{fig:sbstill}
\end{figure}

 Towards this end, we proceed in a manner similar to \cite{sra10} and decompose the global simulation domain into an ensemble of small regions that are comparable to shearing box models as described in detail in \S\ref{sec:diag}.  To provide a sense of the structure of the subdomain, ${\mathcal{S}}$, we employ the transformation to a Cartesian domain through the mapping $(R,\phi,z) \rightarrow (x,y,z) = (R-R_{0},R_{0}\phi,z)$.  To further facilitate the comparison to a shearing box, we consider the quantity
\begin{equation}
v_{y} = v_{\phi} - v_{K}(R_{0}).
\end{equation}   
Figure~\ref{fig:sbstilla} shows the quantity $v_{y}/c_{s}$ in the transformed subdomain, $\mathcal{S}$, at 100 orbits in simulation \zzw{64}.  Similarly, Figure~\ref{fig:sbstillb}, illustrates the toroidal magnetic field for the same time and simulation.  We note the visual similarity between these images and those presented from shearing box simulations.

\subsection{Evolution of Local Ensemble}
\label{sec:lvgevol}

\begin{figure}
\centering
\subfloat[Accretion Efficiency (ZF)]{\label{fig:locevola} \includegraphics[width=0.5\textwidth]{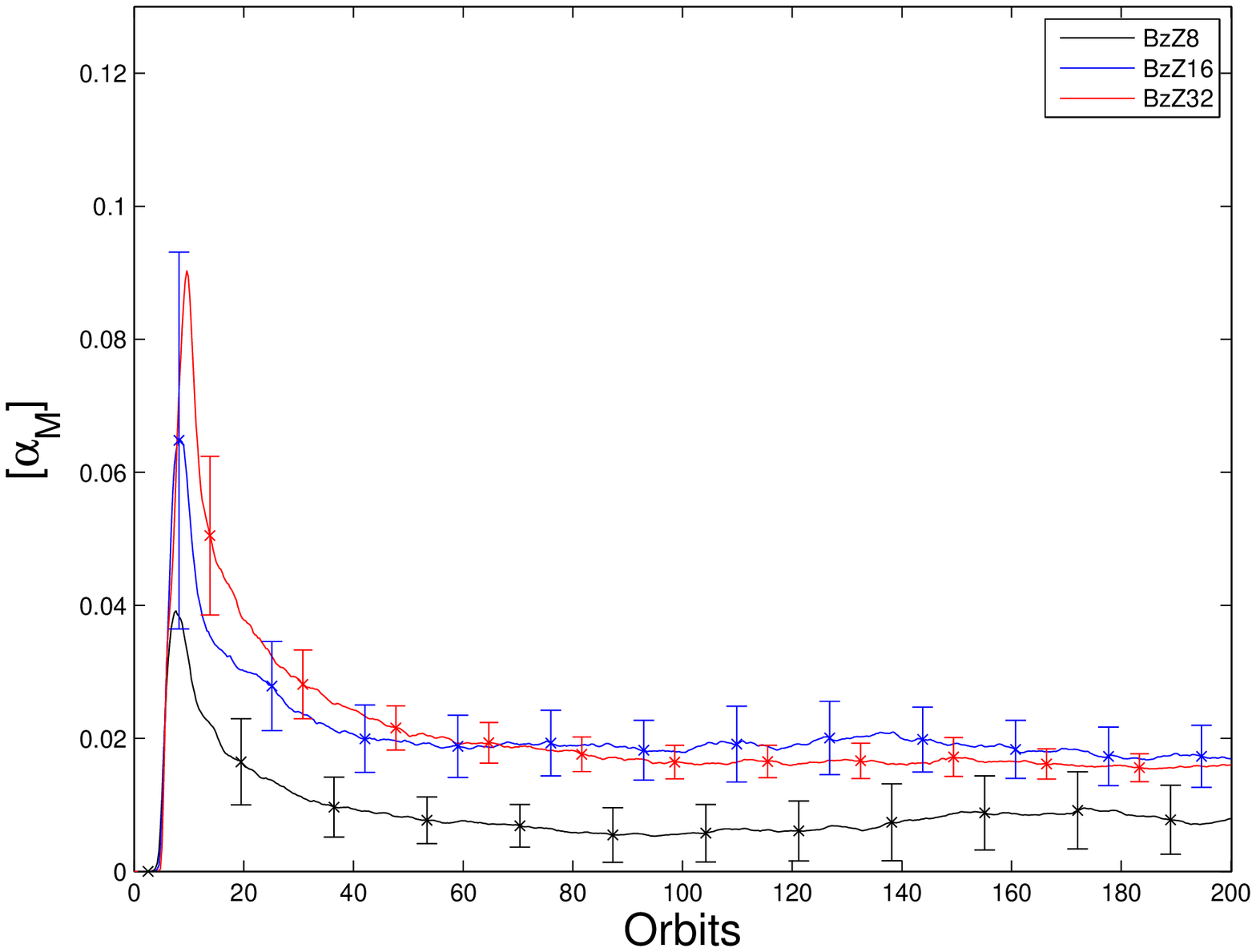}}                
\subfloat[Magnetic Energy (ZF)]{\label{fig:locevolb} \includegraphics[width=0.5\textwidth]{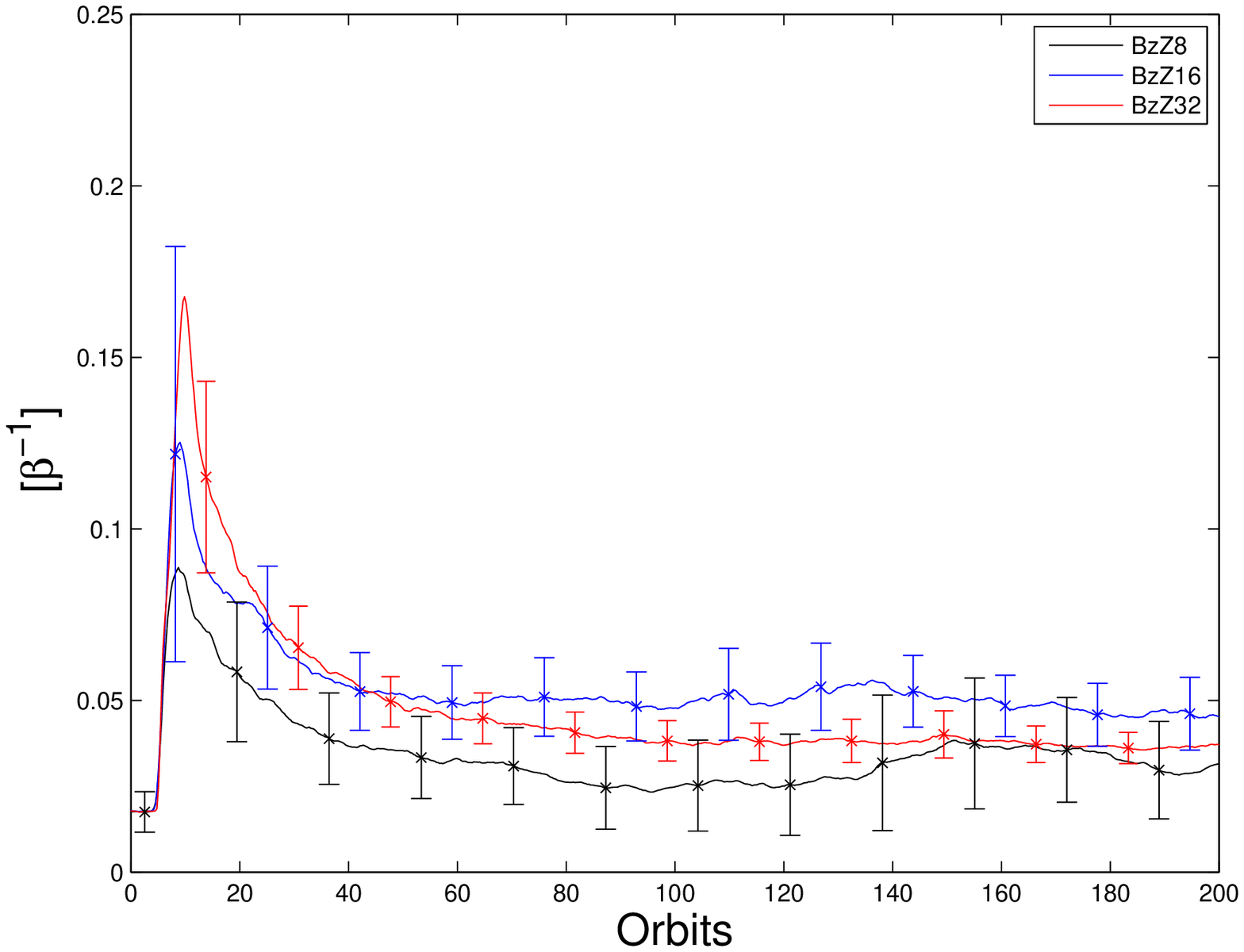}}\\
\subfloat[Accretion Efficiency (NF)]{\label{fig:locevolc} \includegraphics[width=0.5\textwidth]{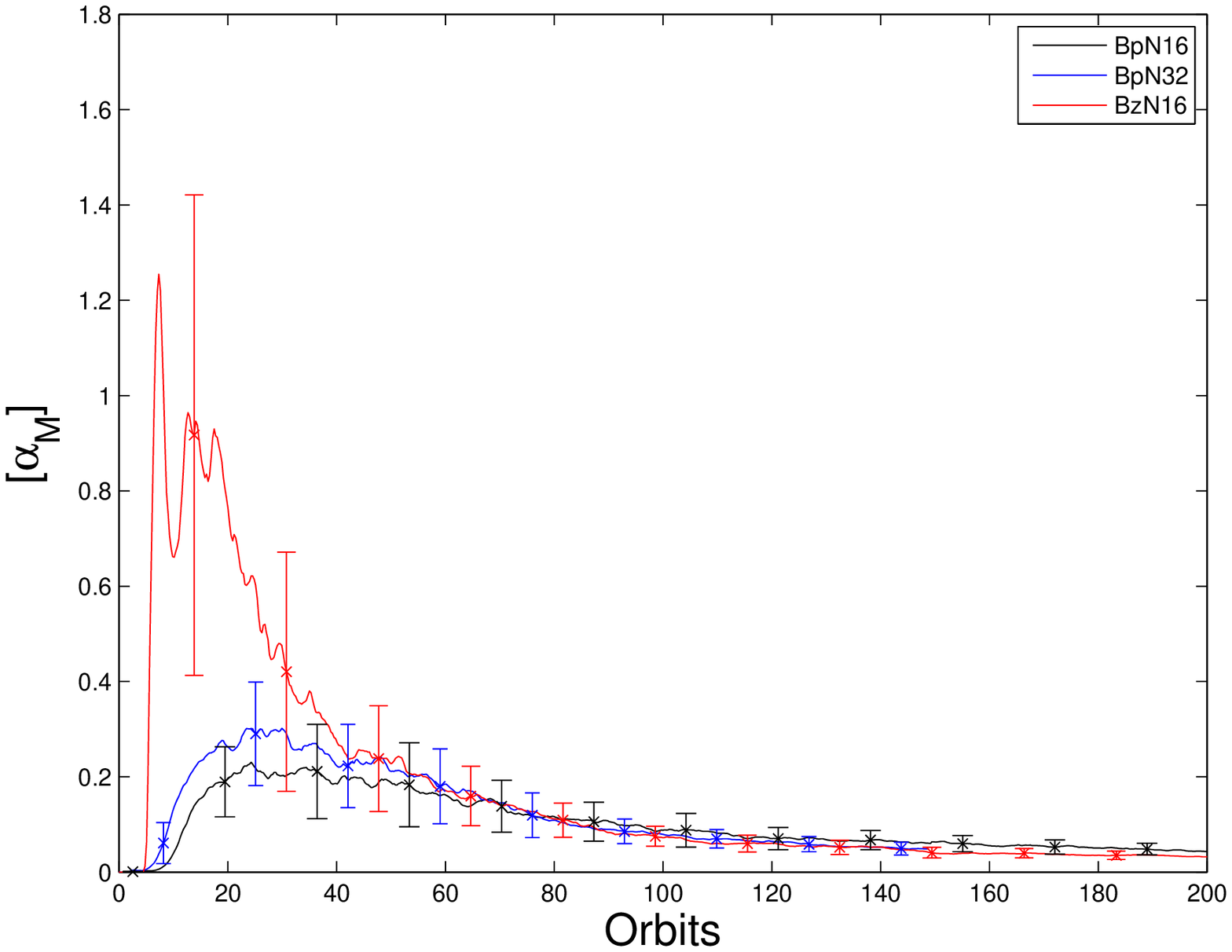}}                
\subfloat[Magnetic Energy (NF)]{\label{fig:locevold} \includegraphics[width=0.5\textwidth]{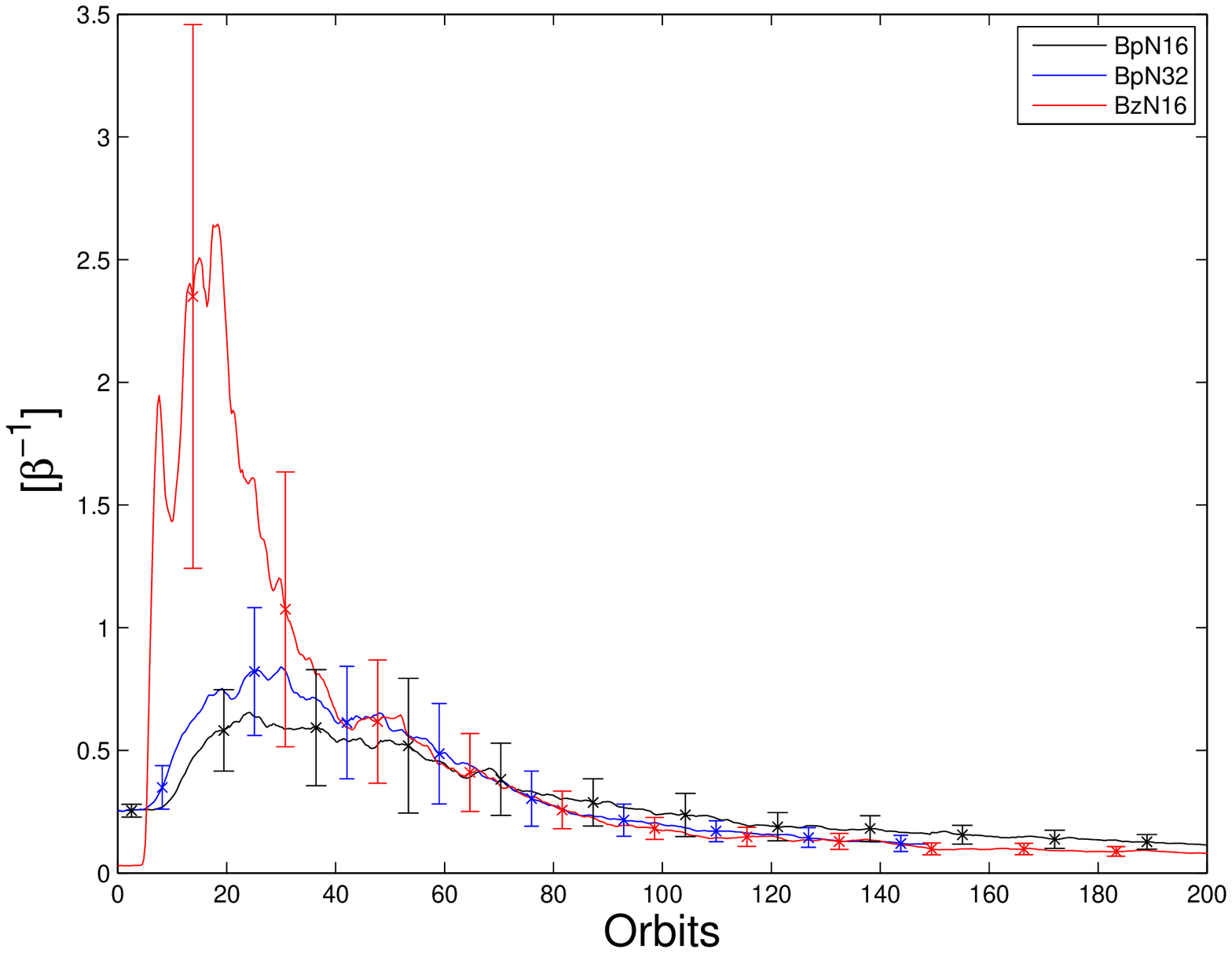}}\\  
\caption{Evolution of the local ensemble.}
\label{fig:locevol}
\end{figure}

The benefit of treating a global simulation as a local ensemble is that it allows us to study the local dynamics of accretion disk turbulence over a large range of parameter space simultaneously.  Additionally, it offers the opportunity to directly test the importance of curvature terms and the degree to which local flux is truly conserved.  Local statistics can be used not only to study the average value of a quantity but also the variation in the quantity and how it evolves.  Figure~\ref{fig:locevol} illustrates the time evolution of $[\alpha_{M}]$ and $[\beta^{-1}]$ for a selection of ZF and NF simulations.  Comparing $[\alpha_{M}]$ and $[\beta^{-1}]$ in the ZF simulations, Figures~\ref{fig:locevola} and~\ref{fig:locevolb}, illustrates the resolution dependence analogous to the physical metrics considered above, however the variability of these quantities is inversely related to the resolution.  The behavior of run \zz{8} is statistically separated from the behavior of the higher-resolution simulations, with runs \zz{16} and \zz{32} demonstrating more comparable values of these quantities.  

The behavior and evolution of the accretion efficiency and scaled magnetic energy for a selection of NF runs is given in Figures~\ref{fig:locevolc} and~\ref{fig:locevold}.  Most striking is the extreme initial transient associated with run \zn{16}, characterized by values of $\alpha_{M}$ well in excess of unity and strongly magnetized regions with Alfven speeds several times the sound speed.  This transient, however, is short lived with the long-term evolution comparable to that of runs \pn{16} and \pn{32}.  The toroidal runs, like the ZF simulations, display an inverse relationship between variability and resolution.  The resolution dependence of the variability is itself interesting, while the global treatments of these quantities discussed above can demonstrate their convergence that the variability weakens with resolution suggests a separate resolution requirement.

\subsection{Instantaneous Correlation of Flux and Stress}
\label{sec:lvgfs}

\begin{figure}
 \centering
  \subfloat[Vertical flux, ZF runs]{\label{fig:zfsz} \includegraphics[width=0.5\textwidth]{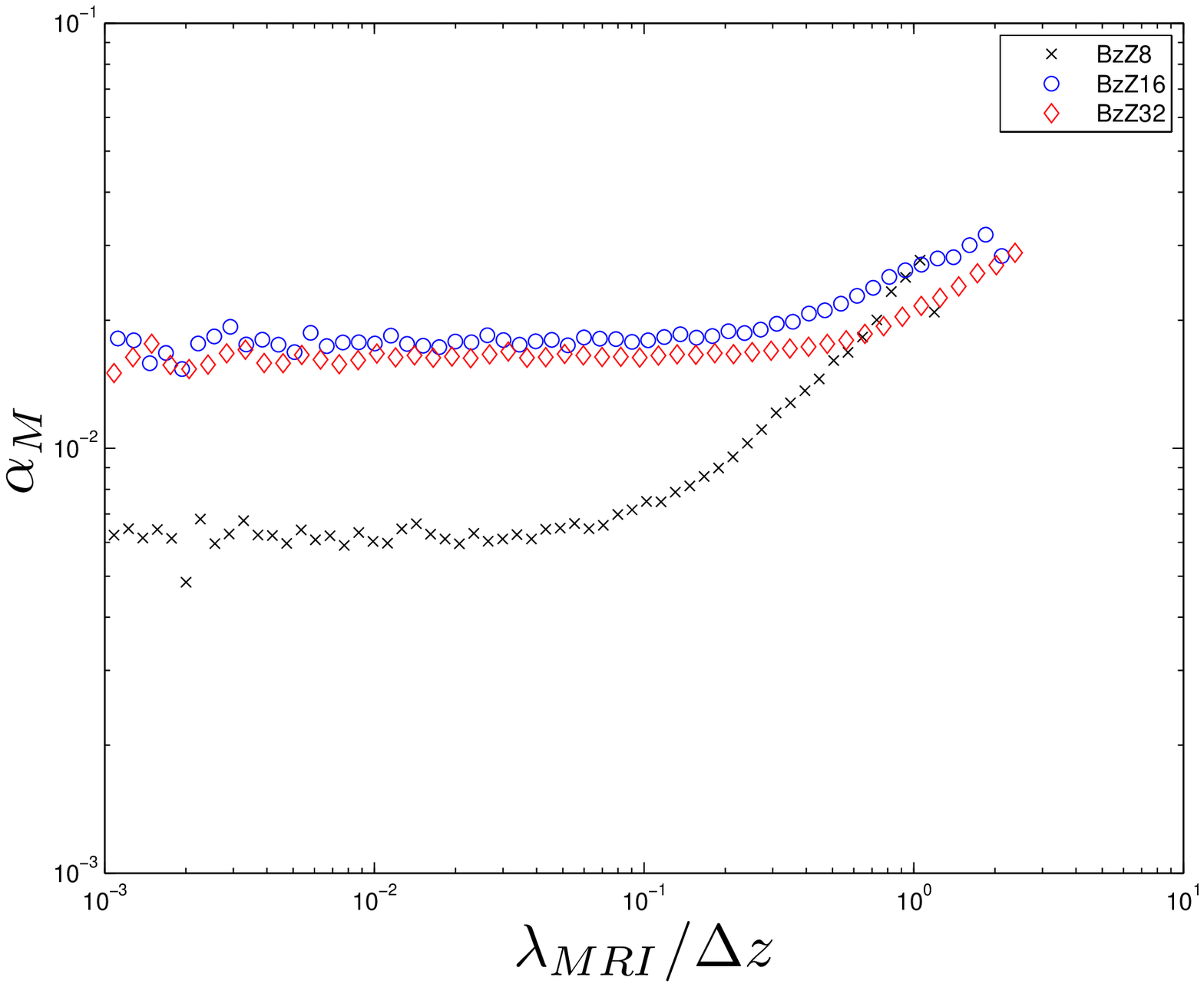}}                
  \subfloat[Toroidal flux, ZF runs]{\label{fig:zfsp} \includegraphics[width=0.5\textwidth]{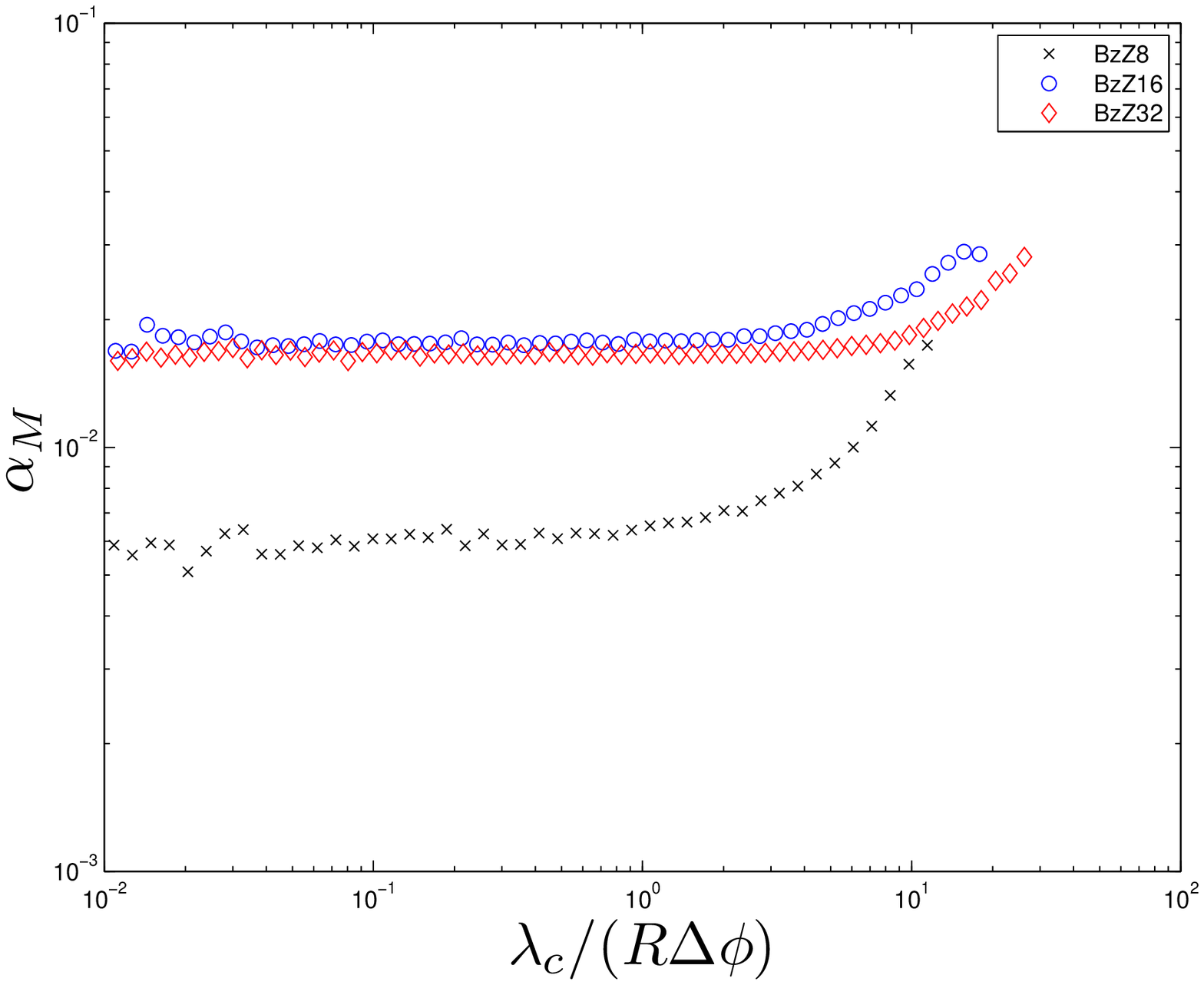}}   \\
  \subfloat[Vertical flux, NF runs]{\label{fig:nfsz} \includegraphics[width=0.5\textwidth]{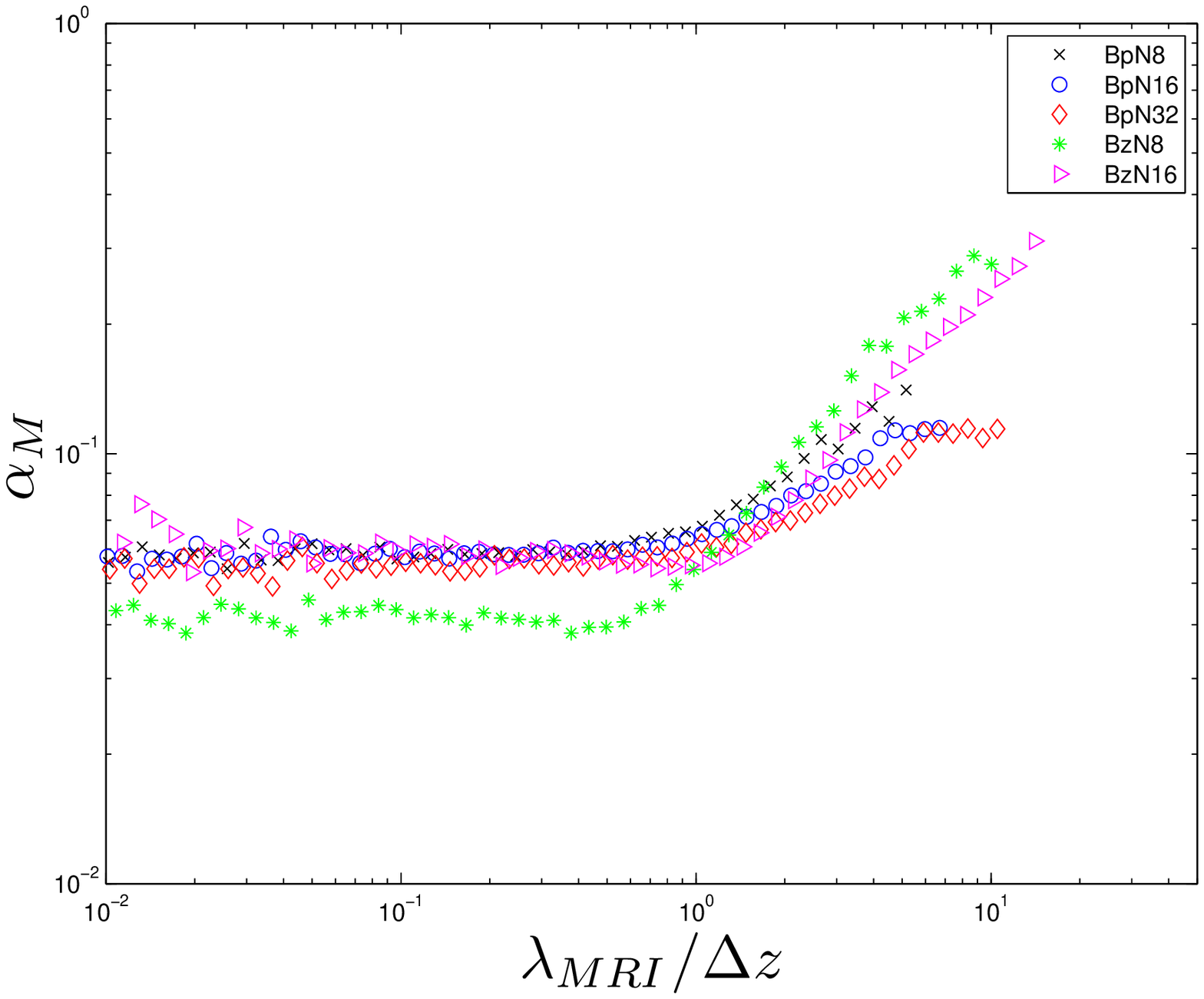}}                
  \subfloat[Toroidal flux, NF runs]{\label{fig:nfsp} \includegraphics[width=0.5\textwidth]{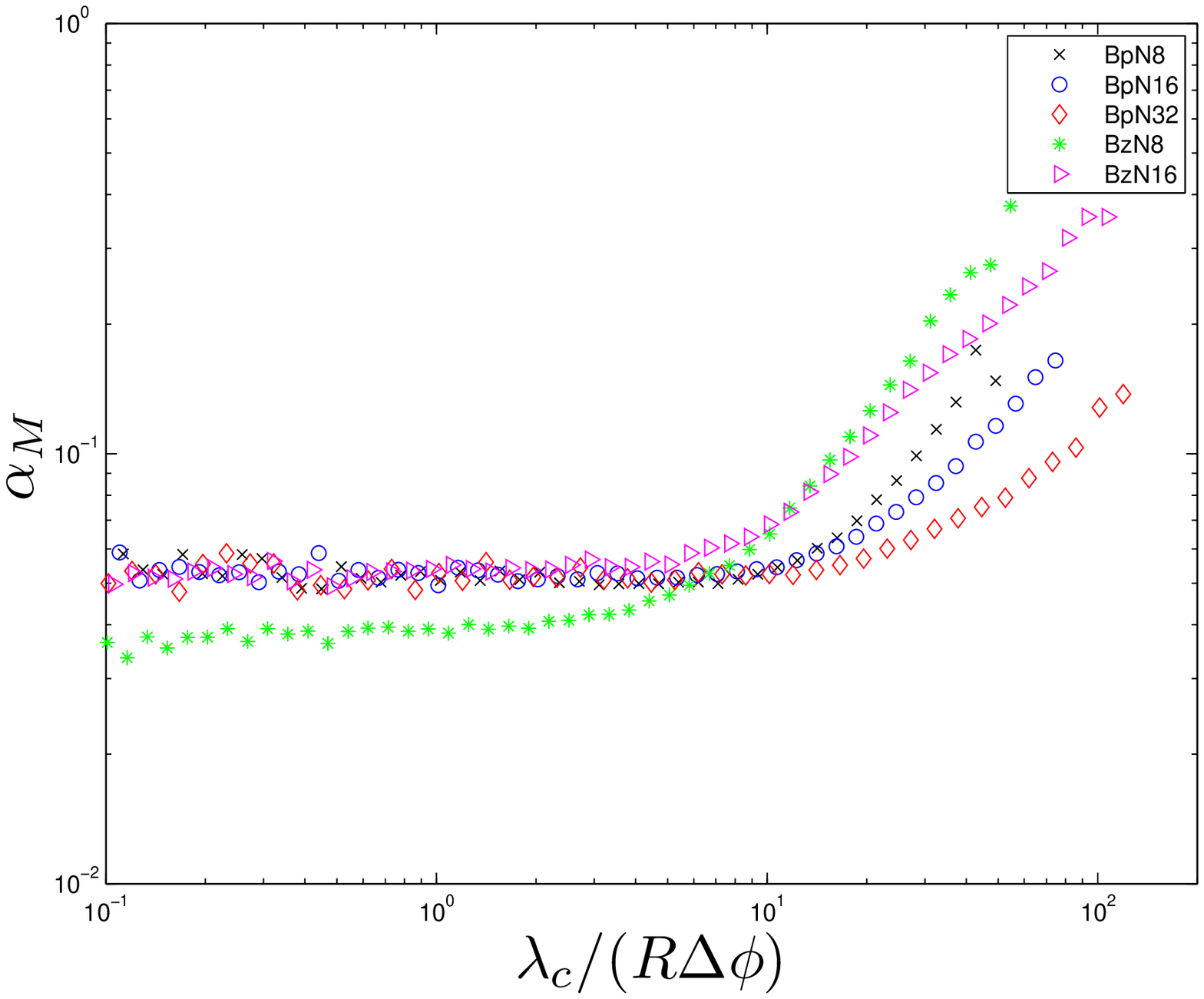}}       
  \caption{Correlations between vertical and toroidal magnetic flux and stress, ZF and NF runs.}
  \label{fig:fs}
\end{figure}

In \citet{sra10}, we make a preliminary effort towards connecting local and global models of accretion disks using the method described in \S\ref{sec:diag} to generate ``local'' statistics.  The simulations considered were useful but suffered due to the added difficulty of separating effects of stratification and resolvability from the desired comparison.  However that the resolvability of the linear MRI in saturated turbulence is important or sufficient is not clear from analytic theory.  The simulations considered in this work are designed from first principles to be more directly applicable to the comparisons we would like to make.  We have previously considered the instantaneous correlation between vertical flux and stress; here we also study the relationship between toroidal flux and stress.  That there is a structure to the relationship between toroidal flux and stress is interesting (Figures~\ref{fig:zfsp} and \ref{fig:nfsp}).  At first glance, one might not expect a transition in the correlation, since the comparison is between toroidal field and stress, a term dominated by the toroidal field.  The nature of this correlation is unclear, as in contrast to the vertical MRI which grows quite rapidly and would be expected to correlate to the presence of field, it is not clear that the toroidal MRI would be associated with timescales small enough to correlate to the presence of field.  Assuming that the connection between toroidal flux and stress is causal, as is believed to be the case in the vertical flux-stress correlation, it may suggest that toroidal flux generates stress in the non-linear regime in a manner much faster than it would in the linear.  An alternative possibility is that the correlation between toroidal flux and stress is a consequence of the vertical flux-stress relationship.  In this case, the presence of vertical flux would drive stress with toroidal flux amplification as an intermediate step.  This possibility would suggest that the relationship between the turning points in the two flux-stress relationships is meaningful.  We note, that while the fluxes are scaled to the grid resolution these are not the same as the quality factors presented earlier as the flux is averaged over each wedge as opposed to taken on a cell by cell basis.  This averaging reduces the strength of the net field which results in lower values than the quality factors.   

The calculated flux-stress relationship for net-zero fields is given for both vertical flux (Figure~\ref{fig:zfsz}) and toroidal flux (Figure~\ref{fig:zfsp}).   It was noted in \citet{sra10} that the transition between the flat unresolved region and the linear resolved region suggested a seeming super-resolvability, specifically that we saw this transition at $\lambda_{MRI} \approx \Delta z/20$.  We note similar behavior in run \zz{8}, in Figure~\ref{fig:zfsz}, in which the transition occurs at $\lambda_{MRI} \approx \Delta z/10$.  The remaining simulations exhibit a transition at $\lambda_{MRI} \approx \Delta z$, in agreement with the scaling law given by Eqn~\ref{eqn:mriscalez} (\citeinp{pcp07}) and the simulations of \citet{bas11}.    Related to this, we see a qualitatively similar resolvability threshold in the toroidal flux-stress relationship (Figure~\ref{fig:zfsp}).  While the two highest resolution simulations exhibit a transition at approximately $10R\Delta \phi$ , the transition for \zz{8} occurs at approximately half of that.  In \citet{sra10}, this super-resolvability was attributed to the effect of slower-growing unstable MRI modes and while this still may explain the placement of the transition point in the higher resolution simulations it is likely not the reason for the behavior of run \zz{8}.  The cause of this discrepancy between \zz{8} and the other simulations is unclear; however it is clear that this super-resolvability of the transition seems to be a characteristic of unresolved turbulence. 

Next we consider the flux-stress relationship for the net field simulations for both vertical flux (Figure~\ref{fig:nfsz}) and toroidal flux (Figure~\ref{fig:nfsp}).  We note that in general the net flux runs exhibit a very similar general structure to the net-zero flux runs.  Specifically, the transition points between the flat and linear stress-response regimes are independent of initial field topology.  None of the net-field simulations considered exhibit the super-resolvability seen in \zz{8} which suggests that even low resolution net-field runs are more resolved than their zero-flux counterparts.  The general trend is that, with increasing resolution, the stress response to unresolved flux is generally increased and the slope of the linear regime, in which flux is resolved, becomes shallower.  An interesting feature is seen in the stress response to vertical flux for the net toroidal field runs.  For vertical flux $\lambda_{MRI}/\Delta z \approx 6$ we see a flat stress response.  This region occurs well below what we would expect for the MRI-stable region in which the most unstable vertical mode exceeds the vertical domain of the simulation, and we also note that the net vertical field runs maintain a linear slope in this region.  

A further comparison we can make between local and global models is based not just on the structure of the flux-stress relationships, but on the precise manner in which increased flux generates an increased stress response.  While the dependence of local saturation predictors on box-size prevent direct comparisons, we can work backwards and use the slope of the resolved region to define an effective local box size.  The physical meaning of this length scale is unclear, and is presented here as merely a convenient manner in which to quantify the slope of the linear-response regime.  Formally, we use the saturation predictors given by Eqn~\ref{eqn:mriscalez} and Eqn~\ref{eqn:mriscalep} to constrain the value of the box-size to fit the instantaneous correlations we find.  The values of box-size from the vertical flux-stress and toroidal flux-stress are denoted $\ell_{z}$ and $\ell_{\phi}$ respectively.
  
The values of this effective box-size are given in Table~\ref{tab:ell}.  Both the vertical and azimuthal length scales are monotonic with resolution, with the net field simulations corresponding to larger effective box-sizes.  Of interest is the approximate relationship, $\ell_{z} \approx 2\pi \ell_{\phi}$, in particular due its connection with commonly used local domain sizes.  Also of note is that by reducing the azimuthal domain we see a significant drop in $\ell_{z}$, while $\ell_{\phi}$ remains roughly constant.

\begin{table}[htdp]
\begin{center}
\begin{tabular}{|c|c|c|}
\hline
Simulation & $\ell_{z}/H_{0}$ & $\ell_{\phi}/H_{0}$ \\
\hline
\zz{8} & 0.348 & 1.838\\
\zz{16} & 0.899 & 4.956\\
\zz{32} & 0.964 & 5.040 \\
\zzw{32} & 0.283 & 5.970\\
\zzw{64} & 0.372 & 7.696 \\
\hline
\pn{8} & 0.487 & 3.377 \\
\pn{16} & 0.674 & 4.612 \\
\pn{32} & 0.905 & 4.953 \\
\zn{8} & 0.472 & 5.552 \\
\zn{16} & 0.731 & 8.255 \\
\hline
\end{tabular}
\end{center}
\caption{Effective box-size for global simulations.}
\label{tab:ell}
\end{table}%

\begin{figure}
\centering
\subfloat[\zz{32}]{\label{fig:fluxa} \includegraphics[width=0.5\textwidth]{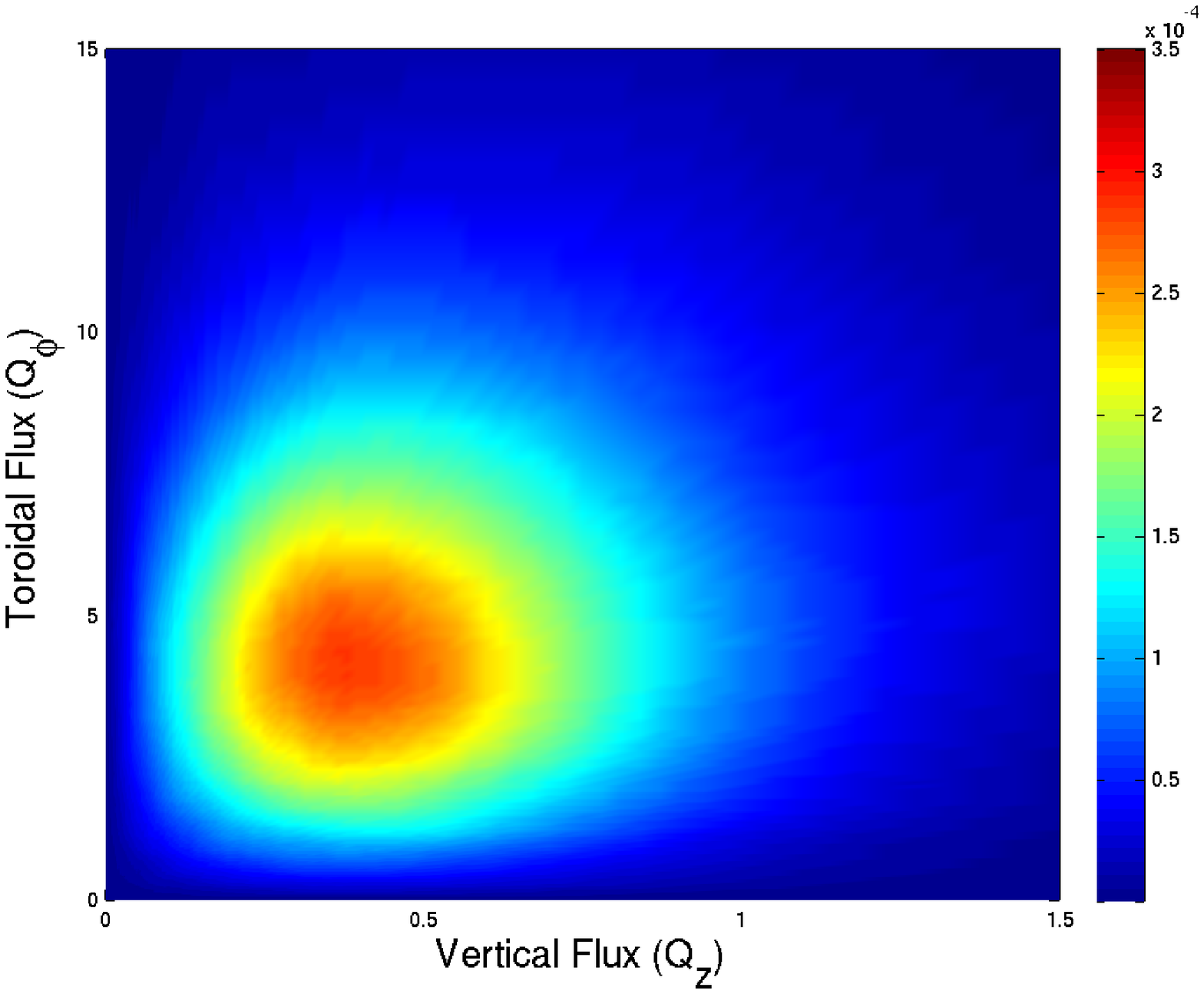}}                
\subfloat[\pn{32}]{\label{fig:fluxb} \includegraphics[width=0.5\textwidth]{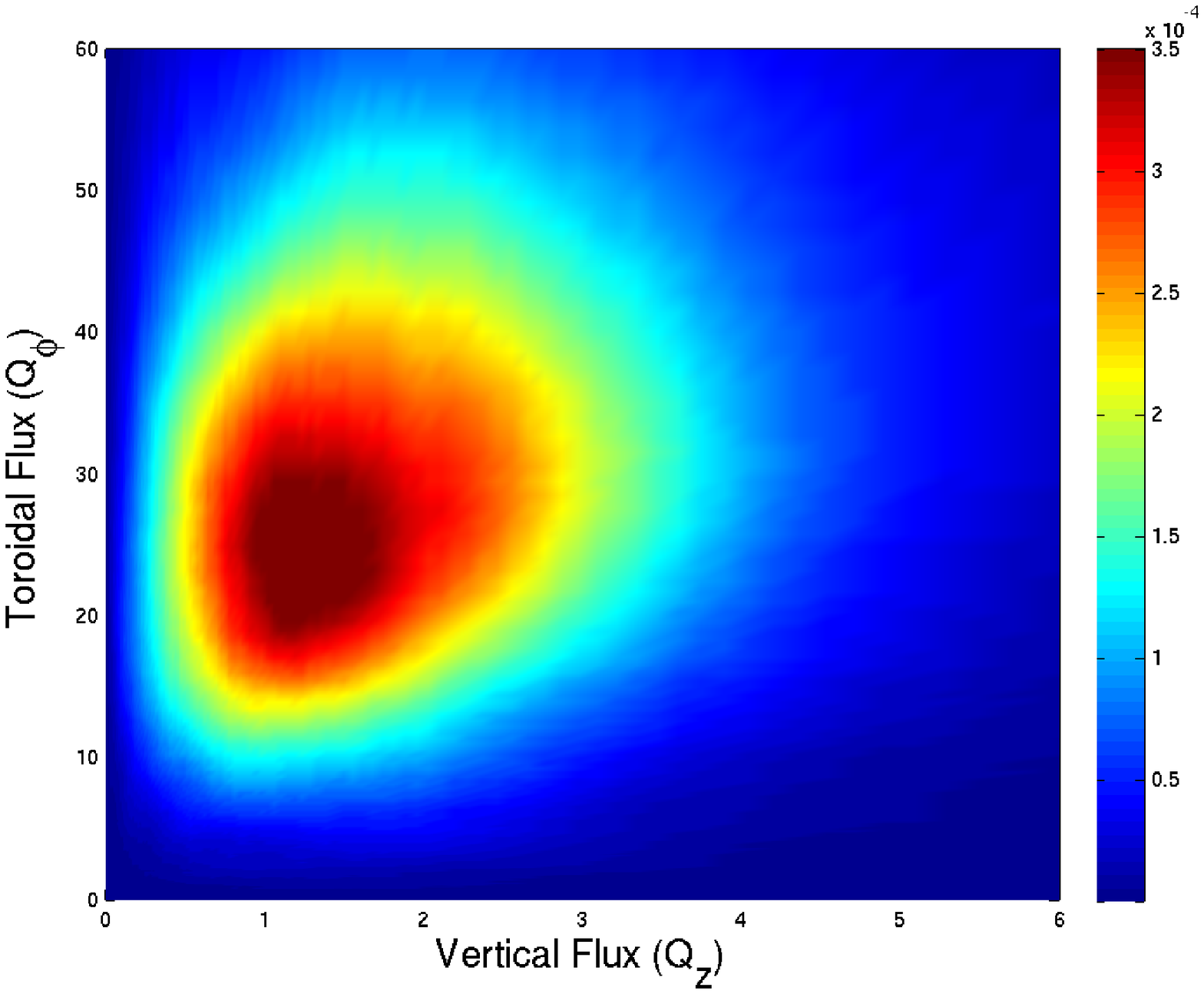}}\\
\subfloat[\zn{16}]{\label{fig:fluxc} \includegraphics[width=0.5\textwidth]{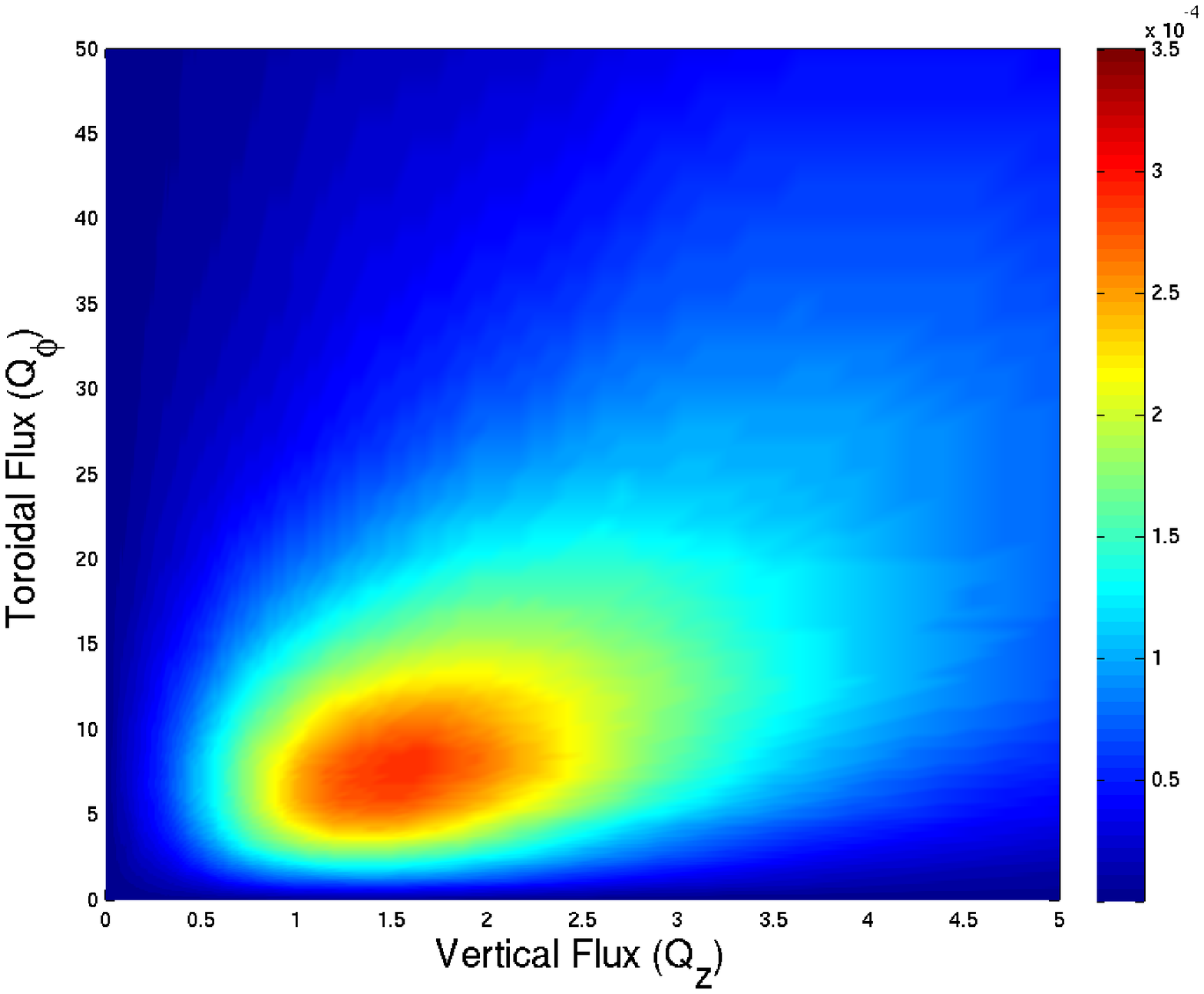}}
\caption{Distribution of magnetic flux in the QSS.}
\label{fig:flux}
\end{figure}

The manner in which a localized region of the disk responds to the instantaneous presence of flux makes up an important part of the dynamics of the disk, the natural analog to this is the distribution of magnetic flux through the localized regions of the disk.  Here, we focus on the latter.  We compute a distribution function of the vertical and toroidal magnetic flux through each wedge during the QSS.  As in the flux-stress diagrams, these fluxes are scaled to the grid resolution.  The distribution is calculated using logarithmic binning in the flux (80 bins per decade) and the result is smoothed using a ten bin moving window.  The results for simulations \zz{32}, \pn{32}, and \zn{16} are displayed in Figures~\ref{fig:fluxa}, \ref{fig:fluxb}, and \ref{fig:fluxc}.

Of interest is the fact that for simulation \zz{32}, the majority of the disk is below the threshold in which there is a linear stress response to flux.  In contrast to this, simulations \pn{32} and \zn{16} have the majority of the disk well above the resolvability threshold of $(Q_{z},Q_{\phi}) \approx (1,10)$.  Related to the initial conditions, \pn{32} is biased towards toroidal flux and \zn{16} is biased towards vertical.  Forming an intermediate case, \zz{32} possesses a more even distribution of flux albeit weaker than the net field cases.

\begin{figure}
\centering
\includegraphics[width=0.8\textwidth]{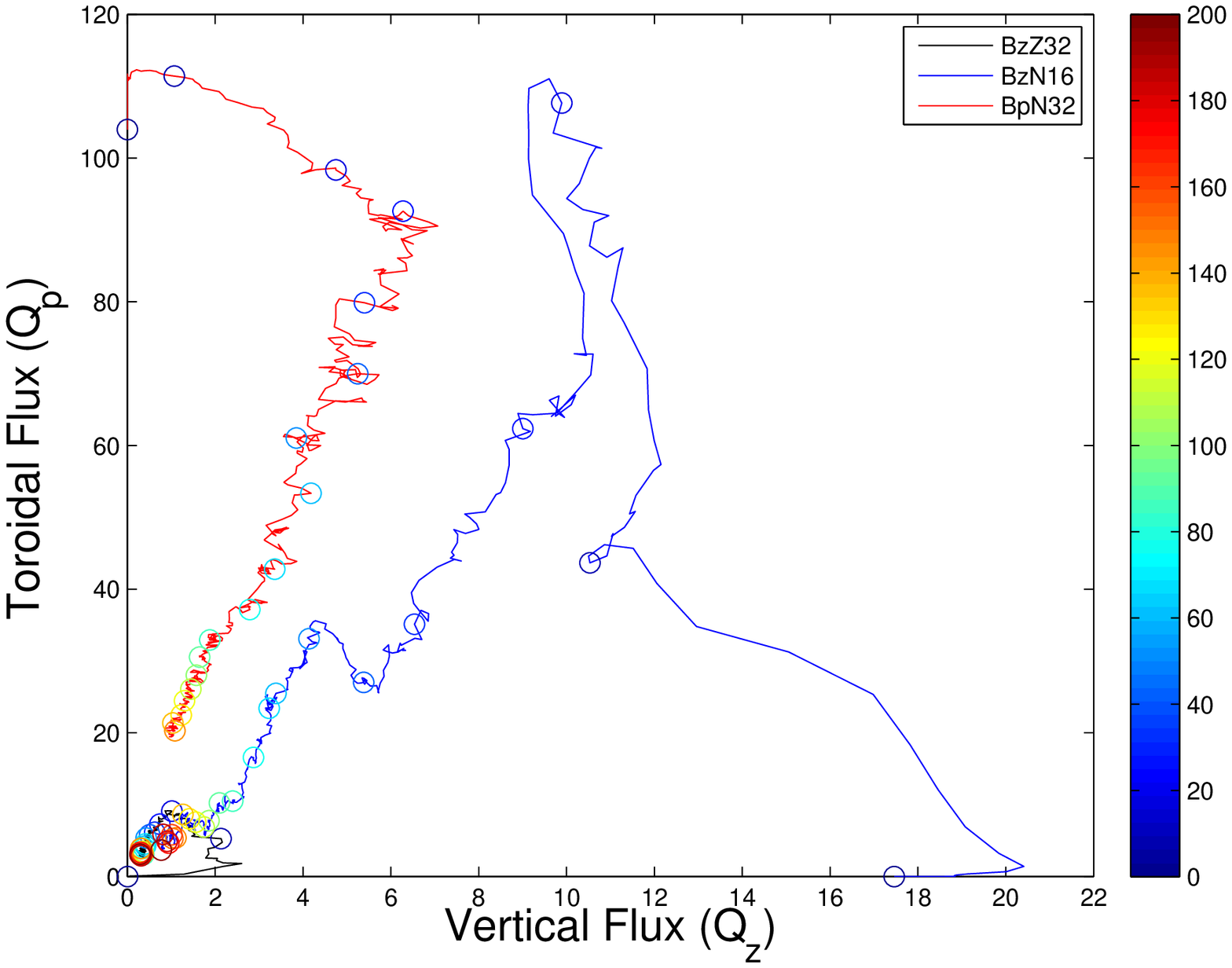}                
\caption{Evolution of mean quality factors of the local ensemble.  Circular markers placed approximately every $8$ orbits, with color determined by the orbit associated with the marker.}
\label{fig:qspace}
\end{figure}

An alternative manner by which to consider the distribution of magnetic flux is through the consideration of the local statistics as representative of the trajectory of an ensemble.  That is, to consider the path $([Q_{z}](t),[Q_{\phi}](t))$ and its temporal evolution.  Figure~\ref{fig:qspace} illustrates the trajectory for simulations \zz{32}, \pn{32}, and \zn{16} where markers are placed on the trajectory approximately every $8$ orbits.  For the net field simulations the evolution of the flux during the initial transient is quite stark, resulting in migration from the axes to large values of both azimuthal and vertical flux, specifically large enough to exceed the transition point to linear stress response.  After the initial transient, the net flux simulations are characterized by a general decrease in the mean flux.  In contrast to the net flux simulations, the trajectory of \zz{32} is constrained to a much smaller region of the flux-space.  The trajectory only briefly crosses into the region associated with a linear stress response, unlike the net flux simulations whose trajectories spend the majority of the simulation in this region.  Compared to truly local simulations, in which the boundary conditions prevent migration in the $Q_{z} \times Q_{\phi}$ space, global simulations are characterized by significant evolution particularly those initialized with a net flux.  This suggests the possibility that local simulations initialized with a net flux could result in an artificially-maintained accretion efficiency due to the boundary conditions preventing the advection of flux off the grid.  The trajectory of simulations \zz{32} and \zn{16} evolve to possess comparable fluxes despite exhibiting significantly different behavior during much of the simulation.  The limited temporal domain for which simulation \pn{32} was evolved prevents making a similar comparison, however the behavior is consistent with the possibility of evolving to a state of magnetic flux comparable to simulations \zz{32} and \zn{16}.  Run \pn{16}, while run for $50$ orbits more than \pn{32}, displays similar behavior that is consistent with this possibility but does not achieve it during the time domain simulated.

\section{Conclusions}
\label{sec:con}

In this paper, we have presented results from a series of simulations utilizing, for the first time, orbital advection in global cylindrical coordinates.  The use of orbital advection, and the order of magnitude performance enhancement it provides, has allowed the exploration of global disks at resolutions not only comparable to local models, but in many cases exceeding the resolution of shearing box simulations in the literature.  Removing the constraint of the Keplerian timestep has also allowed the exploration of simulations with isotropic resolution ($\Delta z = \Delta R \approx R \Delta \phi$) without the significant computational cost normally associated with azimuthal resolution.

The primary distinction between local and global simulations is the significant degree of temporal evolution of the latter.  Simple estimates of accretion relying on the viscous timescale from anomalous viscosity disk theory vastly underestimate the degree of accretion observed during the initial transient of global simulations.  However, a more sophisticated treatment utilizing a one-dimensional model (equation~\ref{eqn:alphamod}) based on the measured stress in the simulation accurately reproduce the temporal evolution and radial distribution of the mass.  Estimates of $\alpha$ based on the viscous timescale inferred from the initial transient result in values well over an order of magnitude above the values of $\alpha$ that characterize the simulation in the QSS and even in excess with the measured values of $\alpha$ associated with the transient.  This discrepancy may have potentially important astrophysical implications regarding observational estimates of $\alpha$ based on the viscous timescale.

Understanding and minimizing the grid-scale dependance exhibited by simulations of accretion disk turbulence is a vital precursor to constructing computational models that can be confidently compared against observational data.  The ability to quantify complex, spatially and temporally varying turbulence into a simple scalar, the magnetic tilt angle, and use that to verify the presence of converged turbulence is an important step towards that goal.  While several convergence metrics have been explored here focusing on either the physical, numerical, or spectral nature of the simulations, the magnetic tilt angle stands out as being the most important indicator of convergence.  In the context of unstratified global simulations presented here, it appears that numerically-converged MRI-driven MHD turbulence is characterized by $\theta_{B} \approx \indeg{13}$.  While all the metrics considered here are useful towards understanding the physics and numerics of MRI-driven turbulence, the magnetic tilt angle alone possesses all the qualities we desire in a convergence metric.  The tilt angle is monotonic with resolution, and is independent of initial field topology, while exhibiting weak dependence on stratification and local versus global formalism.  The data presented here suggests the following resolution requirements for convergence: a vertical resolution of $H/\Delta z \geq 32$, $\Delta R = \Delta z$, and an azimuthal resolution that satisfies $R \Delta \phi \leq 2\Delta z$.  This final requirement is particularly constraining given the computational difficulty associated with nearly isotropic resolution in simulations without orbital advection.  The importance of orbital advection in future simulations is clear.

The use of shearing box models, while necessary to explore turbulence at small scales, is predicated on a series of significant assumptions.  Testing that the predictions of local simulations hold in the context of global models is an important validation of the local model.  Towards this end, we have demonstrated that saturation predictors from the local regime correspond to instantaneous correlations between magnetic flux and stress in the global regime.  This, combined with an understanding of flux distributions in global simulations would allow simpler statistical treatments of disk turbulence.  However, the conservation of magnetic flux on small scales implicit in local models is not realized in global disks.  Indeed, local statistics derived from global simulations seeded with a magnetic field topology possessing net flux are characterized by significant migrations in the space $Q_{z} \times Q_{\phi}$ in contrast to the assumption of conserved flux implicit in the shearing box formalism.

The next stage of this work will focus on a study of the small-scale turbulent structure of the simulations presented here through an exploration of two-dimensional correlation functions in the manner of \citet{ggsj09} and \citet{bas11}.  This study will focus on the structure of the fluid variables as well as the turbulent energetics, through consideration of the terms in the magnetic energy equation.  Future work will utilize an extension of orbital advection to adiabatic MHD to conduct a similar convergence study of stratified accretion disks.  Finally, the use of orbital advection to augment simulations of more directly applicable astrophysical systems will be explored.    

\section*{Acknowledgements}
The authors would like to thank Sean O'Neill, Aaron Skinner, Phil Armitage, and Julian Krolik for useful discussions during the completion of this work.  KAS and CSR acknowledge support from NASA under the Astrophysics Theory Program grant NNX10AE41G, JMS acknowledges support from NSF grant AST-0908269, and KB acknowledges support by the NSF under grant numbers AST-0807471 and AST-0907872 and by NASA under grant numbers NNX09AB90G and NNX11AE12G.  This research was supported in part by the NSF through TeraGrid resources provided by the Texas Advanced Computing Center under grant number TG-ASTAST090105 and TG-AST090106. The authors acknowledge the Texas Advanced Computing Center at The University of Texas at Austin for providing HPC and visualization resources that have contributed to the research results reported within this paper.

\appendix
\section{Orbital Advection}
\label{sec:oa}

The equations of ideal MHD are represented by a system of hyperbolic conservation laws and as such numerical methods to solve them are limited by the CFL condition.  The condition, roughly, states that the maximum timestep a numerical method can take is limited by the time it takes for a signal to cross one cell.  When simulating weakly magnetized disks rotating about a central object, it is generally the case that the maximum sound speed is dominated by the Keplerian rotation.  In the frame rotating at the Keplerian velocity, the signal speeds are significantly smaller and thus the effective CFL condition in this frame would be relaxed.  The goal of orbital advection, first introduced for hydrodynamic systems by \citet{fargo}, is to solve the equations of MHD in the frame rotating at the Keplerian velocity.  

We begin by noting that the equations of MHD are close to linear in the velocity (the violations of this lead to source terms), and so in our goal to separate the rotational dynamics we consider the velocity decomposition
\begin{equation}
\vv = v_{K} \hat{\phi} + \vv'.
\end{equation}
In the above equation, $v_{K}$ represents the Keplerian velocity.  The Keplerian velocity is defined to satisfy 
\begin{equation}
\label{eqn:vkep}
\frac{v_{K}^2}{R} = \pd{\Phi}{R},
\end{equation}
where $R$ represents the cylindrical radius in the coordinate system ($R,\phi,z$) and $\Phi$ is the static gravitational potential of the central object (assumed axially symmetric).  Note, that this assumes the disk is unstratified.  In the case of stratified disks, the Keplerian velocity is defined to satisfy Eqn~\ref{eqn:vkep} in the midplane and there is an additional source term accounting for the discrepancy between centrifugal force and radial gravitation away from the midplane.

We further define the Keplerian angular velocity as, $\Omega_K = v_{K}/R$, and the shear parameter as, 
\begin{equation}
q(R) = - \myfrac{1}{2} \myfrac{d\ln \Omega_{K}^{2}}{d \ln R}.  
\end{equation}

Applying this velocity decomposition and rearranging leads to the new system,
\begin{mathletters}
\begin{eqnarray}
\dt \rho + \div (\rho \vv') + \Omega_{K} \dphi \rho &= 0 \\
\dt (\rho \vv') + \div (\rho \vv' \vv' - \vb \vb + P^{*}\vI) + \Omega_{K} \dphi (\rho \vv') &=  \vFc \\
\vFc = 2\Omega_{K} \rho v_{\phi}' \hat{R} + \Omega_{K}(q-2)\rho v_{R}' \hat{\phi} -\rho \partial_{z}\Phi \hat{z}\\
\dt \vb  - \nabla \times (\vv' \times \vb) - \nabla \times (v_{K} \hat{\phi} \times \vb) &= 0 .
\end{eqnarray} 
\end{mathletters}

The structure of this system is suggestive, in that each of the initial conservation laws is now in the form of a conservation law with a flux set by the perturbation velocity, $\vv'$, and an additional linear advection term.  The nature of the new term, $\vFc$, is clear from a physical perspective as being the Coriolis and centrifugal forces associated with our transformation into the non-inertial rotating frame.  The induction equation is modified in an analogous manner, with the evolution of the magnetic field coming from the EMF in the rotating frame (associated with $\vv' \times \vb$) and a Keplerian EMF ($v_{K} \hat{\phi} \times \vb$).

To numerically solve the resulting system, we consider an operator-split method that decomposes the equations into a system of linear advection equations and an MHD system (with Coriolis and centrifugal source terms) with a characteristic velocity given by $\vv'$.  The advection system is given by,
\begin{mathletters}
\label{eqn:advect}
\begin{eqnarray}
\dt \rho + \Omega_{K} \dphi \rho &= 0 \\
\dt (\rho \vv')  + \Omega_{K} \dphi (\rho \vv') &=  0 \\
\dt \vb   - \nabla \times (v_{K} \hat{\phi} \times \vb) &= 0 .
\end{eqnarray} 
\end{mathletters}
While the MHD system is given by,
\begin{mathletters}
\label{eqn:mhdvp}
\begin{eqnarray}
\dt \rho + \div (\rho \vv')  &= 0 \\
\dt (\rho \vv') + \div (\rho \vv' \vv' - \vb \vb + P^{*}\vI)  &=  \vFc \\
\dt \vb  - \nabla \times (\vv' \times \vb) &= 0 .
\end{eqnarray} 
\end{mathletters}

Formally, the operator-split method has a timestep constraint set only by the non-inertial MHD system as the linear advection system can be solved analytically.  Care must be taken when solving the advection system to ensure that the algorithm is conservative and preserves the solenoidal constraint of the magnetic field.  Solving these two sets of systems of equations is done by modifying existing code, and in particular is an extension of the method presented in \citet{athfargo} to the cylindrical integrator described in \citet{athcyl}.  As this is an extension, the description here will focus on how this method deviates from the shearing box implementation.  

Orbital advection in the shearing box is based on the linearization of the Keplerian velocity about the midpoint of the domain and thus the radial orbital velocity profile is uniquely specified by the shear parameter.  Transitioning to global simulations requires more flexibility, as certain choices of gravitational potential (e.g. the Paczynski-Wiita potential) may involve values of $q$ that have radial variation.  Therefore, the solution of the advection equation is modified to allow an arbitrary user-specified orbital velocity profile, $\Omega_{K}(R)$.    

Solution of the advection system, Eqn~\ref{eqn:advect}, must be solved in a manner consistent with constrained transport and must maintain the conservative property of the algorithm.  This is described in detail in \citet{athfargo}, however we will briefly summarize the method here.  To ensure that the solution is conservative, the azimuthal profile at each cylindrical ring in the simulation domain is reconstructed and this reconstruction is integrated upstream to calculate a flux at each cell interface.  Updating the hydrodynamic quantities using this flux ensures the conservation of quantities to machine precision.  Similarly, the evolution of the magnetic field must be done in a way to ensure the solenoidal constraint, this is handled using constrained transport as in the main integrator.  The magnetic field is updated using the Keplerian EMF calculated by integrating the upstream Keplerian EMF of the appropriate cell edges.  Modifications to this routine to extend this to cylindrical geometry are minor and involve the inclusion of the appropriate geometric scaling terms.  The simulations discussed here use a third-order reconstruction of the azimuthal profiles.  

Modifying the cylindrical integrator to solve the system given in Eqn~\ref{eqn:mhdvp} requires the addition of the appropriate source terms.  This must be done consistent with the additional geometric source terms associated with the curvilinear geometry as described in \citet{athcyl} and at second-order accuracy.  This is done for the shearing box implementation using Crank-Nicholson, however the quadratic geometric source terms make this approach cumbersome in cylindrical geometry.  Instead, this is accomplished using a two-stage Runge-Kutta method and is amenable to the simple inclusion of additional physics in the future.

The method presented here was tested extensively, reproducing all of the relevant (isothermal, rotating systems) tests presented in \citet{athcyl} and comparing the results with those computed solely using the cylindrical integrator.  These tests include: static force balance problems; the advection of a field loop; verification of the Rayleigh stability criterion for hydrodynamic differentially rotating systems, and reproducing stability for $q=1.99$ and instability for $q=2.01$; and comparing the growth and saturation of MRI-driven turbulence in accretion disk models using both a Newtonian and Paczynski-Wiita to that produced using the standard cylindrical integrator.  In the interests of brevity, the full details of the tests will not be presented here but we will present two relevant examples.  Our interest is instabilities in differentially rotating systems, and so we will present two tests: verification of the Rayleigh stability criterion for hydrodynamic systems, and a comparison of the growth and saturation of the MRI in an unstratified, Newtonian disk.

\subsection{Rayleigh Stability Criterion}
Here we consider differentially rotating systems with $\Omega(R) = \Omega_{0} R^{-q}$.  Rayleigh's criterion states that systems of this form will be stable for $q<2$ and unstable otherwise.  Simulations of differentially rotating systems at the cusp of stability are used to demonstrate that the momentum source terms are being correctly implemented.  Additionally, the Keplerian case ($q=1.5$) is demonstrated to be stable due to its particular importance.  To this end, the values of $q$ considered are, $q \in \{1.5,1.95,1.99,2.01,2.05\}$, using both orbital advection and the standard cylindrical integrator.

The physical domain simulated is given by $(R,\phi,z) \in [3,7] \times [0,\pi/2] \times [-0.5,0.5]$, using a resolution of $(N_{R},N_{\phi},N_{z}) = (128,200,32)$.  The simulation is initialized using a constant density $\rho_{0} = 200$, $\Omega_{0} = 2\pi$ and an isothermal sound speed of $c_{s} = 0.1$ which results in a flow that is highly rotationally dominated.  Note that this is just a three-dimensional extension of the Rayleigh stability test presented in \citet{athcyl}.  As in \citet{athcyl} the instability is seeded using perturbations to the azimuthal velocity of the form, $v_{\phi} = v_{K} (1+\Delta)$, where $\Delta \in [-10^{-4},10^{-4}]$ and is uniformly distributed.  The simulations are evolved to $t=300$, the number of cycles taken and the wall-time to complete each simulation is given in Table~\ref{tab:ray}.  Taking as an example the case $q=3/2$, corresponding to a Newtonian disk, the simulation without orbital advection takes approximately $24$ times as long to run.  Calculating the cycles/minute, the full operator-split method takes roughly $1.5$ times as long for each cycle.  Even in the case of an unstable disk, \eg $q=2.05$, where mass is driven off the grid there is an almost order of magnitude performance increase.

\begin{figure}
 \centering
  \subfloat[Evolution of the scaled stress (Eqn~\ref{eqn:sstress}) with (OA) and without (CYL) orbital advection.]{\label{fig:oaa} \includegraphics[width=0.5\textwidth]{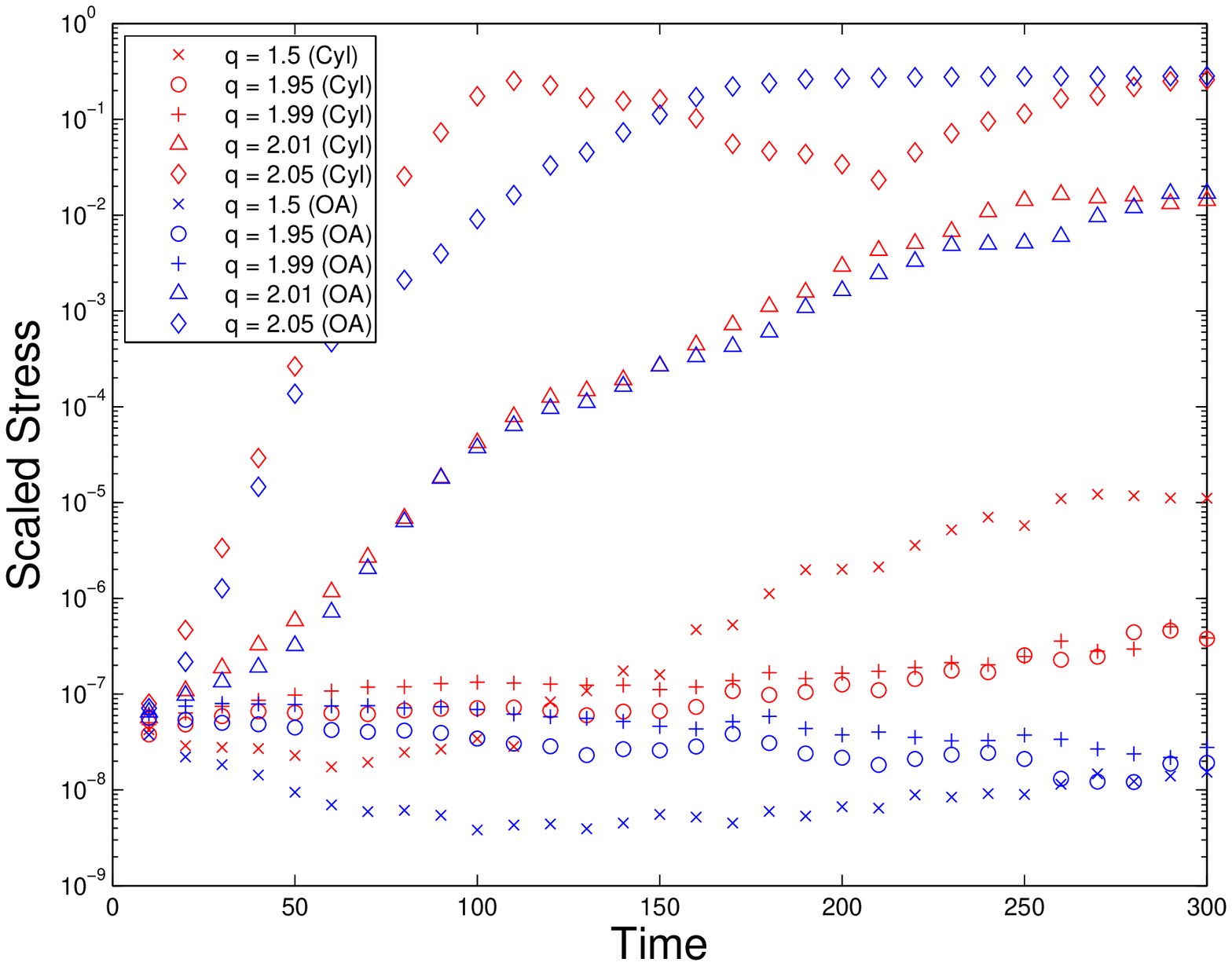}}                
  \subfloat[Comparison of $\alpha_{M}$ with and without orbital advection.]{\label{fig:oab} \includegraphics[width=0.5\textwidth]{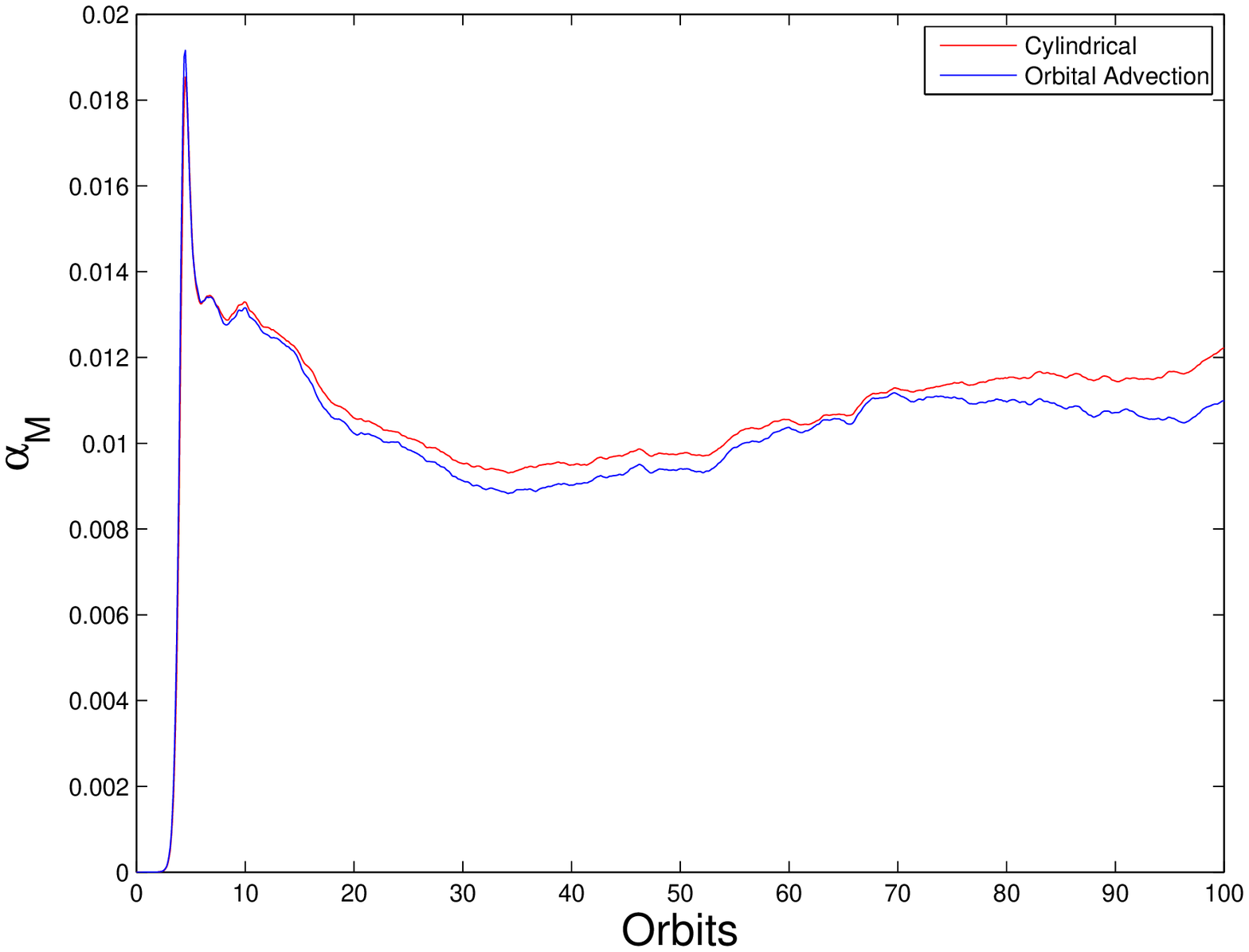}}   \\
  \subfloat[Speedup using orbital advection.]{\label{fig:oac} \includegraphics[width=0.5\textwidth]{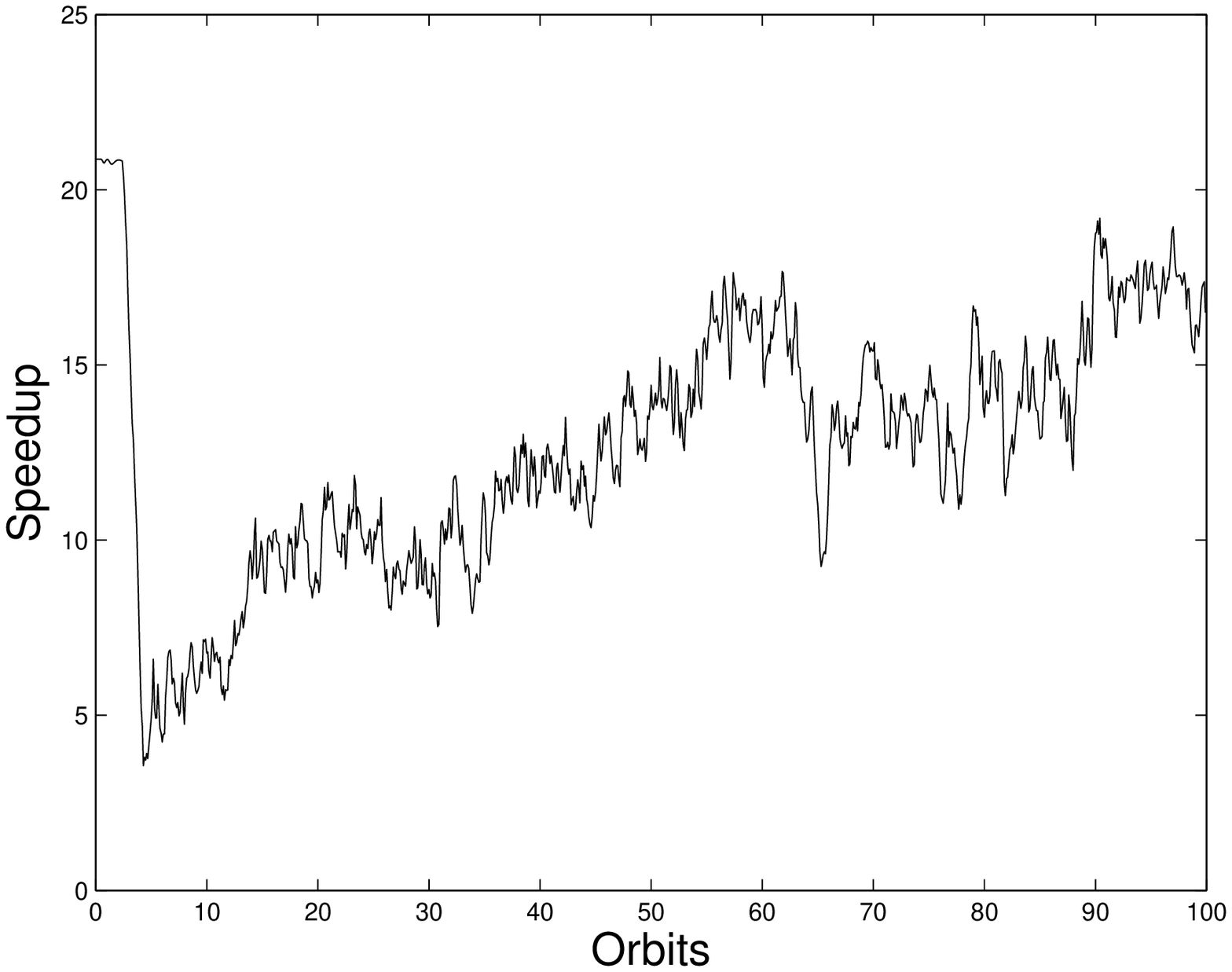}}                
  \caption{Testing the accuracy and performance of orbital advection.}
  \label{fig:oa}
\end{figure}

To measure the instability of these systems, we consider a radial scaling of the Reynolds stress normalized to the gas pressure given by,
\begin{equation}
\myfrac{<R\rho v_{R} v_{\phi}'>}{<RP>} = \myfrac{ \int R\rho v_{R}(v_{\phi}-v_{K})dV}{c_{s}^{2} \int R \rho dV },
\label{eqn:sstress}
\end{equation}
where the integrals are performed over the entire domain.  The time evolution of this quantity for all simulations is shown in Figure~\ref{fig:oaa}.  As expected, there is a clear difference in the evolution of the stress depending on Rayleigh's criterion.  For $q>2$ (unstable) we see that the scaled stress grows by many orders of magnitude, whereas in the case of $q<2$ (stable) we see that the scaled stress remains fairly flat.
\begin{table}
\begin{center}
\begin{tabular}{|c|c|c|}
  \hline
  $q$ & Cylindrical Integrator & Orbital Advection\\
  & (Cycles : Wall-time) & (Cycles : Wall-time) \\
  \hline
  $q = 1.5$ & $117.7 : 5.3$h & $3.1 : 13.2$m \\  
  $q = 1.95$ & $72.9 : 4.7$h & $3.2 : 11.7$m \\
  $q = 1.99$ & $69.9 : 4.6$h & $3.2 : 12$m \\
  $q = 2.01$ & $68.6 : 4.5$h & $4.5 : 17.4$m \\
  $q = 2.05$ & $65.4 : 4.9$h & $9.6 : 35.8$m \\
  \hline
\end{tabular}
\caption{Comparison of the performance of the cylindrical integrator versus orbital advection.  Specifically shown are cycles (in thousands) and wall-time (in h[ours] and m[inutes]).  The simulations were run on Ranger using 64 processors and an MPI topology of $(T_{R},T_{\phi},T_{z}) = (8,4,2)$.}
\label{tab:ray}
\end{center}
\end{table}

\subsection{Growth and Saturation of the MRI}

The final test we consider is perhaps the most relevant.  Here we compare the behavior of two identical Newtonian MRI simulations using both orbital advection and the standard cylindrical integrator.  We compare the results of run \zz{8} with and without orbital advection.  Figure~\ref{fig:oab} illustrates the time evolution of $\alpha_{M}$ in both simulations.  We note that the initial growth and saturation of the MRI are in excellent agreement.  Further, the agreement is quite strong well into the highly-nonlinear regime.  At late times there is a discrepancy between the cylindrical and orbital advection run, with the cylindrical run exhibiting slightly larger values of stress.  This resembles the discrepancy in the behavior of runs \zz{32} and \zzr{32}{R}, in which a reduction of azimuthal resolution results in an increased stress.  This resemblance is not entirely surprising as, in many ways, the use of orbital advection allows a more precise treatment of azimuthal structure.

Finally, and perhaps most importantly, we discuss the quantitative benefit of orbital advection.  To address this point we compare the timesteps used in the evolution of each simulation.  The timesteps are sampled ten times per orbit, and denoted $\Delta t_{C}$ and $\Delta t_{F}$ for the cylindrical and orbital advection runs, respectively.  We define the speedup as
\begin{equation}
S(t) = \frac{\Delta t_{F}}{\Delta t_{C}} .
\end{equation}             

The time dependance of the speedup is shown in Figure~\ref{fig:oac}.  The timestep associated with orbital advection is significantly more volatile than without, as the timestep is associated with turbulent quantities as opposed to the time-invariant Keplerian rotation.  In particular, the timestep drops significantly during the initial linear growth of the MRI due to the impulsive accretion associated with exponential growth phase.  Upon the saturation and breakup into turbulence, the timestep increases again steadily and overall maintains an order of magnitude increase in the allowable timestep.     
 
\bibliographystyle{apj}
\bibliography{GlobalI}

\begin{thebibliography}{28}
\expandafter\ifx\csname natexlab\endcsname\relax\def\natexlab#1{#1}\fi

\bibitem[{Armitage(1998)}]{armitage98}
Armitage, P.~J. 1998, Astrophysical Journal Letters v.501, 501, L189

\bibitem[{Balbus \& Hawley(1991)}]{mri1}
Balbus, S.~A., \& Hawley, J.~F. 1991, Astrophysical Journal, 376, 214

\bibitem[{Balbus \& Hawley(1998)}]{bh98}
---. 1998, Reviews of Modern Physics, 70, 1

\bibitem[{Beckwith {et~al.}(2011)Beckwith, Armitage, \& Simon}]{bas11}
Beckwith, K., Armitage, P.~J., \& Simon, J.~B. 2011, arXiv.org, 1105, 1789, 24
  pages and 25 figures. MNRAS in press. Version with high resolution figures
  available from
  http://jila.colorado.edu/~krb3u/Thin{\_}Disk/thin{\_}disk{\_}turbulence.pdf

\bibitem[{Blackman {et~al.}(2008)Blackman, Penna, \&
  Varni{\`e}re}]{Blackman:2008fe}
Blackman, E.~G., Penna, R.~F., \& Varni{\`e}re, P. 2008, New Astronomy, 13, 244

\bibitem[{Brandenburg {et~al.}(1995)Brandenburg, Nordlund, Stein, \&
  Torkelsson}]{brandenburg95}
Brandenburg, A., Nordlund, A., Stein, R.~F., \& Torkelsson, U. 1995, \apj, 446,
  741

\bibitem[{Davis {et~al.}(2010)Davis, Stone, \& Pessah}]{dsp10}
Davis, S.~W., Stone, J.~M., \& Pessah, M.~E. 2010, The Astrophysical Journal,
  713, 52

\bibitem[{Fromang(2010)}]{fromang10}
Fromang, S. 2010, Astronomy and Astrophysics, 514, L5

\bibitem[{Fromang \& Papaloizou(2007)}]{fp07}
Fromang, S., \& Papaloizou, J. 2007, Astronomy and Astrophysics, 476, 1113

\bibitem[{Goodman \& Xu(1994)}]{gx94}
Goodman, J., \& Xu, G. 1994, Astrophysical Journal, 432, 213

\bibitem[{Guan {et~al.}(2009)Guan, Gammie, Simon, \& Johnson}]{ggsj09}
Guan, X., Gammie, C.~F., Simon, J.~B., \& Johnson, B.~M. 2009, Astrophysical
  Journal, 694, 1010

\bibitem[{Hawley(2001)}]{hawley01}
Hawley, J.~F. 2001, The Astrophysical Journal, 554, 534

\bibitem[{Hawley {et~al.}(1995)Hawley, Gammie, \& Balbus}]{hgb}
Hawley, J.~F., Gammie, C.~F., \& Balbus, S.~A. 1995, Astrophysical Journal
  v.440, 440, 742

\bibitem[{Hawley {et~al.}(2011)Hawley, Guan, \& Krolik}]{hgk11}
Hawley, J.~F., Guan, X., \& Krolik, J.~H. 2011, arXiv.org, 1103, 5987,
  submitted to the Astrophysical Journal

\bibitem[{Johnson {et~al.}(2008)Johnson, Guan, \& Gammie}]{jgg08}
Johnson, B.~M., Guan, X., \& Gammie, C.~F. 2008, The Astrophysical Journal
  Supplement Series, 177, 373

\bibitem[{King {et~al.}(2007)King, Pringle, \& Livio}]{kpl07}
King, A.~R., Pringle, J.~E., \& Livio, M. 2007, Monthly Notices of the Royal
  Astronomical Society, 376, 1740

\bibitem[{Masset(2000)}]{fargo}
Masset, F. 2000, Astronomy and Astrophysics Supplement, 141, 165

\bibitem[{Noble {et~al.}(2010)Noble, Krolik, \& Hawley}]{nkh10}
Noble, S.~C., Krolik, J.~H., \& Hawley, J.~F. 2010, The Astrophysical Journal,
  711, 959

\bibitem[{Pessah {et~al.}(2007)Pessah, Chan, \& Psaltis}]{pcp07}
Pessah, M.~E., Chan, C.-k., \& Psaltis, D. 2007, The Astrophysical Journal,
  668, L51

\bibitem[{Pessah \& Goodman(2009)}]{pg09}
Pessah, M.~E., \& Goodman, J. 2009, The Astrophysical Journal Letters, 698, L72

\bibitem[{Pringle(1981)}]{pringle81}
Pringle, J.~E. 1981, In: Annual review of astronomy and astrophysics. Volume
  19. (A82-11551 02-90) Palo Alto, 19, 137, a{\&}AA ID. AAA030.064.057

\bibitem[{Sano {et~al.}(2004)Sano, Inutsuka, Turner, \& Stone}]{sits04}
Sano, T., Inutsuka, S.-i., Turner, N.~J., \& Stone, J.~M. 2004, The
  Astrophysical Journal, 605, 321

\bibitem[{Simon {et~al.}(2011)Simon, Hawley, \& Beckwith}]{shb11}
Simon, J.~B., Hawley, J.~F., \& Beckwith, K. 2011, The Astrophysical Journal,
  730, 94

\bibitem[{Skinner \& Ostriker(2010)}]{athcyl}
Skinner, M.~A., \& Ostriker, E.~C. 2010, The Astrophysical Journal Supplement,
  188, 290

\bibitem[{Sorathia {et~al.}(2010)Sorathia, Reynolds, \& Armitage}]{sra10}
Sorathia, K.~A., Reynolds, C.~S., \& Armitage, P.~J. 2010, Astrophysical
  Journal, 712, 1241

\bibitem[{Stone \& Gardiner(2010)}]{athfargo}
Stone, J.~M., \& Gardiner, T.~A. 2010, The Astrophysical Journal Supplement,
  189, 142

\bibitem[{Stone {et~al.}(2008)Stone, Gardiner, Teuben, Hawley, \&
  Simon}]{athena}
Stone, J.~M., Gardiner, T.~A., Teuben, P., Hawley, J.~F., \& Simon, J.~B. 2008,
  The Astrophysical Journal Supplement Series, 178, 137

\bibitem[{van Ballegooijen(1989)}]{vb89}
van Ballegooijen, A.~A. 1989, in IN: Accretion disks and magnetic fields in
  astrophysics; Proceedings of the European Physical Society Study Conference,
  Harvard-Smithsonian Center for Astrophysics, Cambridge, MA,
  Harvard-Smithsonian Center for Astrophysics, Cambridge, MA, 99--106

\end{thebibliography}
\end{document}